\documentclass[12pt]{article}
\pdfoutput=1
\usepackage{makeidx}
\makeindex
\usepackage[a4paper]{geometry}
\usepackage{jheppub,amsmath,amssymb,amsfonts,amsxtra,mathrsfs,graphics,graphicx,amsthm,epsfig,ytableau,bm,longtable,float,placeins,color,tikz,mathtools,xfrac,footnote,rotating,lscape}
\restylefloat{table}
\usepackage{dsfont}
\usepackage{multicol}
\usepackage{lipsum}
\usepackage{textgreek}
\usetikzlibrary{decorations.pathmorphing}
\usetikzlibrary{decorations.markings}
\usetikzlibrary{quotes,arrows.meta}
\usetikzlibrary{arrows, decorations.markings, calc, fadings, decorations.pathreplacing, patterns, decorations.pathmorphing, positioning}
\usepackage{tikz-cd}
\usepackage[inline]{enumitem}

\usepackage{fixmath} % for \mathbold
\usepackage{scalerel}
\newlength\bshft
\def\fakebold#1{\ThisStyle{\ooalign{$\SavedStyle#1$\cr%
  \kern-\bshft$\SavedStyle#1$\cr%
  \kern\bshft$\SavedStyle#1$}}}

\usetikzlibrary{positioning,shapes}
\usetikzlibrary{chains}
\usetikzlibrary{arrows,fit,decorations.pathreplacing}
\tikzstyle{every picture}+=[remember picture]
\tikzstyle{na} = [baseline=-.5ex]

\usepackage{empheq}
\usepackage{multirow}
\usepackage{booktabs}
\usepackage[american]{babel}

\usepackage[latin1]{inputenc}

\usepackage{array,booktabs}

\makeatletter
\newcommand{\vast}{\bBigg@{1}}
\newcommand{\Vast}{\bBigg@{5}}
\makeatother

\usepackage{hyperref}

\numberwithin{equation}{section}

\newcommand{\cf}{\textit{cf.}}
\newcommand{\eg}{\textit{e.g.}}

\newcommand{\etal}{\textit{et al.}}
\newcommand{\ie}{\textit{i.e.}}

\newcommand{\ii}{\mathrm{i}}
\newcommand{\?}{\;\!}

\numberwithin{equation}{section}

\newcommand{\be}{\begin{equation}} \newcommand{\ee}{\end{equation}}
\newcommand{\bea}{\begin{equation} \begin{aligned}} \newcommand{\eea}{\end{aligned} \end{equation}}

\def\U{\mathrm{U}}
\def\SO{\mathrm{SO}}

\def\SU{\mathrm{SU}}

\def\USp{\mathrm{USp}}

\newcommand{\rd}{\mathrm{d}}

\newcommand{\vol}{\mathrm{vol}}

\DeclareMathOperator{\Tr}{Tr}

\DeclareMathOperator{\sign}{sign}

\DeclareMathOperator{\re}{\mathbb{R}e}
\DeclareMathOperator{\im}{\mathbb{I}m}

\DeclareMathOperator{\Li}{Li}

\newcommand{\pd}{\partial}

\newcommand{\cA}{\mathcal{A}}
\newcommand{\cB}{\mathcal{B}}
\newcommand{\cC}{\mathcal{C}}

\newcommand{\cF}{\mathcal{F}}

\newcommand{\cH}{\mathcal{H}}
\newcommand{\cI}{\mathcal{I}}
\newcommand{\cJ}{\mathcal{J}}

\newcommand{\cM}{\mathcal{M}}
\newcommand{\cN}{\mathcal{N}}
\newcommand{\cO}{\mathcal{O}}
\newcommand{\cP}{\mathcal{P}}

\newcommand{\cV}{\mathcal{V}}
\newcommand{\cW}{\mathcal{W}}

\newcommand{\cZ}{\mathcal{Z}}

\newcommand{\bC}{\mathbb{C}}

\newcommand{\bR}{\mathbb{R}}
\newcommand{\bZ}{\mathbb{Z}}

\newcommand{\fg}{\mathfrak{g}}

\newcommand{\fh}{\mathfrak{h}}

\newcommand{\fm}{\mathfrak{m}}

\newcommand{\fn}{\mathfrak{n}}

\newcommand{\fp}{\mathfrak{p}}

\newcommand{\fR}{\mathfrak{R}}
\newcommand{\fs}{\mathfrak{s}}
\newcommand{\ft}{\mathfrak{t}}

\usepackage[Symbol]{upgreek}
\usepackage{bm}

\usepackage{calligra}
\DeclareMathAlphabet{\mathcalligra}{T1}{calligra}{m}{n}

\newcommand{\mathbbm}[1]{\text{\usefont{U}{bbm}{m}{n}#1}}

\theoremstyle{plain}

  \theoremstyle{definition}

\providecommand{\examplename}{Example}
\providecommand{\theoremname}{Theorem}

\makeatletter
\g@addto@macro\bfseries{\boldmath}
\makeatother

\makeatletter
\newcommand*{\rom}[1]{\expandafter\@slowromancap\romannumeral #1@}
\makeatother

\title{The joy of factorization at large $N$: five-dimensional indices and AdS black holes}

\author[a,b]{Seyed Morteza Hosseini,}
\author[c]{Itamar Yaakov}
\author[c,d]{and Alberto Zaffaroni}
\affiliation[a]{Department of Physics, Imperial College London, London, SW7 2AZ, UK}
\affiliation[b]{Kavli IPMU (WPI), UTIAS, The University of Tokyo, Kashiwa, Chiba 277-8583, Japan}
\affiliation[c]{INFN, sezione di Milano-Bicocca, I-20126 Milano, Italy}
\affiliation[d]{Dipartimento di Fisica, Universit\`a di Milano-Bicocca, I-20126 Milano, Italy}
\emailAdd{s.hosseini@imperial.ac.uk}
\emailAdd{itamar.yaakov@mib.infn.it}
\emailAdd{alberto.zaffaroni@mib.infn.it}
\abstract{We discuss the large $N$ factorization properties of five-dimensional supersymmetric partition functions for CFT with a holographic dual. We consider partition functions on manifolds of the form $\mathcal{M}= \mathcal{M}_3 \times S^2_\epsilon$, where $\epsilon$ is an equivariant parameter for rotation. We show that, when $\mathcal{M}_3$ is a squashed three-sphere, the large $N$ partition functions can be obtained by gluing elementary blocks associated with simple physical quantities. The same is true for various observables of the theories on $\mathcal{M}_3=\Sigma_\mathfrak{g} \times S^1$, where $\Sigma_\mathfrak{g}$ is a Riemann surface of genus $\mathfrak{g}$, and, with a natural assumption on the form of the saddle point, also for the partition function, corresponding to either the topologically twisted index or a mixed one. This generalizes results in three and four dimensions and correctly reproduces the entropy of known black objects in AdS$_6 \times_{w} S^4$ and AdS$_7\times S^4$. We also provide the supersymmetric background and explicitly perform localization for the mixed index on $\Sigma_\mathfrak{g} \times S^1 \times S^2_\epsilon$, filling a gap in the literature.
}

\begin{document}

\setcounter{tocdepth}{2}
\maketitle

%*******************************************************************************
%
%     Main body
%
%*******************************************************************************

\date{Dated: \today}

% \hypersetup{
% colorlinks,breaklinks,
%             linkcolor=black
% }

% \tableofcontents

% \hypersetup{
% colorlinks,breaklinks,
%             linkcolor=[rgb]{0,0,0.7}
% }

\section{Introduction}
\label{sec:intro}

There has been recently a lot of progress in the microscopic derivation of the entropy of
anti de Sitter (AdS) black holes, starting with the magnetically charged and topologically twisted ones in AdS$_4\times S^7$ \cite{Benini:2015eyy}
and followed by the Kerr-Newman (KN) black holes in AdS$_5\times S^5$ \cite{Choi:2018hmj,Cabo-Bizet:2018ehj,Benini:2018ywd}. This analysis has been extended to many other black objects in AdS$_d$, with $d=4,5,6,7$
with different types of electromagnetic charges and rotations and with various amounts of supersymmetry. For a (partial) review, see \cite{Zaffaroni:2019dhb}.
A general entropy functional that captures the large $N$ entropy of black holes
and black strings in general AdS compactifications with a holographic dual was written in \cite{Hosseini:2019iad}.
In this picture, the entropy functional is  a sum of universal contributions called gravitational blocks,
which can be related to some simple field theory quantity, either the central charge  or the sphere-partition function of the dual conformal field theory (CFT).
The proposal has been  successfully tested for all known examples in maximally supersymmetric
compactifications and in many other cases with less supersymmetry.
The microscopic counting of states for black objects in AdS is usually done by a large $N$ saddle point  analysis
of a supersymmetric index in the dual CFT. Black hole physics then suggests that, in the large $N$ limit,  the corresponding field theory partition functions should factorize  and the CFT free energy 
\be
 F_{\cM} \equiv - \log Z_{\cM} \, ,
\ee
where $\cM$ is the Euclidean boundary of the black object, should  be the sum of universal contributions associated to a geometric description of $\cM$ as the gluing of elementary pieces.
The form  of the entropy functional in \cite{Hosseini:2019iad} was indeed suggested by the decomposition  of the supersymmetric partition functions in holomorphic blocks \cite{Beem:2012mb}.%
\footnote{The idea of ``gluing'' or ``sewing'' building blocks to compose field theory observables is old and it has been successfully used in many different contexts. See \cite{Nekrasov:2003vi,Pasquetti:2011fj,Hwang:2012jh,Imamura:2013qxa,Yoshida:2014ssa,Hwang:2015wna,Benini:2015noa,Nieri:2015yia,Pasquetti:2016dyl,Gukov:2017kmk,Closset:2018ghr,Festuccia:2018rew} for  developments related to our context.} 
In this paper, we call this property {\it large $N$ factorization}. For a more precise definition see below and the beginning of section \ref{sec:2}.

All recent field theory computations for black objects in AdS$_4$ and AdS$_5$ have confirmed  the above picture. The factorization properties are particularly nontrivial 
when magnetic charges and rotation are simultaneously present, see for example  \cite{Hosseini:2019lkt}   for rotating black strings in AdS$_5\times S^5$, and \cite{Choi:2019dfu}  for a class of  rotating black holes in AdS$_4$.

In this paper we analyse the large $N$ factorization properties of five-dimensional partition functions on various manifolds of the form $\cM= \cM_3 \times S^2_\epsilon$, where $\epsilon$ is an equivariant parameter for rotations along $S^2$ and  a chemical potential for the angular momentum of the dual black hole. We will consider three examples. The first   is the partition function on $S^3_{b} \times S^2_{\epsilon}$, with a squashing along $S^3$ and a topological twist along $S^2$ \cite{Crichigno:2018adf}. Although there is no associated dual black object, we expect factorization to hold nevertheless. The other two examples correspond to partition functions on $(S^2_\epsilon \times S^1)\times \Sigma_\fg$, with a topological twist along the genus $\fg$ Riemann surface $\Sigma_\fg$, and a different type of supersymmetry. When $S^2$ is twisted this is the (partially) refined five-dimensional topologicallly twisted index discussed in \cite{Hosseini:2018uzp,Crichigno:2018adf}. The case where $S^2$ is not twisted corresponds to  a mixed index, first considered in \cite{Jain:2021sdp}.

All these partition functions can be written, for genus $0$, by gluing copies of the  Nekrasov's partition functions \cite{Nekrasov:2002qd}, in the spirit of  \cite{Nekrasov:2003vi}. Since a proper derivation of the mixed index is lacking in the literature we  provide it here, by identifying the supergravity background the five-dimensional field theory should be coupled to in order to preserve supersymmetry and by performing an explicit localization around this background.

For simplicity, we will focus on two specific field theories,  the so-called $E_{N_f+1}$ Seiberg theory, the UV fixed point of a $\cN = 1$ $\USp(2N)$ gauge theory coupled to $N_f$ fundamental hypermultiplets and an antisymmetric one  \cite{Seiberg:1996bd}, and the $\cN = 2$ $\SU(N)$ super Yang-Mills (SYM) theory,
which is supposed to flow at strong coupling to the six-dimensional $\cN = (2,0)$ theory of type $A_{N-1}$.
The two theories are holographically dual to the warped AdS$_6\times_{w} S^4$ solution in massive type IIA \cite{Brandhuber:1999np},
and to AdS$_7\times S^4$ in M-theory, respectively.
Black holes and black strings in these compactifications have been found in
\cite{Benini:2013cda,Benini:2015bwz,Hosseini:2018usu,Suh:2018tul,Suh:2018szn,Hosseini:2020vgl,Hosseini:2020wag}
and all the corresponding entropy functionals  satisfy factorization. In this paper we investigate the factorization properties from the field theory point of view. We will verify that
various quantum field theory observables can be written as a sum over two contributions associated with the two hemispheres of $S^2$
\be
 \label{fact5d}
\sum_{\sigma=1}^2 \cB(\Delta^{(\sigma)}, \epsilon^{(\sigma)})= \frac{ \cF \left( \Delta_i + \frac{\epsilon}2 \ft_i \right)}{\epsilon} \pm  \frac{ \cF \left( \Delta_i - \frac{\epsilon}2 \ft_i \right)}{\epsilon} \, ,
\ee
where $\Delta_i$, $i=1,2$   are (constrained) chemical potentials for a $\U(1)$ flavor symmetry associated with a rotation on $S^4$, $\ft_i$ are the corresponding (constrained) flavor magnetic fluxes on the sphere $S^2$, and $\cF$ has a natural interpretation in terms of the central charge/sphere partition function of the dual CFT.  Factorization takes the simple form \eqref{fact5d} only when written in terms  of constrained variables. If there is a topological twist on $S^2$, corresponding to dual magnetically charged and twisted black holes, we need to use the minus sign in \eqref{fact5d} and the constraints\footnote{In some examples, $S^3_{b} \times S^2_{\epsilon}$ for instance, we will normalize the first constraint as   $\Delta_1+ \Delta_2 =2$ to facilitate comparison with literature.}
\be \Delta_1+ \Delta_2 =2 \pi \, ,\qquad \ft_1+\ft_2=2 \, , \ee
while, if $S^2$ is untwisted, corresponding to dyonic KN black holes, we need to use the plus sign in \eqref{fact5d} and the constraints
\be \Delta_1+ \Delta_2 =2 \pi + \epsilon \, ,\qquad \ft_1+\ft_2=0 \, . \ee
The expression \eqref{fact5d} is clearly reminiscent of an equivariant  localization formula.

In the various examples $\cF$ is proportional to%
\footnote{For the explicit formulae and normalization factors see  \eqref{FS3:USp(2N)} and \eqref{FS3:SYM}, for $S^3_{b} \times S^2_{\epsilon}$,
\eqref{RTTI:final:USp(2N)} and \eqref{RTTI:sym} for the twisted index, and \eqref{SCI:final:USp(2N)} and \eqref{SCI:fina:sym} for the mixed one.} 
\begin{table}[H]
 % \vskip 0.5truecm
\label{freeenergyN}
\centering
\begin{tabular}{lc l  c | c | }   \noalign{\smallskip} 
& $S^3_{b} \times S^2_{\epsilon}$ & $(S^2_\epsilon \times S^1)\times \Sigma_\fg$ \\  \noalign{\smallskip}
\toprule
 \hskip -0.0truecm $E_{N_f+1}$  Seiberg theory & \hskip -0.0truecm $F_{S^5} ( \Delta)$  & \hskip -0.0truecm $\displaystyle \sum_{i=1}^2 \fs_i  \frac{ \partial F_{S^5} ( \Delta)}{\partial \Delta_i}$ \\ \noalign{\smallskip}
  \noalign{\smallskip}  
  \hskip -0.0truecm 6d $\cN = (2,0)$ theory  & \hskip -0.0truecm  $\cA_{6\rd}(\Delta)$ & \hskip -0.0truecm $\displaystyle \sum_{i=1}^2 \fs_i  \frac{ \partial \cA_{6\rd}(\Delta) }{\partial \Delta_i}$ \\\noalign{\smallskip}
 \toprule \noalign{\smallskip} 
 
\end{tabular}
\end{table}

\noindent where $\fs_i$, normalized with $\fs_1+\fs_2= 2 - 2 \fg$, are the magnetic fluxes through the Riemann surface $\Sigma_\fg$.
The quantities in the table have the following interpretations.
\be
 \label{F1}
 F_{S^5} ( \Delta_i ) = - \frac{9 \sqrt{2}}{5 \pi^2} \frac{N^{5/2}}{\sqrt{8 - N_f}} (\Delta_1 \Delta_2)^{\frac{3}{2}} \, ,
\ee
is the large $N$ five-sphere free energy of the Seiberg theory in a convenient parameterization, see \cite[(3.38)]{Chang:2017mxc} and \cite[(2.37)]{Hosseini:2019and}, and 
\be\label{F2}
 \cA_{6\rd}(\Delta) = \frac{N^3}{24} ( \Delta_1 \Delta_2 )^2 \, ,
\ee
can be read off from  the anomaly polynomial of the $\cN =(2,0)$ theory at large $N$, in a way that we will discuss. The expressions  for  $(S^2_\epsilon \times S^1)\times \Sigma_\fg$, on the other hand,  can be interpreted as the $S^3$ partition function of the three-dimensional CFT obtained by compactifying the Seiberg theory on $\Sigma_\fg$, and the trial central charge of the four-dimensional CFT obtained by compactifying  the $\cN=(2,0)$ theory on $\Sigma_\fg$, respectively.

A word of caution is in order. While the  evaluation of  the partition functions on $S^3_{b} \times S^2_{\epsilon}$ in the large $N$ limit can be done
with stardard methods, that for $(S^2_\epsilon \times S^1)\times \Sigma_\fg$ is harder and we will not be able do it from first principles.
For $\epsilon=0$, it was conjectured in  \cite{Hosseini:2018uzp,Crichigno:2018adf} that the twisted index localizes at the critical points of an effective
twisted superpotential $\cW_{(S^2 \times S^1) \times \bR^2}$, similarly to the three-dimensional case \cite{Benini:2015eyy}.
Under this working assumption, the entropy of static black objects in AdS$_6$ and AdS$_7$ was correctly reproduced
\cite{Hosseini:2018uzp,Hosseini:2018usu,Suh:2018tul,Suh:2018szn}.
We will use the same logic here. A similar analysis for the twisted and mixed partition functions on $(S^2_\epsilon \times S^1)\times \Sigma_\fg$
was first done in \cite{Jain:2021sdp} for the Seiberg theory, with the somehow surprising conclusion that factorization holds only in the $\epsilon=0$ limit.
This is in contradiction with the results for rotating twisted black objects that exhibit factorization \cite{Hosseini:2020vgl,Hosseini:2020wag}
and we reconsider the problem here. As we will discuss in detail, the dependence of $\cW_{(S^2_\epsilon \times S^1) \times \bR^2}$ on
gauge magnetic fluxes can be modified by contact terms for $\epsilon\ne 0$.
We will show that, both in the case of twisted and mixed index, there is a {\it natural} definition of an effective twisted superpotential
$\cW_{(S^2_\epsilon \times S^1) \times \bR^2}$ for $\epsilon\ne 0$ that leads to factorization.%
\footnote{Our twisted superpotential is then different from the one used in  \cite{Jain:2021sdp}.}
If we assume that the partition functions localize at the critical points of $\cW$ we find the free energy quoted above.
As we will show, this result correctly reproduces the entropy of all known black objects in AdS$_6$ and AdS$_7$.
It would be nice to have a first principle derivation of $\log Z_\cM$ that does not involve the assumptions made in \cite{Hosseini:2018uzp,Crichigno:2018adf,Jain:2021sdp}.

One other interesting result of our analysis is  that the on-shell superpotential  $\cW_{(S^2_\epsilon \times S^1) \times \bR^2}$ itself factorizes with  blocks given   by \eqref{F1}  or \eqref{F2}.\footnote{See \eqref{W:RTTI:USp(2N)} and \eqref{W:RTTI:SYM:factorized} for the twisted index, and \eqref{W:SCI:USp(2N)} and \eqref{W:SCI:SYM:factorized} for the mixed one.} Indeed, we have the interesting relation
\be
 \label{index}
 \log Z_{(S^2_\epsilon \times S^1)\times \Sigma_g} (\Delta, \ft, \epsilon, \fs) =
 \ii \sum_{i=1}^2 \fs_i  \frac{ \partial \cW_{(S^2_\epsilon \times S^1) \times \bR^2}(\Delta, \ft, \epsilon)}{\partial \Delta_i} \, ,
\ee
which is the five-dimensional generalization of the index theorem  proved in \cite{Hosseini:2016tor,Hosseini:2016cyf}.

The paper is organized as follows. In section \ref{sec:2},  we will discuss general facts about factorization and compare results in different dimensions. This will also serve to  fix notations and a common ground for the rest of the paper.  In section \ref{sec:3}, we discuss the partition function on $S^3_{b} \times S^2_{\epsilon}$ and we show that it factorizes. In section \ref{sec:4} we discuss the refined topologically twisted index. We introduce various equivalent definitions of the effective twisted superpotential $\cW_{(S^2_\epsilon \times S^1) \times \bR^2}$, and discuss the ambiguity in these definitions. We then show that, for a natural symmetric choice of $\cW$, its on-shell value is factorized. Moreover, we will verify that, if the partition function localizes at the critical points of $\cW$, it also has a factorized form. We will also show that this result correctly reproduces the entropy of the known rotating twisted black holes in massive IIA \cite{Hosseini:2020wag} and the density of states of black strings in AdS$_7 \times S^4$ \cite{Hosseini:2020vgl}. In section \ref{sec:5}, we first write the rigid supergravity background for the mixed index, and then explicitly perform localization.  Using  the same assumptions as in section \ref{sec:4}, we show that both the on-shell twisted superpotential and the partition function factorizes. We  also verify that this result correctly reproduces the entropy of a class of KN  black holes in massive IIA \cite{Hosseini:2020wag}. In section \ref{sect:index theorem} we shed light on the relation between the on-shell $\cW$ and $\log Z$ for the twisted and mixed index by proving the generalization of the index theorem discussed in \cite{Hosseini:2016tor}. We conclude in section \ref{sect:conclusions} with a discussion and open problems.

\section{Generalities about factorization}\label{sec:2} 

In this section we discuss some general facts about factorization in three dimensions, that we will use also in five dimensions.

\subsection{The entropy functional}
In the recent and successful approach to microscopic counting, the entropy of supersymmetric AdS black holes is obtained by extremizing the {\it entropy functional} 
\be
 \label{ef}
 \cI ( \Delta_i , \epsilon_a ) \equiv \log Z_\cM(\Delta_i, \epsilon_a) - \ii \sum_{i} \Delta_i Q_i - \ii \sum_a \epsilon_a J_a \, ,
\ee
where  $Q_i$ are the electric charges and $J_a$ the angular momenta.
Here, $Z_\cM(\Delta_i, \epsilon_a)$ is a supersymmetric partition function on the compact manifold $\cM$ that depends on a set of chemical potentials $\Delta_i$ and $\epsilon_a$ conjugate to $Q_i$ and $J_a$, respectively. Extra conserved charges of the black hole, for example magnetic charges, are encoded in the explicit form of the function $Z_\cM(\Delta_i, \epsilon_a)$. The entropy functional \eqref{ef} should be extremized with respect to $\Delta_i$ and $\epsilon_a$. This is the familiar fact that the entropy at zero temperature can be obtained by taking the Legendre transform of the partition function. 

For all known black holes in AdS, supersymmetry imposes a constraint  among the electric charges and the angular momenta. There is a similar linear constraint on the magnetic charges, if present.
It is however convenient to include {\it all}  possible electric charges and  angular momenta in \eqref{ef} and perform a constrained extremization.  In this picture, the variables $\Delta_i$ and $\epsilon_a$  in \eqref{ef} satisfy a linear constraint.  This is consistent with the fact that supersymmetric indices can be only refined with fugacities for symmetries that commute with a particular  supercharge  and it allows for a complete field theory description.    We could obviously solve explicitly the constraint and write the entropy functional in terms of independent variables. However, \eqref{ef} takes a simple form only when written in terms of constrained variables. In particular, in all known cases, it is a homogeneous function of degree one of the constrained $\Delta_i$ and $\epsilon_a$.  

For all known KN or topologically twisted black holes in maximally supersymmetric AdS compactifications, with or without magnetic charges or rotation, the  function $Z_\cM(\Delta_i, \epsilon_a)$ in \eqref{ef} can be written as a sum of contributions from the \emph{gravitational blocks} \cite{Hosseini:2019iad}
\be
 \label{logB}
 \log Z_\cM(\Delta_i, \epsilon_a) =  \sum_\sigma \cB (\Delta_i^{(\sigma)}, \epsilon_a^{(\sigma)}) \, ,
\ee
where
\be
 \label{log}
 \cB (\Delta_i^{(\sigma)}, \epsilon_a^{(\sigma)}) = -  \frac{\cF (\Delta_i^{(\sigma)}) }{\prod_a \epsilon_a^{(\sigma)}} \, ,
\ee
and each $\epsilon_a^{(\sigma)}$ is a linear combination of the rotational chemical potentials $\epsilon_a$ while
$\Delta_i^{(\sigma)} = \Delta_i \pm \ii \fp_i \epsilon^{(\sigma)}$, where $\fp_i$ are magnetic fluxes through $\cM$.\footnote{One can verify that a straightforward generalization covers also spindle black objects \cite{Ferrero:2020laf,Ferrero:2020twa,Hosseini:2021fge,Boido:2021szx,Ferrero:2021wvk}, see \cite{Hosseini:2021fge} for an explicit example.}
Experimentally, $\sigma$ runs over the elementary pieces into which the boundary manifold $\cM$ is decomposed,
in the factorization of the partition function in holomorphic blocks \cite{Beem:2012mb,Pasquetti:2016dyl}. Each black hole corresponds to a different gluing. The function $\cF(\Delta_i)$ is instead universal, related to the prepotential or on-shell action of the relevant supergravity, or, more physically, to the large $N$ limit of the central charge of the dual field theory, in the case of  even-dimensional CFTs,  or of the sphere free-energy for odd-dimensional ones, fully refined with respect to the global symmetries. For the maximal supersymmetric compactifications, $\cF(\Delta)$ is proportional to the values given in table \ref{MagEntropy}.
\begin{table}[H]
  \caption{The structure of the block for the AdS backgrounds with maximal allowed supersymmetry in each dimension. AdS$_6\times_w S^4$ is the background dual to the $\USp(2 N)$ five-dimensional theory considered in this paper \cite{Seiberg:1996bd,Brandhuber:1999np}. The chemical potentials $\Delta_i$ are associated with the Cartan of the internal sphere isometry in all dimensions.}
%  \vskip 0.5truecm
\label{MagEntropy}
\centering
\begin{tabular}{lc l|  c }  \toprule \noalign{\smallskip}
 \hskip -0.0truecm AdS$_4\times S^7$ & \hskip -0.0truecm $\cF(\Delta_a)\propto  N^{3/2}\sqrt{\Delta_1\Delta_2\Delta_3\Delta_4}$  \\ \noalign{\smallskip}
 \toprule \noalign{\smallskip}  
  \hskip -0.0truecm AdS$_5\times S^5$  & \hskip -0.0truecm  $\cF(\Delta_a)\propto N^2 \Delta_1\Delta_2\Delta_3$  \\\noalign{\smallskip}
 \toprule \noalign{\smallskip} 
 \hskip -0.0truecm  AdS$_6\times_w S^4$  & \hskip -0.0truecm $\cF(\Delta_a)\propto N^{5/3}(\Delta_1\Delta_2)^{3/2}$  \\
 \noalign{\smallskip}
 \toprule \noalign{\smallskip} 
  \hskip -0.0truecm AdS$_7\times S^4$  & \hskip -0.0truecm $\cF(\Delta_a)\propto N^3 (\Delta_1\Delta_2)^{2}$ \\
 \noalign{\smallskip}
 
 \toprule
\end{tabular}
\end{table}

Let us look at an example which will be also useful in the future.
Consider rotating four-dimensional black holes with a spherical horizon.
We can decompose the sphere into two hemispheres. We then use the \emph{$A$-gluing} 
for topologically twisted black holes 
\bea\label{Agluing}
 \Delta_i^{(1)} & =\Delta_i +\frac{ \epsilon}{2} \? \fp_i \, , \qquad && \epsilon^{(1)} = \epsilon \, , \\
\Delta_i^{(2)} & = \Delta_i - \frac{ \epsilon}{2}\? \fp_i \, , && \epsilon^{(2)} = - \epsilon  \, ,
\eea
associated with constraints of the form
\be
 \sum_{i = 1}^4 \Delta_i = 2 \, , \qquad \sum_{i = 1}^4 \fp_i=2\, ,
\ee
and  the \emph{identity gluing} (\emph{id}-gluing) 
\bea\label{identitygluing}
\Delta_i^{(1)} & = \Delta_i +\frac{ \epsilon}{2} \? \fp_i\, , \qquad && \epsilon^{(1)} = \epsilon \, , \\
\Delta_i^{(2)} & = \Delta_i -\frac{ \epsilon}{2} \? \fp_i \, , && \epsilon^{(2)} = \epsilon  \, ,
\eea
associated with constraints of the form
\be
 \sum_{i = 1}^4 \Delta_i -  \epsilon= 2 \pi\, , \qquad \sum_{i = 1}^4 \fp_i=0\, ,
\ee
for KN ones, including dyonic ones. The chemical potentials $\Delta_i$ are conjugate to the $\U(1)^4 \subset \SO(8)$ isometry for AdS$_4\times S^7$ and to a basis of $\U(1)$ symmetries $R_i$ in which we can decompose  the general R-symmetry of the model viewed as a generic $\cN=2$ theory. The constraints on magnetic charges are dictated by supersymmetry: the R-symmetry magnetic flux must be 2 for the topological twist and zero for KN black holes.

There are two cases where the entropy functional simplifies. The first is the case of KN black holes in AdS$_d$, $d=4,5,6,7$  with no magnetic charges. The corresponding entropy functional for the maximal supersymmetric compactifications has been written in \cite{Hosseini:2017mds,Hosseini:2018dob,Choi:2018fdc}. With zero magnetic fluxes, the gluing formula degenerates and the entropy functional is the sum of equal contributions, hiding the factorization properties. These become manifest for dyonic black holes \cite{Hristov:2019mqp,Hosseini:2019iad}.

The second is the case of black objects topologically twisted on $S^2$ but with zero angular momentum on the sphere.
The solution depends on magnetic fluxes $\fs_i$ on $S^2$ normalized as $\sum_i \fs_i =2$.
We assume that the boundary manifold $\cM=\cN \times S^2$ factorizes into two blocks of the form $\cN \times D_2$,
where the disks $D_2$ correspond to the two hemispheres. Here, we have $\epsilon=0$.
We can still  use the $A$-gluing defined above, and send $\epsilon$ to zero at the end of the computation.
We obtain
\be
 \label{dimredfree}
 \log Z_\cM (\Delta) = \sum_i \fs_i \frac{\partial \cF(\Delta)}{\partial \Delta_i} \, .
\ee
This is indeed the general structure of the unrefined topologically twisted index at large $N$ in a variety of situations \cite{Benini:2015eyy,Hosseini:2016tor,Hosseini:2016cyf,Hosseini:2018uzp}.
The sphere $S^2$ can be  replaced with a Riemann surface $\Sigma_\fg$ of genus $\fg$ with no major changes and the only difference that now  integer fluxes are normalized as
\be \sum_i \fs_i = 2 - 2 \fg \, .\ee

In the interesting case where the compactification of the original CFT  on the Riemann surface $\Sigma_\fg$ becomes conformal in the IR,%
\footnote{Holographically, there exists a domain wall interpolating between AdS$_{p+1}$ and AdS$_{p-1}\times \Sigma_\fg$, where $p$ is the dimensionality of the original CFT.}
\eqref{dimredfree} will express the large $N$ sphere partition function/central charge
of the lower-dimensional CFT in terms of those of the higher-dimensional one.
This relation is for example satisfied for the large $N$ sphere partition function of
the twisted compactification $\Sigma_\fg$ of the $E_{N_f+1}$ Seiberg theory   \cite{Bah:2018lyv,Hosseini:2018uzp}.
It is also a general relation between the large $N$ central charges of theories related by twisted compactifications,
as it can be easily proved by integrating the anomaly polynomial  on $\Sigma_\fg$ \cite{Hosseini:2016cyf,Hosseini:2018uzp,Hosseini:2019use}.

The factorization properties are really nontrivial when magnetic charges and rotation are simultaneously present.
We will focus on this case in the following and in the rest of the paper.

\subsection{Factorization in field theory}
The gravitational block picture  is expected to be a consequence of  analogous factorization properties of the quantum field theory observables. Let us consider first three dimensions. 
Most of the three-dimensional supersymmetric partition functions, and in particular the topologically twisted index and the superconformal one, can be written by gluing two holomorphic blocks $B(a, \Delta, \epsilon)$  according to the formula \cite{Beem:2012mb}
\be
 \label{Zholfac}
 Z(\Delta, \epsilon) = \int \rd a \,  B ( a^{(1)},  \Delta^{(1)}, \epsilon^{(1)} ) B ( a^{(2)}, \Delta^{(2)}, \epsilon^{(2)} ) \, ,
\ee
where $a^{(\sigma)}$, $\sigma = 1,2,$ are gauge fugacities. Another useful expression is  \cite{Beem:2012mb,Pasquetti:2016dyl}
\be
 \label{Zhol}
 Z(\Delta, \epsilon) = \sum_\alpha   B^\alpha ( \Delta^{(1)}, \epsilon^{(1)} ) B^\alpha ( \Delta^{(2)} , \epsilon^{(2)} ) \, ,
\ee
where $\alpha$ labels a choice of Bethe vacuum for the two-dimensional theory obtained by reducing the theory on a circle and $B^\alpha(\Delta, \epsilon)=\oint \rd a \, B(a, \Delta, \epsilon)$ is a suitable contour integral passing through the Bethe vacuum $a^{\alpha}$. In applications to holography, we typically work in a saddle point approximation where  one particular Bethe vacuum  dominates the sum \eqref{Zhol} \cite{Benini:2015eyy}. We then expect some form of factorization also   at large $N$. In a slightly different but equivalent context, the explicit  analysis has been performed in  \cite{Choi:2019dfu} confirming factorization in the form discussed in the previous section for the topologically twisted index, the superconformal one, and the sphere partition function, and exploring various relations among all these quantities.%
\footnote{The analysis in  \cite{Choi:2019dfu} has been only done, for simplicity, for the $\cN=8$ theory coupled to a fundamental hypermultiplet,  which is supposed to flow to ABJM in the IR, but we might expect the results to hold in general.}  

To understand  the form of \eqref{log} for generic rotation and magnetic charges, it is convenient to expand the holomorphic blocks in the  limit of small $\epsilon$. In this limit, the holomorphic blocks are singular (see \eg\;\cite[(2.22)]{Beem:2012mb} and \cite[(F.15)]{Closset:2018ghr}) 
\bea\label{Bethehol}
B ( a, \Delta, \epsilon ) \underset{\epsilon \to 0}{\sim} \exp \bigg( \! -\frac{1}{\epsilon} \cW ( a , \Delta ) + \ldots \bigg) \, , \\
 B^\alpha ( \Delta, \epsilon ) \underset{\epsilon \to 0}{\sim} \exp \bigg( \! -\frac{1}{\epsilon} \cW ( a^\alpha , \Delta ) + \ldots \bigg) \, ,
\eea
where $\cW (a , \Delta )$ is the effective twisted superpotential of the two-dimensional theory and the Bethe vacua $a^\alpha$ are its critical points. 
An important point is that, at large $N$, the on-shell twisted superpotential $\cW ( a^\alpha , \Delta )$ is related to the $S^3$ free energy  \cite{Hosseini:2016tor,Hosseini:2016cyf,Choi:2019dfu},%
\footnote{The on-shell twisted superpotential of many three-dimensional $\cN=2$ Chern-Simons-matter gauge theories with holographic duals were computed in \cite{Hosseini:2016ume,Jain:2019euv,Jain:2019lqb}.}
and the explicit form of the gravitational block \eqref{log} follows from \eqref{Zhol}.
It might seem that this argument holds only in the strict  limit  $\epsilon \to 0$,
while in reality \eqref{log} and the associated factorization are valid also for $\epsilon \ne 0$.
However, a careful analysis shows that, at large $N$, the subleading terms in \eqref{Bethehol}
vanish except for the first one, whose  only role is to enforce the form of the gluing and the constraint among chemical potentials.
A similar situation holds in other dimensions, including five, as we will see. 

Analogous results hold for the refined topologically twisted index in four dimensions, which can be obtained by gluing two copies of $T^2 \times D_2$. The index for $\cN=4$ SYM captures the density of states of rotating black strings in AdS$_5\times S^5$ and it has be shown to factorize both in gravity and field theory in \cite{Hosseini:2019lkt,Hosseini:2019iad}. The field theory computation has been done by explicitly evaluating the partition function, but the same result would be obtained using the arguments in \cite{Choi:2019dfu}. For future reference, let us notice that computation for black strings are usually done in the Cardy limit where the modular parameter $\tau$ of the torus $T^2$ is small. In this limit the on-shell twisted superpotential of the two-dimensional theory becomes proportional to the trial central charge of the four-dimensional CFT \cite{Hosseini:2016cyf},  
\be
 \cW ( \Delta ) \sim \frac{a(\Delta)}{\tau} \, ,
\ee
in agreement with the general discussion. The Cardy limit on $T^2$  is appropriate for studying the physics
of the  black holes obtained by compactifying the string on a circle, and leads to the charged Cardy formula \cite{Hosseini:2020vgl}.
We will encounter a similar setting in section \ref{sssect:RTTI:logZ:sym}.
For recent results at finite $\tau$ see \cite{Hong:2021bzg}.

In five dimensions, the holomorphic blocks $B(a, \Delta, \epsilon_1, \epsilon_2)$  are given by the K-theoretic Nekrasov's instanton partition function on $\bR^2_{\epsilon_1}\times\bR^2_{\epsilon_2}\times S^1$ \cite{Nekrasov:2002qd,Nekrasov:2003rj,Nakajima:2005fg}. The two chemical potentials $\epsilon_1$ and $\epsilon_2$ are equivariant parameters for the rotations on $\bR^4$ and are conjugate to the two possibile angular momenta for black holes in AdS$_6$ and black strings in AdS$_7$. All the partition functions we will consider can be written in a similar form to \eqref{Zholfac} by gluing 
Nekrasov's partition functions \cite{Hosseini:2018uzp,Crichigno:2018adf,Festuccia:2018rew,Jain:2021sdp}.  There are many similarities with three and four dimensions.  The role of the twisted superpotential is played by the Seiberg-Witten prepotential $\cF_{{\rm SW}}(a, \Delta)$ and the expansion of the holomorphic block  is given by \cite{Nekrasov:2002qd}
\be
 \label{Bethehol5d}
 B ( \Delta, \epsilon_1,\epsilon_2 )
 \underset{\epsilon_1, \epsilon_2 \to 0}{\sim} \exp \bigg( \! -\frac{1}{\epsilon_1\epsilon_2} \cF_{{\rm SW}} ( a , \Delta ) + \ldots \bigg) \, .
\ee 
Analogy with lower dimensions motivated the conjecture made in \cite{Hosseini:2018uzp,Crichigno:2018adf} that the (unrefined) five-dimensional partition functions localize at the critical point 
of $\cF_{{\rm SW}}(a, \Delta)$. The on-shell value of the Seiberg-Witten prepotential for both the Seiberg theory and the $\cN=(2,0)$ theory has ben computed in   \cite{Hosseini:2018uzp,Crichigno:2018adf} and are proportional to the $S^5$ free energy \eqref{F1} and the anomaly coefficient \eqref{F2}, respectively. All these analogies suggest that factorization holds with  the blocks given in table \ref{MagEntropy}.
Unfortunately, decompositions similar to \eqref{Zhol}, which would help in setting these statements on a firmer ground,
are not fully understood in five dimensions. We will try to attack directly the five-dimensional matrix models.

\section{Factorization of the $S^3_b \times S^2_{\epsilon}$ partition function}\label{sec:3}

We are interested in evaluating the partition function of five-dimensional $\cN=1$ gauge theories on $S^3_{b} \times S^2_{\epsilon}$ in the limits that are appropriate for holography. Here $b$ is the squashing parameter of the three-sphere and $\epsilon$ is an $\Omega$-deformation on the twisted two-sphere. We label the (gauge, flavor) fluxes on $S^2_{\epsilon}$ by $(\fn , \ft)$, respectively.
Following \cite{Crichigno:2018adf}, for an $\cN = 1$ gauge theory with gauge group $G$, $I$ hypermultiplets in a representation $\oplus ( \fR_I \oplus \bar \fR_I )$ of the gauge group,
and vanishing Chern-Simons contributions, we can write down the \emph{refined} perturbative part of the partition function as (see App.\?\ref{app:CJW})%
\footnote{One can switch between the conventions used here and those of \cite{Crichigno:2018adf} by setting $g_5^{\text{there}} = \frac12 g_{\text{YM}}$, $\fg_{\text{there}} = 0$, $u_{\text{there}} = - \ii Q a$, $\nu_{\text{there}} = - \ii Q ( 1 - \Delta )$, and $\fm_{\text{there}} - r_{\text{there}} + 1 = \fn + \ft - 1$.}
\be
 \label{Z:S3xS2:general}
 Z_{S^3_b \times S^2_\epsilon} ( a , \fn ; \Delta , \ft , \epsilon | b ) = \frac{1}{|W|} \sum_{ \fn \in \Gamma_\fh}
 \oint \prod_{i = 1}^{\text{rk}(G)}\frac{\rd x_i}{2 \pi \ii x_i} e^{- \cF_{S^3_b \times S^2_\epsilon} (a_i, \fn_i ; \Delta, \ft , \epsilon | b ) } \, ,
\ee
where the exponent $\cF_{S^3_b \times S^2_\epsilon}$ is given by
\bea
 \label{FS3bxS2e:main}
 \cF_{S^3_b \times S^2_\epsilon} ( a_i, \fn_i ; \Delta, \ft , \epsilon | b ) & =
 \frac{16 \pi^2 Q^2}{g_{\text{YM}}^2} \Tr_{\text{F}} ( \fn a ) \\
 & + \sum_{\alpha \in G} \sum_{\ell = - \frac{| B^{\alpha} | - 1}{2}}^{\frac{| B^{\alpha} | - 1}{2}}
 \sign (B^\alpha) \log S_2 \Big( \! - \ii Q \left( \alpha(a) - 1 + \ell \epsilon \right) \Big| b \Big) \\
 & - \sum_{I} \sum_{\rho_I \in \fR_I} \sum_{\ell = - \frac{| B^{\rho_I} | - 1}{2}}^{\frac{| B^{\rho_I} | - 1}{2}}
 \sign (B_{\rho_I}) \log S_2 \Big( \! - \ii Q \left( \rho_I (a) + 1 - \nu_I ( \Delta ) + \ell \epsilon \right) \Big| b \Big) \, .
\eea
Here, $\alpha$ are the roots of the gauge group, $\rho$, $\nu$ denote the weights of the hypermultiplets under the gauge and flavor symmetry groups, respectively, and $| W|$ is the order of the Weyl group of $G$.
Moreover, $g_{\text{YM}}$ is the Yang-Mills coupling constant and $x = e^{\ii a}$.
Finally, $S_2 ( z | b )$ is the double sine function defined by
\be
 \label{def:S2}
 S_2 ( z | b ) \equiv \prod_{m , n \in \bZ_{\geq 0}} \frac{m b + n b^{-1} + Q - \ii z}{m b + n b^{-1} + Q + \ii z} \, , \qquad Q \equiv \frac{1}{2} \left( b + b^{-1} \right) ,
\ee
and
\be
 B^{\rho} \equiv \rho (\fn) + \nu (\ft) - 1 \, \qquad  B^{\alpha} \equiv \alpha (\fn) + 1 \, .
\ee

Before moving forward, let us note the following asymptotic relation for the logarithm of the double sine function, see \cite[App.\,A]{Imamura:2011wg},%
\footnote{Recall that $\sign ( 0 ) = 0$.}
\be
 \label{S2:asymp1}
 \begin{aligned}
  & f_b ( z ) \equiv \log S_2 ( z | b ) \sim \ii \pi \left( \frac{z^2}{2} + \frac{Q^2}{6} - \frac{1}{12} \right) \sign \left[ \re ( z ) \right] \, , \quad \text{as } | \re ( z ) | \to \infty \, ,
 \end{aligned}
\ee
that becomes useful when we study the large $N$ limit of the $S^3_b \times S^2_\epsilon$ partition function.
Note also that, for $a \in \ii \bR$,
\be
 \label{prod:f}
 \sign(B) \sum_{\ell = - \frac{|B| - 1}{2}}^{\frac{|B|-1}{2}} f_b \left( - \ii Q ( a + \ell \epsilon) \right) =
 B f_b ( - \ii Q a ) - \frac{\ii \pi ( Q \epsilon)^2}{24} B ( B^2 - 1 ) \sign ( \im a ) \, .
\ee

The full partition function on $S^3_{b} \times S^2_{\epsilon}$ is a sum over instantonic contributions. In the large $N$ limit, instantons are suppressed and we can restrict to the perturbative part of the partition function given above.

\subsection{$\USp(2N)$ gauge theory with matter}
\label{ssect:USp(2N):free energy}

Let us consider an $\cN = 1$ $\USp(2N)$ gauge theory coupled to $N_f$ hypermultiplets in the fundamental representation
and one hypermultiplet $\mathcal{A}$ in the antisymmetric representation of $\USp(2N)$ \cite{Seiberg:1996bd}.
The global symmetry of the theory is $\SU(2)_R \times \SU(2)_\cA \times \SO(2 N_f) \times \U(1)_I$ where the first factor is the R-symmetry
while the other three factors are flavor symmetries: $\SU(2)_\cA$ acts on $\cA$ as a doublet, $\SO(2 N_f)$ rotates the fundamental hypermultiplets,
and $\U(1)_I$ is the topological symmetry associated to the conserved instanton number current $j = \ast \Tr (F \wedge F)$. The global symmetry is enhanced to $\SU(2)_R \times \SU(2)_\cA \times E_{N_f+1}$ at the UV fixed point  \cite{Seiberg:1996bd}.
The theory arises on the intersection of $N$ D4-branes and $N_f$ D8-branes and orientifold planes, and
is holographically dual to the AdS$_6 \times_w S^4$ background of massive type IIA supergravity \cite{Brandhuber:1999np}
(see also \cite{Intriligator:1997pq,Bergman:2012kr,Morrison:1996xf}). 

The partition function on $S^3_{b} \times S^2_{\epsilon=0}$ at large $N$ was computed in \cite{Crichigno:2018adf} and scales
as $\cO(N^{5/2})$. Here we are interested in the dependence on $\epsilon$ and the factorization properties.

Denote the Cartan elements of $\USp(2N)$ by $a_i$, $i = 1, \ldots, N$,
and normalize the weights of the fundamental representation of $\USp(2N)$ to be $\pm e_i$.
The antisymmetric representation then has weights $\pm e_i \pm e_j$ with $i > j$ and $N-1$ zero weights,
and the roots are $\pm e_i \pm e_j$ with $i > j$ ($\mathcal{V}_1$) as well as $\pm 2 e_i$ ($\mathcal{V}_2$).
Vector and hypermultiplets then contribute to the $\cF_{S^3_b \times S^2_\epsilon}$ functional 
\be
 \label{functional:USp(2N)}
 \cF_{S^3_b \times S^2_\epsilon} (a_i, \fn_i ; \Delta_K, \ft_K , \epsilon | b ) = \cF_{\cA + \cV_1} (a_i, \fn_i ; \Delta_m, \ft_m , \epsilon | b ) + \cF_{\cF + \cV_2} (a_i, \fn_i ; \Delta_f, \ft_f , \epsilon | b ) \, ,
\ee
with
\be
 \begin{aligned}
 \cF_{\cA + \cV_1} & = \sum_{i > j}^{N}
 \big[ \cF_{\Delta_m,\?\ft_m} (\pm a_i \pm a_j )
 - \cF_{\Delta_K=2, \?\ft_K=2} (\pm a_i \pm a_j) \big]
 + (N - 1) \cF_{\Delta_m,\?\ft_m} (0) \, , \\
 \cF_{\cF + \cV_2} & = \sum_{i = 1}^{N}
 \bigg[ \sum_{f = 1}^{N_f} \cF_{\Delta_f,\?\ft_f} (\pm a_i )
 - \cF_{\Delta_K=2,\? \ft_K=2} (\pm 2 a_i) \bigg] \, ,
 \end{aligned}
\ee
where
\be
 \label{block:S3S2}
 \cF_{\Delta_K,\?\ft_K} ( a )  \equiv - \sign (B_K) \sum_{\ell = - \frac{|B_K| - 1}{2}}^{\frac{|B_K| - 1}{2}} \log S_2 \Big( \! - \ii Q \left( a + 1 - \Delta_K + \ell \epsilon \right) \Big| b \Big) \, , 
\ee
with $B_K = \fn + \ft_K - 1$. Here, the index $K$ labels all the matter fields in the theory and we introduced the notation
\be
 \cF_{\Delta_K , \? \ft_K} (\pm a_i ) \equiv \cF_{\Delta_K , \? \ft_K} (a_i , \fn_i ) + \cF_{\Delta_K , \? \ft_K} ( - a_i, - \fn_i ) \, .
\ee
Notice that the vector multiplet contribution is equal to  minus the contribution of a hypermultiplet with $\Delta_K=2$
and $\ft_K=2$. 

As we will see, the dependence on $\Delta_f$ is subleading at large $N$ and we will be interested in the chemical potential $\Delta_m$ and the flux $\ft_m$ for the Cartan subgroup of $\SU(2)_\cA$. As mentioned in the introduction, the free energy will take a nice form when written in terms of constrained variables.
We then define
\bea
 \label{democ:USp(2N):S3xS2}
 \ft_1 & \equiv \ft_m \, , \qquad && \ft_2 \equiv 2 - \ft_m \, ,  &&& \text{ s.t. } \sum_{i = 1}^2 \ft_i = 2 \, , \\
 \Delta_1 & \equiv \Delta_m \, , && \Delta_2 \equiv 2 - \Delta_m \, , &&& \text{ s.t. } \sum_{i = 1}^2 \Delta_i = 2 \, .
\eea

Observe that the last term in $\cF_{\cA + \cV_1}$ is of order $\cO(N)$  in the large $N$ limit and, given the expected $N^{5/2}$ scaling of the free energy, subleading.
We will also make a few assumptions regarding the gauge variables that are true for the solution at $\epsilon=0$ \cite{Crichigno:2018adf} and 
that we will verify afterwards. Assuming  that $| \im a_i | $ scales with some positive power of  $N$, and using \eqref{S2:asymp1} and \eqref{prod:f}, we obtain%
\footnote{We do not include subleading terms in the rest of this calculation.}
\bea
 \label{FHV1:mid}
 \cF_{\cA + \cV_1} & = \ii \pi Q^2 \sum_{i > j}^N \left[ ( \Delta_1 \ft_2 + \Delta_2 \ft_1 ) a_{ i j} - \frac14 (4 \Delta_1 \Delta_2 + \epsilon^2 \ft_1 \ft_2) \fn_{i j} \right] \sign \left( \im a_{ i j} \right) \\
 & + \ii \pi Q^2 \sum_{i > j}^N \left[ ( \Delta_1 \ft_2 + \Delta_2 \ft_1 ) a_{ i j}^+ - \frac14 (4 \Delta_1 \Delta_2 + \epsilon^2 \ft_1 \ft_2) \fn_{i j}^+ \right] \sign \left( \im a_{ i j}^+ \right) .
\eea
For the ease of notation, we  defined $a_{i j} \equiv a_i - a_j$, $a_{i j}^+ \equiv a_i + a_j$ and the same for the gauge fluxes $\fn_i$.
Because of the Weyl reflections of the $\USp(2N)$ group, we restrict to $\im a_i > 0$.
Assuming also that the eigenvalues  are ordered by increasing imaginary part, \ie\; $\im a_i > \im a_j$ for $i > j$, and using 
\be
 \label{sum:sign}
\begin{aligned}
 & \sum_{i , j = 1}^N (a_i - a_j ) \sign ( i - j) = 2 \sum_{j = 1}^N (2 j - 1 - N) a_j \, , \\
 & \sum_{i , j = 1}^N (a_i + a_j ) = 2 N \sum_{j = 1}^N a_j \, ,
 \end{aligned}
\ee
\eqref{FHV1:mid} is simplified to
\be
 \label{FHV1:final}
 \cF_{\cA + \cV_1} = \ii \pi Q^2 \sum_{k = 1}^N (2 k - 1) \left[ ( \Delta_1 \ft_2 + \Delta_2 \ft_1 ) a_k - \frac14 (4 \Delta_1 \Delta_2 + \epsilon^2 \ft_1 \ft_2) \fn_k \right] .
\ee
Similarly, the contribution of $\mathcal{F}_{\cF + \cV_2}$ to the large $N$ free energy can be computed using \eqref{S2:asymp1} and \eqref{prod:f}. It is natural to assume that $a_i$ and $\fn_i$ scale with the same positive power of $N$.
Then, neglecting lower powers of $a_i$ and $\fn_i$ that are subleading, we find
\be
 \label{FFV2:final}
 \cF_{\cF + \cV_2} = - \ii \pi Q^2 (8 - N_f) \sum_{k = 1}^N \left( a_k^2 + \frac{\epsilon^2}{12} \fn_k^2 \right) \fn_k \, .
\ee
Under the same hypothesis also the classical term is subleading, see \eqref{FS3bxS2e:main}.
Putting \eqref{FHV1:final} and \eqref{FFV2:final} together we get the final expression for the $\cF_{S^3_b \times S^2_\epsilon}$ functional
\bea
 \label{FS3xS2:USp(2N):mid}
 \cF_{S^3_b \times S^2_\epsilon} ( a , \fn ; \Delta, \ft , \epsilon | b) & =
 - \ii \pi Q^2 (8 - N_f) \sum_{k = 1}^N \left( a_k^2 + \frac{\epsilon^2}{12} \fn_k^2 \right) \fn_k \\
 & + \ii \pi Q^2 \sum_{k = 1}^N (2 k - 1) \left[ ( \Delta_1 \ft_2 + \Delta_2 \ft_1 ) a_k - \frac14 (4 \Delta_1 \Delta_2 + \epsilon^2 \ft_1 \ft_2) \fn_k \right] ,
\eea
that remarkably can be recast in the following form
\be
 \cF_{S^3_b \times S^2_\epsilon} ( a , \fn ; \Delta, \ft , \epsilon | b) = \frac{4 \ii Q^2 }{\pi} \sum_{\sigma = 1}^2 \frac{\cF_{\text{SW}} \big( a_k^{(\sigma)} ; \Delta_i^{(\sigma)} \big)}{\epsilon^{(\sigma)}} \, ,
\ee
where we used the $A$-gluing parameterization
\bea
 a_k^{(1)} & \equiv a_k - \frac{\epsilon}{2} \fn_k \, , \qquad \Delta_i^{(1)} \equiv \Delta_i + \frac{\epsilon}{2} \ft_i \, , \qquad \epsilon^{(1)} = \epsilon \, , \\
 a_k^{(2)} & \equiv a_k + \frac{\epsilon}{2} \fn_k \, , \qquad \Delta_i^{(2)} \equiv \Delta_i - \frac{\epsilon}{2} \ft_i  \, , \qquad \epsilon^{(2)} = - \epsilon \, .
\eea
Here, $\cF_{\text{SW}}$ is the Seiberg-Witten prepotential of the four-dimensional theory obtained by compactifying the five-dimensional $\cN = 1$ theory on $S^1$
and it receives contributions from all the Kaluza-Klein (KK) modes on $S^1$ \cite{Nekrasov:1996cz}.
In the large $N$ limit, it reads \cite[(3.67)]{Hosseini:2018uzp},
\be
 \label{SW:sum2:USp(2N)}
 \cF_{\text{SW}} ( a_k ; \Delta_i ) = \frac{\pi^2}{4} \sum_{k = 1}^N \left( \frac{8 - N_f}{3} a_k^3 + ( 2 k - 1)  \Delta_1 \Delta_2 a_k \right) .
\ee

Extremizing \eqref{FS3xS2:USp(2N):mid} over the gauge variables $(a_k , \fn_k)$, we find the saddle point equations
\be
 \label{FS3xS2:a:saddle}
 \frac{\pd \cF_{S^3_b \times S^2_\epsilon} (a , \fn)}{\pd a_k} \Big|_{a_k = \mathring{a}_k} = 0 \qquad \Rightarrow \qquad \mathring{a}_k \fn_k = \frac{2 k - 1}{2 (8 - N_f) } ( \Delta_1 \ft_2 + \Delta_2 \ft_1 ) \, ,
\ee
and 
\be
 \begin{aligned}
 \label{FS3xS2:n:saddle}
 & \frac{\pd \cF_{S^3_b \times S^2_\epsilon} (a , \fn)}{\pd \fn_k} = 0 \Big|_{a_k = \mathring{a}_k , \, \fn_k = \mathring{\fn}_k}  \\
 & \Rightarrow \qquad \mathring{\fn}_k = - \frac{\ii}{\epsilon} \sqrt{\frac{2 k-1}{8-N_f}}
 \left( \sqrt{\Big( \Delta_1 + \frac{\epsilon}{2} \ft_1 \Big) \Big( \Delta_2 + \frac{\epsilon}{2} \ft_2 \Big)}
 - \sqrt{\Big( \Delta_1 - \frac{\epsilon}{2} \ft_1 \Big) \Big( \Delta_2 - \frac{\epsilon}{2} \ft_2 \Big)} \right) .
 \end{aligned}
\ee
Observe that \eqref{FS3xS2:a:saddle} and \eqref{FS3xS2:n:saddle} are equivalent to
\be
 \frac{\pd \cF_{\text{SW}} ( a_k^{(\sigma)} ; \Delta_i^{(\sigma)} )}{\pd a_k^{(\sigma)}} = 0 \qquad \Rightarrow \qquad
 \mathring{a}_k^{(\sigma)} = \frac{\ii}{\sqrt{8 - N_f}} \sqrt{ ( 2 k - 1 ) \Delta_1^{(\sigma)} \Delta_2^{(\sigma)} } \, ,
\ee
for $ \sigma = 1, 2$.
We see that both $\mathring{a}_k$ and $\mathring{\fn}_k$ scale as $N^{1/2}$ and are purely imaginary  for generic values of the parameters.
Plugging the saddle points $( \mathring{a}_k , \mathring{\fn}_k )$ back into the partition function \eqref{Z:S3xS2:general} we can write down the large $N$ version of the  $S^3_b \times S^2_\epsilon$ free energy as
\be
 \label{FS3:USp(2N)}
 \boxed{
 F_{S^3_b \times S^2_\epsilon} ( \Delta_i , \ft_i , \epsilon | b )
 = \frac{8}{27} \frac{Q^2}{\epsilon} \left[ F_{S^5} \left( \Delta_i + \frac{\epsilon}2 \ft_i \right) - F_{S^5} \left( \Delta_i - \frac{\epsilon}2 \ft_i \right) \right] ,
 }
\ee
with $F_{S^5} (\Delta_i)$ being the free energy of the theory on $S^5$,
\be
 \label{FS5:USp(2N)}
 F_{S^5} ( \Delta_i ) = - \frac{9 \sqrt{2}\, \pi}{5} \frac{N^{5/2}}{\sqrt{8 - N_f}} (\Delta_1 \Delta_2)^{\frac{3}{2}} \, , \qquad \sum_{\i = 1}^2 \Delta_i = 2 \, .
\ee
Note that
\be
 \label{sum:N52}
 \sum_{k = 1}^N ( 2k - 1)^{3/2} \sim \frac{4 \sqrt{2}}{5} N^{5/2} \, , \quad \text{ for } N \gg 1 \, .
\ee
In the limit $\epsilon \to 0$, our expression for the \emph{refined} free energy \eqref{FS3:USp(2N)} reduces to
\be
 \label{FS3xS2:unrefined}
 F_{S^3_b \times S^2} ( \Delta_i , \ft_i | b )
 = - \frac{4 \sqrt{2} \pi Q^2 N^{5/2}}{5 \sqrt{8 - N_f}} ( \Delta_1 \Delta_1 )^{1/2} ( \Delta_1 \ft_2 + \Delta_2 \ft_1 ) \, ,
\ee
which agrees with \cite[(3.17)]{Crichigno:2018adf} upon the following change of variables
\be
 \label{CJW:map}
 \Delta_1 = 1 + \tilde{\nu}_{\text{there}} \, , \qquad \Delta_2 = 1 - \tilde{\nu}_{\text{there}} \, , \qquad \ft_1 = 1 + \hat \fn_{\text{there}} \, , \qquad \ft_2 = 1 - \hat \fn_{\text{there}} \, .
\ee

\subsection{$\cN = 2$ super Yang-Mills}
\label{ssect:SYM:free energy}

Consider five-dimensional $\cN = 2$ super Yang-Mills with gauge group $\SU(N)$. In $\cN=1$ notations, the theory contains one vector multiplet and one hypermultiplet transforming in the adjoint representation of the gauge group. We introduce a  fugacity $\Delta$ and a flux $\ft$ associated with the flavor symmetry acting on the hypermultiplet.
We are interested in evaluating the $S^3_b \times S^2_\epsilon$ free energy in the 't Hooft limit
\be
 N \gg 1 \, \quad \text{ with } \quad \lambda \equiv g_{\text{YM}}^2 N = \text{fixed} \, .
\ee

The $\cF_{S^3_b \times S^2_\epsilon}$ functional reads
\be
 \label{functional:SYM0}
 \cF_{S^3_b \times S^2_\epsilon} (a_i, \fn_i ; \Delta, \ft , \epsilon | b ) = \cF_{\text{YM}} ( a_k , \fn_k ) + \cF_{\cH} (a_i, \fn_i ; \Delta , \ft , \epsilon | b ) + \cF_{\cV} (a_i, \fn_i , \epsilon | b ) \, ,
\ee
with
\be
 \begin{aligned}
 \cF_{\text{YM}} & = \left( \frac{4 \pi Q}{g_{\text{YM}}}\right)^2 \sum_{k = 1}^N a_k \fn_k \, , \\
 \cF_{\cH} & = \sum_{i , j = 1}^{N} \cF_{\Delta , \? \ft} ( a_{i j} ) \, , \qquad
 \cF_{\cV} = - \sum_{i , j = 1}^{N} \cF_{ \Delta = 2 ,\? \ft = 2} ( a_{i j} ) \, ,
 \end{aligned}
\ee
where $\cF( a )$ is given in \eqref{block:S3S2} and $a_{i j} \equiv a_i - a_j$.  As before,  the vector multiplet contribution is equal  to  minus the contribution of the hypermultiplet with $\Delta=2$
and $\ft=2$.

In the strong 't Hooft coupling $\lambda \gg 1$ the eigenvalues are pushed apart, \ie\;$| \im a_{ i j} | \to \infty$, and \eqref{functional:SYM0}, using \eqref{S2:asymp1}, can be approximated as
\bea
 \cF_{S^3_b \times S^2_\epsilon} (a_i, \fn_i ; \Delta, \ft , \epsilon | b ) & =
 \left( \frac{4 \pi Q}{g_{\text{YM}}}\right)^2 \sum_{k = 1}^N a_k \fn_k \\
 & + \frac{\ii \pi Q^2}{2} \sum_{i , j = 1}^N \left[ ( \Delta_1 \ft_2 + \Delta_2 \ft_1 ) a_{i j} - \frac14 (4 \Delta_1 \Delta_2 + \epsilon^2 \ft_1 \ft_2) \fn_{i j} \right] \sign (\im a_{i j}) \, ,
\eea
where we introduced, as before,  a set of constrained variables
\bea
 \ft_1 & \equiv \ft \, , \qquad && \ft_2 \equiv 2 - \ft \, ,  &&& \text{ s.t. } \sum_{i = 1}^2 \ft_i = 2 \, , \\
 \Delta_1 & \equiv \Delta \, , && \Delta_2 \equiv 2 - \Delta \, , &&& \text{ s.t. } \sum_{i = 1}^2 \Delta_i = 2 \, .
\eea

Assuming that the eigenvalues are ordered by increasing imaginary part, using \eqref{sum:sign}, we obtain
\bea
 \label{FS3xS2:SYM:offshell}
 \cF_{S^3_b \times S^2_\epsilon} (a_i, \fn_i ; \Delta, \ft , \epsilon | b ) & =
 \left( \frac{4 \pi Q}{g_{\text{YM}}}\right)^2 \sum_{k = 1}^N a_k \fn_k \\
 & + \ii \pi Q^2 \sum_{k = 1}^N (2 k - 1 - N)
 \left[ ( \Delta_1 \ft_2 + \Delta_2 \ft_1 ) a_k - \frac14 (4 \Delta_1 \Delta_2 + \epsilon^2 \ft_1 \ft_2) \fn_k \right] .
\eea
Remarkably, this can be recast as
\be
 \label{FS3xS2:fact:offshell}
 \cF_{S^3_b \times S^2_\epsilon} (a_i, \fn_i ; \Delta, \ft , \epsilon | b ) =
 \frac{4 \ii Q^2}{\pi} \sum_{\sigma = 1}^2 \frac{\cF_{\text{SW}} \big( a_k^{(\sigma)} ; \Delta_i^{(\sigma)} \big)}{\epsilon^{(\sigma)}} \, ,
\ee
where we used the $A$-gluing parameterization
\bea
 a_k^{(1)} & \equiv a_k - \frac{\epsilon}{2} \fn_k \, , \qquad \Delta_i^{(1)} \equiv \Delta_i + \frac{\epsilon}{2} \ft_i \, , \qquad \epsilon^{(1)} = \epsilon \, , \\
 a_k^{(2)} & \equiv a_k + \frac{\epsilon}{2} \fn_k \, , \qquad \Delta_i^{(2)} \equiv \Delta_i - \frac{\epsilon}{2} \ft_i  \, , \qquad \epsilon^{(2)} = - \epsilon \, .
\eea
and the effective Seiberg-Witten prepotential, in the strong 't Hooft coupling limit, is given by \cite[(3.20)]{Hosseini:2018uzp}
\be
 \label{SW:sum2:sym}
 \cF_{\text{SW}} ( a_k ; \Delta_i) = \frac{\pi^2}{4} \sum_{k = 1}^N \left( \frac{8 \pi \ii}{g_{\text{YM}}^2} a_k^2 + (2 k - 1 - N ) \Delta_1 \Delta_2 a_k \right) .
\ee
Extremizing \eqref{FS3xS2:SYM:offshell} over the gauge variables $(a_k , \fn_k)$, we find the saddle points
\bea
 \label{a:n:SYM:S3xS2}
 \mathring{a}_{k} & = \frac{\ii g_{\text{YM}}^2}{64 \pi} (2 k - 1 - N) \left(4 \Delta_1 \Delta_2 + \epsilon^2 \ft_1 \ft_2 \right) , \\
 \mathring{\fn}_k & = - \frac{\ii g_{\text{YM}}^2}{16 \pi } (2 k - 1 - N ) ( \Delta_1 \ft_2 + \Delta_2 \ft_1 ) \, .
\eea
Notice that \eqref{a:n:SYM:S3xS2} is equivalent to
\be
 \frac{\pd \cF_{\text{SW}} ( a_k^{(\sigma)} ; \Delta_i^{(\sigma)} )}{\pd a_k^{(\sigma)}} = 0 \qquad \Rightarrow \qquad
 \mathring{a}_k^{(\sigma)} = \ii \frac{g_\text{YM}^2}{16 \pi} ( 2 k - 1 - N ) \Delta_1^{(\sigma)} \Delta_2^{(\sigma)} \, ,
\ee
for $ \sigma = 1, 2$.
Plugging $( \mathring{a}_k , \mathring{\fn}_k )$ back into the partition function \eqref{Z:S3xS2:general}
we can write down the $S^3_b \times S^2_\epsilon$ free energy as
\be
 \label{FS3:SYM:mid}
 F_{S^3_b \times S^2_\epsilon} ( \Delta_i , \ft_i , \epsilon | b )
 = - \frac{Q^2 g_{\text{YM}}^2}{192} N ( N^2 - 1) ( \Delta_1 \ft_2 + \Delta_2 \ft_1 ) ( 4 \Delta_1 \Delta_2 + \epsilon^2 \ft_1 \ft_2 ) \, ,
\ee
that can be more elegantly rewritten in the factorized form
\be\label{FS3:SYM}
\boxed{
 F_{S^3_b \times S^2_\epsilon} ( \Delta_i , \ft_i , \epsilon | b )
 = - N (N^2 - 1) \frac{Q^2 g_{\text{YM}}^2}{96 \epsilon} \left[ \big(\Delta_1^{(1)} \Delta_2^{(1)} \big)^2 - \big( \Delta_1^{(2)} \Delta_2^{(2)} \big)^2 \right] ,
}
\ee
with blocks associated with the function \eqref{F2}.
Note that
\be
 \label{sum:N3}
 \sum_{k = 1}^N ( 2 k - 1 - N )^2 = \frac13 N (N^2 - 1) \, .
\ee
In the limit $\epsilon \to 0$, our expression for the \emph{refined} free energy agrees with \cite[(4.74)]{Crichigno:2018adf}.

\paragraph*{Bethe approach.}

The $S^3_b \times S^2_\epsilon$ free energy \eqref{FS3xS2:SYM:offshell} is linear in the gauge magnetic fluxes $\fn_k$ so one can explicitly perform the sum $\sum_{\fn \in \Gamma_\fh}$ in \eqref{Z:S3xS2:general},
\be
 Z_{S^3_b \times S^2_\epsilon} = \sum_{\fn \in \Gamma_\fh} \oint \prod_{k = 1}^N
 \frac{\rd x_k}{2 \pi \ii x_k}
 e^{- \ii \pi Q^2 ( 2 k - 1 -N ) ( \Delta_1 \ft_2 + \Delta_2 \ft_1 ) a_k}
 e^{- \frac{\pi}{4} Q^2 \left( \frac{64 \pi}{g_{\text{YM}}^2} a_k - \ii (2 k - 1 - N) (4 \Delta_1 \Delta_2 + \epsilon^2 \ft_1 \ft_2 ) \right) \fn_k} ,
\ee
to obtain
\be
 \label{LogZ:S3xS2:SYM:BA}
 Z_{S^3_b \times S^2_\epsilon} = \sum_{a = \mathring{a}} e^{- \ii \pi Q^2 ( 2 k - 1 -N ) ( \Delta_1 \ft_2 + \Delta_2 \ft_1 ) a_k} \, ,
\ee
where the sum is over all solutions $\mathring{a}$ to the Bethe ansatz equations (BAEs)%
\footnote{In the large $N$ limit only one Bethe solution dominates the partition function \cite{Benini:2015eyy}.}
\be
 \label{BAEs:S3xS2}
 1 = e^{- \frac{\pi}{4} Q^2  \sum_{k = 1}^N \left( \frac{64 \pi}{g_{\text{YM}}^2} a_k - \ii (2 k - 1 - N) (4 \Delta_1 \Delta_2 + \epsilon^2 \ft_1 \ft_2 ) \right)}
 \equiv \exp \bigg( \ii \frac{\pd \cW_{S^3_b \times \bR^2_\epsilon} (a_k ; \Delta, \epsilon | b)}{\pd a_k} \bigg) \, .
\ee
Here, $\cW_{S^3_b \times \bR^2_\epsilon} ( a_k ; \Delta, \epsilon | b )$ is the ``quantum corrected'' effective twisted superpotential of the theory on $S^3_b \times \bR^2_\epsilon$ and it reads
\be
 \label{W:S3xS2:SYM}
 \cW_{S^3_b \times \bR^2_\epsilon} ( a_k ; \Delta, \epsilon | b) =
 \frac{\pi Q^2}{4} \sum_{k = 1}^N\left( \frac{32 \ii \pi}{g_{\text{YM}}^2} a_k + (2 k - 1 - N ) ( 4 \Delta _1 \Delta _2 + \epsilon^2 \ft_1 \ft_2 ) \right) a_k \, .
\ee
The solution to the BAEs \eqref{BAEs:S3xS2} is simply given by
\be
 \label{S3xS2:Bethe:sol}
 \mathring{a}_{k} = \frac{\ii g_{\text{YM}}^2}{64 \pi} (2 k - 1 - N) \left(4 \Delta_1 \Delta_2 + \epsilon^2 \ft_1 \ft_2 \right) .
\ee
Plugging the solution \eqref{S3xS2:Bethe:sol} back into the twisted superpotential \eqref{W:S3xS2:SYM} and the partition function \eqref{LogZ:S3xS2:SYM:BA} we find, respectively,
\be
 \begin{aligned}
 \cW_{S^3_b \times \bR^2_\epsilon} ( \Delta, \epsilon | b) & = \frac{\ii Q^2}{96 g_{\text{YM}}^2} N (N^2 - 1) (4 \Delta_1 \Delta_2 + \epsilon^2 \ft_1 \ft_2 )^2 \, , \\
 F_{S^3_b \times S^2_\epsilon} ( \Delta_i , \ft_i , \epsilon | b ) & =
 \ii \sum_{i = 1}^2 \ft_i \frac{\pd \cW_{S^3_b \times \bR^2_\epsilon} ( \Delta, \epsilon | b)}{\pd \Delta_i} \, ,
 \end{aligned}
\ee
in agreement with \eqref{FS3:SYM:mid}.

\paragraph*{$F_{S^3_b \times S^2_\epsilon} (\Delta, \ft, \epsilon | b)$ and the 4d central charge.} The $\cN = 2$ $\SU(N)$ SYM theory is supposed to flow at strong coupling to the  six-dimensional $\cN = (2,0)$ theory of type $A_{N-1}$.
The eight-form anomaly polynomial of the  $\cN = (2,0)$ theory  at large $N$  is given by
\be
 \cA_{6\rd} = \frac{N^3}{24} p_2 ( R ) \, ,
\ee
where $p_2 (R)= e_1^2 e_2^2$ is the second Pontryagin class of the $\SO(5)$ R-symmetry bundle, with $e_\sigma$, $\sigma=1,2$, being the Chern roots.
Notice that the chemical potentials $\Delta_1$ and $\Delta_2$ are naturally associated with the Cartan of $\SO(5)$ and the block function \eqref{F2} is formally obtained by replacing $e_1 \to \Delta_1$ and $e_2 \to \Delta_2$ in the anomaly polynomial. 
The compactification of the 6d $(2,0)$ theory on a topologically twisted $S^2$ gives rise to a class of
four-dimensional $\cN = 1$ CFTs \cite{Bah:2012dg}. The theories are specified by the internal flux $\ft$ and have an additional global symmetry associated with the $\U(1)$ rotational isometry of $S^2$ and conjugated to the equivariant parameter $\epsilon$. We can read off the conformal anomaly coefficient $a (\Delta, \ft, \epsilon)$
of the four-dimensional $\cN = 1$ theory by integrating  $\cA_{6\rd}$ on an $\Omega$-deformed $S^2_\epsilon$.
The integration can be done most conveniently by the localization formula \cite[Sect.\?3.3.2]{Hosseini:2020vgl} and it yields
\be
 \label{A4d:SYM}
 \cA_{4\rd} = - \frac{N^3}{48} ( \Delta_1 \ft_2 + \Delta_2 \ft_1 ) \left( 4 \Delta_1 \Delta_2 \, c_1 (F)^2 + \ft_1 \ft_2 \, c_1( J)^2 \right) c_1 (F) \, ,
\ee 
where $c_1(F)$ is the first Chern class of the 4d R-symmetry bundle
and $c_1(J)$ is the first Chern class of the background $\U(1)$ gauge field coupled to the rotation of $S^2_\epsilon$.
Setting $c_1 ( J ) = \epsilon c_1 (F)$ and comparing \eqref{A4d:SYM} with the six-form anomaly polynomial, at large $N$,
\be
 \cA_{4\rd} = \frac{16}{27} a (\Delta, \ft , \epsilon) c_1(F)^3 \, ,
\ee
we find the trial $a$ central charge of the 4d $\cN = 1$ theory
\be
 \label{a:SYM}
 a (\Delta, \ft, \epsilon) = - \frac{9 N^3}{256} ( \Delta_1 \ft_2 + \Delta_2 \ft_1 ) ( 4 \Delta_1 \Delta_2 + \epsilon^2 \ft_1 \ft_2 ) \, .
\ee
Remarkably, we observe the following large $N$ relation between the $S^3_b \times S^2_\epsilon$ free energy \eqref{FS3:SYM:mid}
and the $a$ central charge \eqref{a:SYM}
\be
 F_{S^3_b \times S^2_\epsilon} (\Delta, \ft, \epsilon | b) = \frac{4}{27} ( g_{\text{YM}} Q )^2 a (\Delta, \ft,  \epsilon) \, .
\ee
 
\section{Refined topologically twisted index}\label{sec:4}

We now consider the partition functions on $(S^2_\epsilon \times S^1)\times \Sigma_\fg$, with a topological twist both along the genus $\fg$ Riemann surface $\Sigma_\fg$ and on  $S^2$. We also turn on an $\Omega$-background along $S^2$ with equivariant parameter $\epsilon$. This corresponds to  the (partially) refined five-dimensional topologically twisted index introduced in \cite{Hosseini:2018uzp,Crichigno:2018adf}. The index depends on fugacities $y$ and fluxes $(\fs , \ft)$ on $(\Sigma_\fg , S^2_{\epsilon} )$ for the flavor symmetries.
Setting the possible Chern-Simons levels to zero, the perturbative part of the matrix model  reads \cite{Hosseini:2018uzp}
\bea
 \label{5DRTTI}
 Z (\fm , \fn , a ; \fs , \ft , \Delta | \epsilon ) = & \frac{1}{|W|} \sum_{\{ \fm , \fn \} \in \Gamma_{\fh}}
 \oint_{\cC} \prod_{i = 1}^{\text{rk}(G)} \frac{\rd x_i}{2 \pi \ii x_i} \?
 \bigg( \det\limits_{i j} \frac{\partial^2  \cW_{(S^2_\epsilon \times S^1) \times \bR^2} (a , \fn ; \Delta, \ft, \epsilon)}{\partial a_i \partial a_j} \bigg)^{\fg } \\
 & \times \exp \left( \frac{8 \pi^2}{g_{\text{YM}}^2} \Tr_{\text{F}} ( \fm \fn )\right)
  \prod_{\alpha \in G}
 \prod_{\ell = - \frac{| B_2^{\alpha} | - 1}{2}}^{\frac{| B_2^{\alpha} | - 1}{2}}
 \bigg( \frac{1-x^\alpha \zeta^{2 \ell} }{x^{\alpha/2} \zeta^{\ell} } \bigg)^{B_1^\alpha \sign(B_2^\alpha)} \\
 & \times \prod_{I} \prod_{\rho_I \in \fR_I}
  \prod_{\ell = - \frac{| B_2^{\rho_I} | - 1}{2}}^{\frac{| B_2^{\rho_I} | - 1}{2}}
   \bigg( \frac{x^{\rho_I/2} y^{\nu_I /2} \zeta^{\ell}}{1 - x^{\rho_I} y^{\nu_I} \zeta^{2 \ell} } \bigg)^{B_1^{\rho_I} \sign( B_2^{\rho_I} )} \, ,
\eea
where $(\fm , \fn)$ and $(\fs , \ft)$ are the gauge and flavor magnetic fluxes on $(\Sigma_\fg , S^2_{\epsilon} )$, respectively;
$x=e^{\ii a}$, $y = e^{\ii \Delta}$, and $\zeta = e^{\ii \epsilon / 2}$.
We have also defined
\bea
 \label{def:B:TTI}
 & B_1^{\rho} \equiv \rho (\fm) + \nu (\fs) + \fg - 1 \, , \qquad && B_1^\alpha \equiv \alpha (\fm) - \fg + 1 \, , \\
 & B_2^{\rho} \equiv \rho (\fn) + \nu (\ft) - 1 \, ,  && B_2^\alpha \equiv \alpha (\fn) + 1 \, .
\eea
Here, $ \cW_{(S^2_\epsilon \times S^1) \times \bR^2}$ is the effective twisted superpotential of
the two-dimensional theory obtained by compactifying the 5d $\cN = 1$ theory on $S^2_\epsilon \times S^1$ (with infinitely many KK modes).
In particular, the contribution of a hypermultiplet to the twisted superpotential $ \cW_{(S^2_\epsilon \times S^1) \times \bR^2}$
can be written as
\bea
  \cW^G_{(S^2_\epsilon \times S^1) \times \bR^2} ( a , \fn ; \Delta, \ft, \epsilon ) & =
 - \sum_{\ell = - \frac{|B_2| - 1}{2}}^{\frac{|B_2| - 1}{2}} \sum_{k \in \bZ} (a + \Delta + k + \ell \epsilon) \left[ \log ( a + \Delta + k + \ell \epsilon ) - 1 \right] \sign(B_2) \\ &
 = - \sum_{\ell = - \frac{|B_2| - 1}{2}}^{\frac{|B_2| - 1}{2}} \Li_2 \big( e^{\ii ( a + \Delta + \ell \epsilon )} \big) \sign(B_2) \, ,
\eea
where, in the spirit  of \cite{Nekrasov:2009uh,Nekrasov:2014xaa}, we have  resummed the one-loop contribution of the KK modes on $S^1$
and included the $|B_2|$ zero-modes on $S^2$, decomposed according to their charges under the $\U(1)$ isometry of the sphere.
In order to comply with the regularization scheme used in \eqref{5DRTTI},
we add local parity terms \cite{Hosseini:2018uzp}, so that the total contribution of a hypermultiplet to the twisted superpotential can be written as
\be
  \cW^\cH_{(S^2_\epsilon \times S^1) \times \bR^2} ( a , \fn ; \Delta, \ft, \epsilon ) =
 - \sum_{\ell = - \frac{|B_2| - 1}{2}}^{\frac{|B_2| - 1}{2}} \left( \Li_2 \big( e^{\ii ( a + \Delta + \ell \epsilon )} \big) - \frac12 g_2 ( a + \Delta + \ell \epsilon ) \right) \sign(B_2) \, ,
\ee
where the functions $g_s ( a)$, $s \in \bZ_{\geq 0}$, are related to the Bernoulli polynomials by 
\be
 \label{Li:inversion:g}
 \Li_s (e^{\ii a} ) + ( - 1)^s \Li_s (e^{-\ii a} ) = - \frac{(2 \pi \ii)^s}{s!} B_s \left( \frac{a}{2 \pi} \right) \equiv \ii^{s -2} g_s ( a ) \, ,
\ee
for $0 < \re (a) < 2 \pi$.  
In particular,
\be
 g_2 ( a ) = \frac{a^2}{2} - \pi a + \frac{\pi^2}{3} \, , \qquad g_3 (a) = \frac{a^3}{6} - \frac{\pi}{2} a^2 + \frac{\pi^2}{3} a \, .
\ee
The right-hand side of  \eqref{Li:inversion:g} is extended by periodicity to arbitrary values of $\re (a)$. 
In the range $-2 \pi < \re (a) < 0$, we need to use
\be\label{invf}
 g_s (2 \pi - a) = (-1)^s g_s (a) \, .
\ee
The contribution of a vector multiplet can be obtained  via
\be
  \cW^\cV_{(S^2_\epsilon \times S^1) \times \bR^2}  ( a , \fn , \epsilon ) =
 -  \cW^\cH_{(S^2_\epsilon \times S^1) \times \bR^2} ( a , \fn ; 2 \pi, 2, \epsilon ) \, .
\ee
Putting together $ \cW^\cH$ and $ \cW^\cV$, and adding the classical Yang-Mills contribution, 
we can write down the complete effective twisted superpotential as follows
\bea
 \label{W:TTI:general}
  \cW_{(S^2_\epsilon \times S^1) \times \bR^2} & =
 - \frac{8 \pi^2 \ii}{g_{\text{YM}}^2} \Tr_\text{F} (\fn a) \\
 & \hspace{-1cm} + \sum_{\alpha \in G} \sum_{\ell = - \frac{|B_2^\alpha| - 1}{2}}^{\frac{|B_2^\alpha| - 1}{2}}
  \left( \Li_2 \big( e^{\ii ( \alpha (a) + \ell \epsilon )} \big) - \frac12 g_2 ( 2 \pi + \alpha( a ) + \ell \epsilon ) \right) \sign(B_2^\alpha) \\
 & \hspace{-1cm} - \sum_I \sum_{\rho_I \in \fR_I} \sum_{\ell = - \frac{|B_2^{\rho_I}| - 1}{2}}^{\frac{|B_2^{\rho_I}| - 1}{2}}
 \left( \Li_2 \big( e^{\ii ( \rho_I (a) + \nu_I ( \Delta ) + \ell \epsilon )} \big) - \frac12 g_2 ( \rho_I ( a ) + \nu_I ( \Delta ) + \ell \epsilon ) \right) \sign(B_2^{\rho_I}) \, .
\eea

Finally, in studying the large $N$ limit of the topologically twisted index \eqref{5DRTTI} we shall use
the following formulae for the asymptotic behavior of the polylogarithms
\be
 \label{asymp:Li(s)}
 \Li_s (e^{\ii ( a + \Delta )} ) + \frac{\ii^s}{2} g_s ( a + \Delta ) \sim \frac{\ii^{s}}{2} g_s ( a + \Delta ) \sign ( \im a ) \, ,\quad \text{ as } | \im a | \to \infty \, ,
\ee
where $0 < \re ( a+ \Delta ) < 2 \pi$,\footnote{For other ranges of $\re ( a+ \Delta )$, we need to shift the argument of $g_s$ by appropriate multiples of $2 \pi$.}  and
\be
 \label{prod:g2:g3}
  \sign(B) \sum_{\ell = -\frac{|B| - 1}{2}}^{\frac{|B| - 1}{2}} g_2 ( a  + \Delta +  \ell \epsilon ) = B g_2 ( a + \Delta ) + \frac{\epsilon^2}{4 \pi^3} g_3 (\pi (B+1)) \, .
\ee

\subsection{Alternative interpretations for $ \cW_{(S^2_\epsilon \times S^1) \times \bR^2}$}
\label{ssect:W:Bethe:Gluing}
In the following, we will give two \emph{independent} interpretations of the twisted superpotential \eqref{W:TTI:general}. They can be used as alternative definitions and we have checked that both yield \eqref{W:TTI:general}. 

\paragraph*{(i) Bethe approach.}
It is easy to see that the twisted superpotential \eqref{W:TTI:general} appears in the partition function as%
\footnote{We dropped the dependence on the flavor parameters $(\Delta, \fs, \ft)$ to avoid clutter.}
\be
 \label{Wdef} Z_{(S^2_\epsilon \times S^1 ) \times \Sigma_\fg} =\frac{1}{|W|} \sum_{\{ \fm , \fn \} \in \Gamma_{\fh}}
 \oint_{\cC} \prod_{i = 1}^{\text{rk}(G)} \frac{\rd x_i}{2 \pi \ii x_i} \?  \exp \bigg( \ii \sum_k \fm_k \frac{\pd  \cW_{(S^2_\epsilon \times S^1) \times \bR^2} ( a , \fn ; \epsilon)}{\pd a_k} \bigg) Z_{\text{int}} \big|_{\fm = 0} (a , \fn ; \epsilon) \, ,
\ee
where $Z_{\text{int}}$ is the integrand in \eqref{5DRTTI}. 

Resumming the gauge magnetic fluxes $\fm$ on the Riemann surface,  we obtain a set of poles at the Bethe vacua, the critical points $\mathring{a}$ of the twisted superpotential. We still need to perform a sum over the gauge fluxes on $S^2_\epsilon$.
It was conjectured in \cite{Hosseini:2018uzp,Jain:2021sdp} that the partition function localizes at the solutions to the \emph{generalized} BAEs.
In our case, these take the form%
\footnote{The second equation is the natural generalization to $\epsilon\ne 0$ of the condition $\frac{\partial\cF_{\rm SW}}{\partial a} = \frac{\partial \cW}{\partial \fn}$ used in \cite{Hosseini:2018uzp}.}
\bea
 \label{gen:BAEs:RTTI}
 1 & = \exp \bigg( \frac{\pd  \cW_{(S^2_\epsilon \times S^1) \times \bR^2} ( a , \fn ; \epsilon)}{\pd a_k} \bigg) \Big|_{a = \mathring{a} , \, \fn = \mathring{\fn}} \, , \\
 1 & = \exp \bigg( \frac{\pd  \cW_{(S^2_\epsilon \times S^1) \times \bR^2} ( a , \fn ; \epsilon)}{\pd \fn_k} \bigg) \Big|_{a = \mathring{a} , \, \fn = \mathring{\fn}} \, .
\eea
In comparison to the Bethe approach for the three- and four-dimensional indices, see for example \cite{Nekrasov:2014xaa,Benini:2015eyy,Benini:2016hjo,Closset:2018ghr,Hosseini:2016cyf,Benini:2018mlo}, 
the equation in the second line of \eqref{gen:BAEs:RTTI} is a new feature of the five-dimensional indices,
where we have two sets of \emph{physical} gauge magnetic fluxes.

The relation \eqref{Wdef} can be used as a working definition of $ \cW_{(S^2_\epsilon \times S^1) \times \bR^2}$. Notice, however, that this definition is {\it inherently
ambiguous}. We can always add  to $ \cW_{(S^2_\epsilon \times S^1) \times \bR^2}$ a function that depends on $\fn$ but not on $a$ and \eqref{Wdef} would be still true. For example, precisely for this reason,
our twisted superpotential  differs from the one used in \cite{Jain:2021sdp} in a similar context.
\paragraph*{(ii) Gluing $ \cW_{(\bR^2_\epsilon \times S^1) \times \bR^2}$.}
Consider a five-dimensional $\cN = 1$ gauge theory on $(\bR^2_{\epsilon_1} \times S^1 ) \times \bR^2_{\epsilon_2}$ with $\epsilon_1 = \epsilon$ and $\epsilon_2 = 0$.
The twisted superpotential of the two-dimensional effective theory obtained by reducing the five-dimensional theory on the $\Omega$-deformed copy of $\bR^2 \times S^1$ is then defined as \cite{Nekrasov:2009rc}
\be
 \label{NSW:def}
  \cW_{(\bR^2_{\epsilon} \times S^1 ) \times \bR^2} ( a ; \epsilon) \equiv - \ii \lim_{\epsilon_2 \to 0} \epsilon_2 \log Z_{\bC^2 \times S^1} ( a ; \epsilon_1, \epsilon_2) \big|_{\epsilon_1 = \epsilon}\, ,
\ee
where $Z_{\bC^2 \times S^1} (a ; \epsilon_1, \epsilon_2)$ is the K-theoretic Nekrasov partition function
\be\label{ZN}
Z_{\mathbb{C}^{2}\times S^{1}} ( g_{\text{YM}} , k , a ; \Delta , \epsilon_{1} , \epsilon_{2} ) =
Z^{\text{cl}}_{\mathbb{C}^{2}\times S^{1}} Z^{\cH}_{\mathbb{C}^{2}\times S^{1}} Z^{\cV}_{\mathbb{C}^{2}\times S^{1}} \, ,
\ee
with \cite{Nakajima:2005fg,Gottsche:2006bm,Bershtein:2018srt}%
\footnote{The partition function can be also derived using localization \cite{Crichigno:2018adf} and
the result differs from the one given here in the regularization scheme. Various parity prescriptions
start differing at order $\cO(\epsilon^2)$ with constant terms or terms proportional to $g_1 (a)$.
In the theories considered in this paper these differences cancel after gluing when you sum over positive and negative
weights or lead to irrelevant constant terms.}
\bea
 \label{Nekrasov:pf}
 Z^{\text{cl}}_{\mathbb{C}^{2}\times S^{1}} ( g_{\text{YM}} , k, a ; \epsilon_{1},\epsilon_{2} ) & =
 \exp \left( \frac{4 \pi ^2}{g_{\text{YM}}^2 \epsilon_1 \epsilon_2} \Tr_{\text{F}} (a)^2 + \frac{\ii k}
 {6 \epsilon_1 \epsilon_2} \Tr_{\text{F}}(a^3) \right) , \\
 Z^{\cV}_{\mathbb{C}^{2}\times S^{1}} (a, \epsilon_1 , \epsilon_2) & = Z^{\cP_\cV}_{\mathbb{C}^{2}\times S^{1}} (a, \epsilon_1 , \epsilon_2) \prod_{\alpha \in G} ( x^{\alpha}; p, t)_\infty \, , \\
 Z^{\cH}_{\mathbb{C}^{2}\times S^{1}} (a ; \Delta, \epsilon_1 , \epsilon_2) & = Z^{\cP_\cH}_{\mathbb{C}^{2}\times S^{1}} (a ; \Delta, \epsilon_1 , \epsilon_2) \prod_{\rho \in \fR} (x^{\rho} y^{\nu}; p, t)_{\infty}^{-1} \, ,
\eea
where, for completeness, we also included a Chern-Simon term with level $k$. Here, we defined the double $(p,t)$-factorial as
\be
 ( x ; p , t)_\infty = \prod_{i , j = 0}^{\infty} (1 - x p^i t^j ) \, ,
\ee
where $x = e^{\ii a}$, $y = e^{\ii \Delta}$, $p = e^{- \ii \epsilon_1}$ and $t = e^{- \ii \epsilon_2}$.
Moreover, $Z^{\cP_\cV}_{\mathbb{C}^{2}\times S^{1}}$ and $Z^{\cP_\cH}_{\mathbb{C}^{2} \times S^{1}}$ denote the parity contributions
\be
 \begin{aligned}
 & Z^{\cP_\cV}_{\mathbb{C}^{2}\times S^{1}} (a ; \epsilon_1 , \epsilon_2) =
 \prod_{\alpha \in G} \exp \bigg[ \frac{1}{\epsilon_1 \epsilon_2} \bigg( \frac{\ii}{2} g_3 \left( - \alpha( a ) \right)
 - \frac{\ii (\epsilon_1+\epsilon_2)}{4} g_2 \left( - \alpha(a) \right) \\
  & \hspace{3.cm} + \frac{\ii ( \epsilon_1 + \epsilon_2 )^2}{16} g_1 \left( - \alpha( a ) \right)
  - \frac{\ii}{96} (\epsilon_1+\epsilon_2)^3 + \frac{\ii \pi}{48} \left( \epsilon_1^2 + \epsilon_2^2 \right) - \zeta (3) \bigg) \bigg] \, , \\
  & Z^{\cP_\cH}_{\mathbb{C}^{2}\times S^{1}} (a ; \Delta, \epsilon_1 , \epsilon_2) =
  \prod_{\rho \in \fR} \exp \bigg[ \frac{1}{\epsilon_1 \epsilon_2}
  \bigg( \frac{\ii}{2} g_3 \big(  \rho( a ) + \nu ( \Delta ) \big) \\
  & \hspace{3.cm} + \frac{\ii (\epsilon_1+\epsilon_2)}{4} g_2 \big(  \rho( a ) + \nu ( \Delta ) \big)
  + \frac{\ii ( \epsilon_1 + \epsilon_2 )^2}{16} g_1 \big(  \rho( a ) + \nu ( \Delta ) \big) \\
  & \hspace{3.cm} + \frac{\ii}{96} (\epsilon_1+\epsilon_2)^3
  + \frac{\ii \pi}{48} \left( \epsilon_1^2 + \epsilon_2^2 \right) + \zeta (3) \bigg) \bigg] \, .
\end{aligned}
\ee

The classical contribution to the effective twisted superpotential \eqref{NSW:def}, using \eqref{Nekrasov:pf}, thus reads
\be
 \label{W:flat:cl}
  \cW^{\text{cl}}_{(\bR^2_\epsilon \times S^1 ) \times \bR^2} ( g_{\text{YM}} , k, a ; \epsilon ) =
 - \frac{4 \pi^2 \ii}{g_{\text{YM}}^2 \epsilon} \Tr_{\text{F}} ( a )^2 + \frac{k}{6 \epsilon} \Tr_{\text{F}} ( a )^3 \, .
\ee
Next, we can write the following asymptotic expansion for the contribution of a vector multiplet to the twisted superpotential \eqref{NSW:def}
\be
 \label{W:flat:vector}
  \cW^{\cV}_{(\bR^2_{\epsilon} \times S^1 ) \times \bR^2} ( a ; \epsilon) =
  \cW^{\cP_\cV}_{(\bR^2_{\epsilon} \times S^1 ) \times \bR^2} ( a ; \epsilon)
 - \sum_{s = 0}^{\infty} \frac{(- \ii \epsilon)^{s-1} B_s}{s!} \sum_{\alpha \in G} \Li_{3-s}(e^{\ii \alpha (a)}) \, , \quad \text{ as } \epsilon \to 0 \, ,
\ee
where
\bea
 \label{W:flat:vector:P}
  \cW^{\cP_{\cV}}_{(\bR^2_{\epsilon} \times S^1 ) \times \bR^2} ( a ; \epsilon) = &
 - \sum_{s = 0}^{3} \frac{(- \ii \epsilon)^{s-1} B_s}{s!} \, \frac{\ii^{3-s}}{2} \sum_{\alpha \in G} g_{3-s} \left( \alpha (a ) + 2 \pi \right) \\
 & - \sum_{\alpha \in G} \left[ \frac{\epsilon}{48} g_1 \Big( \alpha ( a ) + \pi + \frac{\epsilon}{2} \Big) - \frac{\ii}{\epsilon} \zeta (3) \right] ,
\eea
and $B_s = \left\{1, - \frac{1}{2} , \frac{1}{6} , 0 , - \frac{1}{30} , 0 , \ldots \right\}$ is the $s$th Bernoulli number.
The contribution of a hypermultiplet to the twisted superpotential \eqref{NSW:def}, as $\epsilon \to 0$, is similarly given by
\be
 \label{W:flat:hyper}
  \cW^{\cH}_{(\bR^2_{\epsilon} \times S^1 ) \times \bR^2} ( a ; \Delta, \epsilon) =
  \cW^{\cP_\cH}_{(\bR^2_{\epsilon} \times S^1 ) \times \bR^2} ( a ; \epsilon)
 + \sum_{s = 0}^{\infty} \frac{(- \ii \epsilon)^{s-1} B_s}{s!} \sum_{\rho_I \in \fR} \Li_{3-s}(e^{\ii (\rho_I (a) + \nu_I (\Delta) )} ) \, ,
\ee
where
\bea
 \label{W:flat:hyper:P}
  \cW^{\cP_{\cH}}_{(\bR^2_{\epsilon} \times S^1 ) \times \bR^2} ( a ; \Delta, \epsilon) = &
 \sum_{s = 0}^{3} \frac{(- \ii \epsilon)^{s-1} B_s}{s!} \, \frac{\ii^{3-s}}{2} \sum_{\rho_I \in \fR} g_{3-s} \left( \rho_I ( a ) + \nu_I ( \Delta ) \right) \\
 & +  \sum_{\rho_I \in \fR} \left[ \frac{\epsilon}{48} g_1 \Big( \rho_I ( a ) + \nu_I ( \Delta ) + \pi + \frac{\epsilon}{2} \Big) - \frac{\ii}{\epsilon} \zeta (3) \right] .
\eea

Finally, the effective twisted superpotential of the two-dimensional theory obtained by compactifying the five-dimensional $\cN = 1$
theory on $S^2_\epsilon \times S^1$ is constructed via gluing two copies of $\cW_{( \bR^2_\epsilon \times S^1 ) \times \bR^2} (a ; \Delta, \epsilon)$
according to the $A$-\emph{gluing}%
\footnote{We refer the reader to \cite[(2.138)]{Hosseini:2018uzp} and the discussion around (2.115) therein, 
to understand the shift in the chemical potential, \ie\;$\Delta \to \Delta - \epsilon^{(\sigma)}$, $\sigma = 1, 2$.}
\bea
 a^{(1)}_k & = a_k + \frac{\epsilon}{2} \fn_k \, , \qquad \Delta^{(1)} = \Delta + \frac{\epsilon}{2} ( \ft - 2 ) \, , \qquad \epsilon_{1}^{(1)} = \epsilon \, , \\
 a^{(2)}_k & = a_k - \frac{\epsilon}{2} \fn_k \, , \qquad \Delta^{(2)} = \Delta - \frac{\epsilon}{2}  ( \ft - 2 ) \, , \qquad \epsilon_{1}^{(2)} = - \epsilon \, .
\eea
Explicitly, up to irrelevant constant terms,
\be
 \label{W:RTTI:glue:NS}
  \cW_{(S^2_\epsilon \times S^1) \times \bR^2} ( a , \fn; \Delta, \ft , \epsilon ) =
 \sum_{l = 1}^{2} \cW_{(\bR^2_{\epsilon} \times S^1 ) \times \bR^2} (a^{(l)} ; \Delta^{( l )}, \epsilon^{( l )} ) \, .
\ee
One can check indeed that
\be
 \label{diff:WTTT:Wgule}
 \eqref{W:TTI:general} - \eqref{W:RTTI:glue:NS} = \frac{\epsilon^2}{48} \left( \sum_{\alpha \in G} B_2^\alpha - \sum_{I} \sum_{\rho_I \in \fR_I }B_2^{\rho_I} \right) ,
\ee
with $B_2^\alpha$ and $B_2^\rho$ given in \eqref{def:B:TTI}.
In writing \eqref{diff:WTTT:Wgule} we used that, as $\epsilon \to 0$,
\be
  \Li_{2} ( e^{\ii ( a + \Delta + \ell \epsilon )} ) = \sum_{s = 0}^\infty \frac{(\ii \ell \epsilon)^s}{s!} \Li_{2-s} ( e^{\ii ( a + \Delta )} ) \, ,
\ee
and 
\bea
 \sign (B) \sum_{s = 0}^\infty \sum_{\ell = - \frac{|B|-1}{2}}^{\frac{|B|-1}{2}} \frac{(\ii \ell \epsilon)^s}{s!} \Li_{2-s} ( e^{\ii ( a + \Delta )} ) =
 -\sum_{s = 0}^\infty \frac{(- \ii \epsilon )^{s-1} B_s }{s!}
 \Big(
 \Li_{3-s} \big(e^{ \ii ( a + \Delta + \frac{1}{2} ( B - 1 ) \epsilon )} \big) \\
 - (-1)^s \Li_{3-s} \big(e^{ \ii ( a + \Delta - \frac{1}{2} (B-1) \epsilon )} \big)
 \Big) \, .
\eea
We thus find agreement between  \eqref{W:TTI:general} and \eqref{W:RTTI:glue:NS}
up to an irrelevant constant term and linear terms in $\fn$ that cancel after summing over positive and negative roots and weights for all the theories in this paper. 

On a final note, we observe that the consistency between the gluing and the Bethe approach to the definition of the twisted superpotential  is a consequence of the fact that the topologically twisted index itself
can be obtained by gluing copies of the Nekrasov partition function  \cite{Hosseini:2018uzp}.

\subsection{$\cW_{( S^2_\epsilon \times S^1 ) \times \bR^2}$ and its factorization}

In this section, we consider  the twisted superpotential $ \cW_{(S^2_\epsilon \times S^1)\times \bR^2}$ as
a function of both the gauge variables $a$ and the fluxes $\fn$, and study its critical points,
or, in other words, the solutions to the generalized BAEs. We will show that the on-shell twisted superpotential
factorizes into contributions coming from the North pole and the South pole of the two-sphere $S^2_\epsilon$.
The poles of the sphere are the two fixed points of the rotational symmetry and we will see that to each fixed point
we can associate a block $\cB_5 ( \Delta , \epsilon)$. We consider the usual two examples
\begin{enumerate}[label=(\roman*)]
 \item $\cN = 1$ $\USp(2N)$ gauge theory with matter. In this case, we find that
 \be
  \label{B5:USp(2N)}
  \cB_5 ( \Delta_i , \epsilon) \equiv \frac{4 \ii \pi^2}{27} \frac{F_{S^5} (\Delta_i )}{\epsilon} \, ,
\ee
where $F_{S^5} (\Delta_i ) \propto (\Delta_1 \Delta_2)^{3/2}$ is the free energy of the theory on $S^5$
that depends on constrained parameters $\Delta_{1,2}$ for the $\SU(2)_\cA \times \SU(2)_R$ symmetry.
 \item $\cN = 2$ SYM that decompactifies to the $\cN = (2,0)$ theory of type $A_{N-1}$ in six dimensions.
The block in this case is given by
 \be
  \label{B5:6d}
  \cB_5 ( \Delta_i , \epsilon) \equiv - \frac{\ii \pi^2 g_\text{YM}^2}{8} \frac{\cA_{6\rd} (\Delta_i )}{\epsilon} \, ,
\ee
with $\cA_{6\rd} (\Delta_i ) \propto (\Delta_1 \Delta_2 )^2$ being the anomaly coefficient of the 6d $(2,0)$ theory
that depends on the constrained parameters $\Delta_{1,2}$ for the $\U(1)^2 \subset \SO(5)$ R-symmetry.

\end{enumerate}

It is interesting to observe that a form of factorization holds for the off-shell twisted superpotential, even before extremization.

\subsubsection{$\USp(2N)$ gauge theory with matter}

The effective twisted superpotential has the same structure as \eqref{functional:USp(2N)} and we only need to replace $\cF_{\Delta_k , \? \ft_K} ( a  )$, see \eqref{block:S3S2}, with
\be
 \label{W:block:RTTI}
  \cW_{(S^2_\epsilon \times S^1) \times \bR^2}^{\Delta_K, \? \ft_K } ( a ) = - \sign(B_2^K) \sum_{\ell = - \frac{|B_2^K| - 1}{2}}^{\frac{|B_2^K| - 1}{2}}
 \left( \Li_2 \big( e^{\ii (a + \Delta_K + \ell \epsilon )} \big) - \frac12 g_2 (  a + \Delta_K + \ell \epsilon ) \right) \, .
\ee
We refine the partition function with fugacities $\Delta_m$ and fluxes $(\fs_m,\ft_m)$ for the $\SU(2)_\cA$ symmetry acting on the antisymmetric hypermultiplet. Similarly to section \ref{ssect:USp(2N):free energy}, we assume that $| \im a_i | $ and $| \im \fn_i | $ scale with some positive power of  $N$
and that the eigenvalues $a_i$  are ordered by increasing imaginary part.
We also define
\bea
 a_{i j} & \equiv a_i - a_j \, , \qquad \qquad ~ \fn_{i ,j} \equiv \fn_i - \fn_j \, , \\
 a^+_{i j} & \equiv a_i + a_j \, , \qquad \qquad ~ \fn^+_{i ,j} \equiv \fn_i + \fn_j \, , \\
 B_{2, i j} & \equiv \fn_i -\fn_j + 1 \, , \qquad B_{2, i j}^{m} \equiv \fn_i - \fn_j + \ft_m - 1 \, .
\eea
Consider first the following contribution
\be
 \cW_{(S^2_\epsilon \times S^1) \times \bR^2}^{-} \equiv - \sum_{i > j}^{N} \big[ \cW_{(S^2_\epsilon \times S^1) \times \bR^2}^{\Delta_K=2, \?\ft_K=2} ( \pm a_{ i j} )
  - \cW_{(S^2_\epsilon \times S^1) \times \bR^2}^{\Delta_m,\?\ft_m} ( \pm a_{ i j } ) \big] \, .
\ee
Using \eqref{asymp:Li(s)}, we obtain 
\bea
 \label{W:RTTI:USp(2N):mid}
 \cW_{(S^2_\epsilon \times S^1) \times \bR^2}^{-} & =
 - \frac{1}{2} \sum_{i > j }^N \sum_{\ell = -\frac{|B_{2, i j}|-1}{2}}^{\frac{|B_{2, i j}|-1}{2}} g_2 ( 2 \pi + a_{i j} + \ell \epsilon ) \sign (\im a_{i j}) \sign ( B_{2, i j} ) \\
 & + \frac{1}{2} \sum_{i > j}^N \sum_{\ell = -\frac{|B^{m}_{2 , i j}|-1}{2}}^{\frac{|B^{m}_{2, i j}|-1}{2}} g_2 ( a_{i j} + \Delta_m + \ell \epsilon ) \sign (\im a_{i j})  \sign ( B^{m}_{2, i j} ) \\
 & - \frac{1}{2} \sum_{i > j }^N \sum_{\ell = -\frac{|B_{2, j i}|-1}{2}}^{\frac{|B_{2, j i}|-1}{2}} g_2 ( 2 \pi + a_{j i} + \ell \epsilon ) \sign (\im a_{j i}) \sign ( B_{2, j i} ) \\
 & + \frac{1}{2} \sum_{i > j}^N \sum_{\ell = -\frac{|B^{m}_{2 , j i}|-1}{2}}^{\frac{|B^{m}_{2, i j}|-1}{2}} g_2 ( a_{j i } + \Delta_m + \ell \epsilon ) \sign (\im a_{j i})  \sign ( B^{m}_{2, j i} ) \, .
\eea
The above equation can be simplified further by performing the product over $\ell$ in \eqref{W:RTTI:USp(2N):mid}, using \eqref{prod:g2:g3}.
Employing the constrained chemical potentials $\Delta_{i}$ and fluxes $\ft_i$, $i=1,2$ (we also give the definition of the constrained fluxes $\fs_i$ on $\Sigma_\fg$ for future reference), 
 \bea
 \label{democ:USp(2N):RTTI}
 \fs_1 & \equiv \fs_m \, , \qquad && \fs_2 \equiv 2 (1 - \fg ) - \fs_m \, ,  \qquad \qquad \text{ s.t. } \sum_{i = 1}^2 \fs_i = 2 - 2 \fg \, , \\
 \ft_1 & \equiv \ft_m \, , \qquad && \ft_2 \equiv 2 - \ft_m \, ,  \qquad \qquad \qquad \quad \, \text{ s.t. } \sum_{i = 1}^2 \ft_i = 2 \, , \\
 \Delta_1 & \equiv \Delta_m \, , && \Delta_2 \equiv 2 \pi - \Delta_m \, , \qquad \qquad \quad \, ~ \text{ s.t. } \sum_{i = 1}^2 \Delta_i = 2 \pi \, ,
\eea
we can  write
\be
 \label{WTTIm}
  \cW_{( S^2_\epsilon \times S^1) \times \bR^2}^- (a_i, \fn_i ; \Delta, \ft , \epsilon ) =
 - \frac{1}{2} \sum_{i > j}^N \left[ ( \Delta_1 \ft_2 + \Delta_2 \ft_1 ) a_{i j} + \frac14 (4 \Delta_1 \Delta_2 + \epsilon^2 \ft_1 \ft_2) \fn_{i j} \right] \sign (\im a_{i j}) \, .
\ee
The contribution of
\be
 \cW_{(S^2_\epsilon \times S^1) \times \bR^2}^{+} \equiv - \sum_{i > j}^{N} \big[ \cW_{(S^2_\epsilon \times S^1) \times \bR^2}^{\Delta_K=2, \?\ft_K=2} ( \pm a_{ i j}^+ )
  - \cW_{(S^2_\epsilon \times S^1) \times \bR^2}^{\Delta_m,\?\ft_m} ( \pm a_{ i j }^+ ) \big] \, ,
\ee
can be found similarly. It reads
\be
 \label{WTTIp}
 \cW_{( S^2_\epsilon \times S^1) \times \bR^2}^+ (a_i, \fn_i ; \Delta, \ft , \epsilon ) =
 - \frac{1}{2} \sum_{i > j}^N \left[ ( \Delta_1 \ft_2 + \Delta_2 \ft_1 ) a_{i j}^+ + \frac14 (4 \Delta_1 \Delta_2 + \epsilon^2 \ft_1 \ft_2) \fn_{i j}^+ \right] \sign (\im a_{i j}^+) \, .
\ee
Combining \eqref{WTTIm} and \eqref{WTTIp}, and using \eqref{sum:sign}, we find
\be
 \label{WTTImp}
 \cW_{(S^2_\epsilon \times S^1) \times \bR^2}^{\cA + \cV_1} =
 - \frac12 \sum_{k = 1}^N (2 k - 1) \left[ ( \Delta_1 \ft_2 + \Delta_2 \ft_1 ) a_k + \frac14 (4 \Delta_1 \Delta_2 + \epsilon^2 \ft_1 \ft_2) \fn_k \right] .
\ee
The contribution of $\cW_{(S^2_\epsilon \times S^1) \times \bR^2}^{\cF + \cV_2}$ to the large $N$ twisted superpotential can be computed similarly, using \eqref{asymp:Li(s)} and \eqref{prod:g2:g3}.
Neglecting lower powers of $a_i$ and $\fn_i$ that are subleading, we get
\be
 \label{WTTI:FV2:final}
 \cW_{(S^2_\epsilon \times S^1) \times \bR^2}^{\cF + \cV_2} =
 - \frac{8 - N_f}{2} \sum_{k = 1}^N \left( a_k^2 + \frac{\epsilon^2}{12} \fn_k^2 \right) \fn_k \, .
\ee
Under the same assumption also the classical term in \eqref{W:TTI:general} is subleading.
Putting \eqref{WTTImp} and \eqref{WTTI:FV2:final} together we can finally write down the complete twisted superpotential
\bea
 \label{WTTI:USp(2N):mid}
  \cW_{(S^2_\epsilon \times S^1) \times \bR^2} (a , \fn ; \Delta, \ft , \epsilon) & =
 - \frac{8 - N_f}{2} \sum_{k = 1}^N \left( a_k^2 + \frac{\epsilon^2}{12} \fn_k^2 \right) \fn_k \\
 & - \frac12 \sum_{k = 1}^N (2 k - 1) \left[ ( \Delta_1 \ft_2 + \Delta_2 \ft_1 ) a_k + \frac14 (4 \Delta_1 \Delta_2 + \epsilon^2 \ft_1 \ft_2) \fn_k \right] ,
\eea
that can be more elegantly put in the form
\be
  \cW_{(S^2_\epsilon \times S^1) \times \bR^2} (a , \fn ; \Delta, \ft , \epsilon) = - 2 \pi \sum_{\sigma = 1}^2 \frac{\cF_{\text{SW}} \big( a_k^{(\sigma)} ; \Delta_i^{(\sigma)} \big)}{\epsilon^{(\sigma)}} \, ,
\ee
using the $A$-gluing parameterization
\bea
 \label{A:gluing:W:TTI}
 a_k^{(1)} & \equiv a_k + \frac{\epsilon}{2} \fn_k \, , \qquad \Delta_i^{(1)} \equiv \Delta_i + \frac{\epsilon}{2} \ft_i  \, , \qquad \epsilon^{(1)} = \epsilon \, , \\
 a_k^{(2)} & \equiv a_k - \frac{\epsilon}{2} \fn_k \, , \qquad \Delta_i^{(2)} \equiv \Delta_i - \frac{\epsilon}{2} \ft_i \, , \qquad \epsilon^{(2)} = - \epsilon \, ,
\eea
with $\cF_{\text{SW}} ( a_k ; \Delta_i )$ being the effective Seiberg-Witten prepotential evaluated in the large $N$ limit \cite[(3.67)]{Hosseini:2018uzp}%
\footnote{One needs to rescale $(a_k , \Delta_i ) \to \pi (a_k , \Delta_i )$ to go back to the conventions of \eqref{SW:sum2:USp(2N)} and section \ref{ssect:USp(2N):free energy}. The same remark applies to \eqref{FS5::pi:USp(2N)}.}
\be
 \label{SW:sum2pi:USp(2N)}
 \cF_{\text{SW}} ( a_k ; \Delta_i ) = \frac{1}{4 \pi} \sum_{k = 1}^N \left( \frac{8 - N_f}{3} a_k^3 + (2 k - 1) \Delta_1 \Delta_2 a_k \right) .
\ee
Extremizing \eqref{WTTI:USp(2N):mid} over the gauge variables $(a_k , \fn_k)$, we find the solution to the generalized BAEs
\be
 \label{TTI:a:Bethe}
 \frac{\pd \cW_{(S^2_\epsilon \times S^1) \times \bR^2} (a , \fn)}{\pd a_k} \Big|_{a_k = \mathring{a}_k} = 0 \qquad \Rightarrow \qquad \mathring{a}_k \fn_k = -\frac{2 k - 1}{2 (8 - N_f) } ( \Delta_1 \ft_2 + \Delta_2 \ft_1 ) \, ,
\ee
and 
\be
 \label{TTI:n:Bethe}
 \begin{aligned}
 & \frac{\pd  \cW_{(S^2_\epsilon \times S^1) \times \bR^2} (a , \fn) }{\pd \fn_k} = 0 \Big|_{a_k = \mathring{a}_k , \, \fn_k = \mathring{\fn}_k}  \\
 & \Rightarrow \qquad \mathring{\fn}_k = \frac{\ii}{\epsilon} \sqrt{\frac{2 k-1}{8-N_f}}
 \left( \sqrt{\Big( \Delta_1 + \frac{\epsilon}{2} \ft_1 \Big) \Big( \Delta_2 + \frac{\epsilon}{2} \ft_2 \Big)}
 - \sqrt{\Big( \Delta_1 - \frac{\epsilon}{2} \ft_1 \Big) \Big( \Delta_2 - \frac{\epsilon}{2} \ft_2 \Big)} \right) .
 \end{aligned}
\ee
Notice that \eqref{TTI:a:Bethe} and \eqref{TTI:n:Bethe} are equivalent to 
\be
 \frac{\pd \cF_{\text{SW}} ( a_k^{(\sigma)} ; \Delta_i^{(\sigma)} )}{\pd a_k^{(\sigma)}} = 0 \qquad \Rightarrow \qquad
 \mathring{a}_k^{(\sigma)} = \frac{\ii}{\sqrt{8 - N_f}} \sqrt{ ( 2 k - 1 ) \Delta_1^{(\sigma)} \Delta_2^{(\sigma)} } \, ,
\ee
for $ \sigma = 1, 2$.
Plugging the saddle points $(\mathring{a}_k , \mathring{\fn}_k)$ back into the twisted superpotential \eqref{WTTI:USp(2N):mid} we obtain
\be
 \label{W:RTTI:USp(2N)}
 \boxed{
  \cW_{(S^2_\epsilon \times S^1) \times \bR^2} ( \Delta_i , \ft_i , \epsilon )
 = \frac{4 \pi^2 \ii}{27 \epsilon} \left[ F_{S^5} \left( \Delta_i + \frac{\epsilon}2 \ft_i \right) - F_{S^5} \left( \Delta_i - \frac{\epsilon}2 \ft_i \right) \right] ,
 }
\ee
with $F_{S^5} (\Delta_i)$ being the free energy of the theory on $S^5$,
\be
 \label{FS5::pi:USp(2N)}
 F_{S^5} ( \Delta_i ) = - \frac{9 \sqrt{2}}{5 \pi^2} \frac{N^{5/2}}{\sqrt{8 - N_f}} (\Delta_1 \Delta_2)^{\frac{3}{2}} \, , \qquad \sum_{i = 1}^2 \Delta_i = 2 \pi \, .
\ee
In the limit $\epsilon \to 0$, our expression for the \emph{refined} twisted superpotential \eqref{W:RTTI:USp(2N)} reduces to
\be
 \label{W:RTTI:unrefined}
  \cW_{(S^2_\epsilon \times S^1) \times \bR^2} ( \Delta_i , \ft_i )
 = - \frac{4 \sqrt{2} \, \ii}{15} \frac{N^{5/2}}{\sqrt{8 - N_f}}
 \sum_{i =1}^2 \ft_i \frac{\pd (\Delta_1 \Delta_2)^{3/2}}{\pd \Delta_i} \, ,
\ee
which matches the expression in \cite[(3.88)]{Hosseini:2018uzp}.

\paragraph*{$\cW_{( S^2_\epsilon \times S^1 ) \times \bR^2}$ and the free energy on $S^3_b \times S^2_\epsilon$.}
Comparing \eqref{FS3xS2:USp(2N):mid} with \eqref{WTTI:USp(2N):mid} we find the following remarkable relations
\bea
  \cW_{( S^2_\epsilon \times S^1 ) \times \bR^2} ( \pi a , - \fn ; \pi \Delta_i , \ft_i , \pi \epsilon) & =
 \frac{\ii \pi}{2 Q^2} \cF_{S^3_b \times S^2_\epsilon} ( a , \fn ; \Delta_i , \ft_i , \epsilon | b ) \, , \\
  \cW_{( S^2_\epsilon \times S^1 ) \times \bR^2} ( \pi \Delta_i , \ft_i , \pi \epsilon) \Big|_{a_k = \mathring{a}_k , \, \fn_k = \mathring{\fn}_k}  & =
 \frac{\ii \pi}{2 Q^2} F_{S^3_b \times S^2_\epsilon} (\Delta_i , \ft_i , \epsilon | b ) \, .
\eea

\subsubsection{$\cN = 2$ super Yang-Mills}
\label{sssect:RTTI:W:sym}

 The twisted superpotential of the theory reads
\be
 \label{W:TTI:SYM}
  \cW_{(S^2_\epsilon \times S^1) \times \bR^2} =
  \cW_{\text{YM}} ( a_k , \fn_k ) +  \cW_{\cH} (a_i, \fn_i ; \Delta , \ft , \epsilon ) +  \cW_{\cV} (a_i, \fn_i , \epsilon ) \, ,
\ee
with
\be
 \begin{aligned}
  \cW_{\text{YM}} & = - \frac{8 \pi^2 \ii}{g_{\text{YM}}^2} \sum_{k = 1}^N a_k \fn_k \, , \\
  \cW_{\cH} & = \sum_{i , j = 1}^{N}  \cW_{\Delta , \? \ft} ( a_{i j}  ) \, , \qquad
  \cW_{\cV} = - \sum_{i \neq j}^{N}  \cW_{\Delta = 2 \pi , \? \ft = 2} ( a_{i j} ) \, ,
 \end{aligned}
\ee
and $ \cW( a )$ given in \eqref{W:block:RTTI}. Note that
\be
  \cW_{\cV} (a_i, \fn_i , \epsilon ) = -  \cW_{\cH} (a_i, \fn_i ; 2 \pi , 2 , \epsilon ) \, .
\ee
In the strong 't Hooft coupling $\lambda \gg 1$, \eqref{W:TTI:SYM} can be approximated as
\bea
 \label{W:RTTI:SYM:mid}
  \cW_{( S^2_\epsilon \times S^1) \times \bR^2} (a_i, \fn_i ; \Delta, \ft , \epsilon ) & =
 - \frac{8 \pi^2 \ii}{g_{\text{YM}}^2} \sum_{k = 1}^N a_k \fn_k \\
 & - \frac{1}{2} \sum_{i \neq j }^N \sum_{\ell = -\frac{|B_{i j}|-1}{2}}^{\frac{|B_{i j}|-1}{2}} g_2 ( 2 \pi + a_{i j} + \ell \epsilon ) \sign (\im a_{i j}) \sign ( B_{i j} ) \\
 & + \frac{1}{2} \sum_{i , j = 1}^N \sum_{\ell = -\frac{|B_{i j}^F|-1}{2}}^{\frac{|B_{i j}^F|-1}{2}} g_2 ( a_{i j} + \Delta + \ell \epsilon ) \sign (\im a_{i j})  \sign ( B_{i j}^F ) \, ,
\eea
where we used  \eqref{asymp:Li(s)} to substitute the $\Li_2 ( e^{\ii ( a + \Delta)} )$ with $g_2 ( a + \Delta)$ as $| \im a_{ i j} | \to \infty$ and we defined
\be
 B_{i j} \equiv \fn_ i - \fn_j + 1 \, , \qquad B_{i j}^F = \fn_i - \fn_j + \ft - 1 \, .
\ee
Let us introduce the following democratic parameterization for the $\U(1)^2 \subset \SU(2)_R \times \SU(2)_F$ symmetry
\bea
 \label{democ:SYM:RTTI}
 \fs_1 & \equiv \fs \, , \qquad && \fs_2 \equiv 2 (1 - \fg ) - \fs \, ,  \qquad \qquad \text{ s.t. } \sum_{i = 1}^2 \fs_i = 2 - 2 \fg \, , \\
 \ft_1 & \equiv \ft \, , \qquad && \ft_2 \equiv 2 - \ft \, ,  \qquad \qquad \qquad \quad \, \text{ s.t. } \sum_{i = 1}^2 \ft_i = 2 \, , \\
 \Delta_1 & \equiv \Delta \, , && \Delta_2 \equiv 2 \pi - \Delta \, , \qquad \qquad \quad \, ~ \text{ s.t. } \sum_{i = 1}^2 \Delta_i = 2 \pi \, .
\eea
Then, performing the product over $\ell$ in \eqref{W:RTTI:SYM:mid}, using \eqref{prod:g2:g3}, we can simplify \eqref{W:RTTI:SYM:mid} to
\bea
  \cW_{( S^2_\epsilon \times S^1) \times \bR^2} (a_i, \fn_i ; \Delta, \ft , \epsilon ) & =
 - \frac{8 \pi^2 \ii}{g_{\text{YM}}^2} \sum_{k = 1}^N a_k \fn_k \\
 & - \frac{1}{4} \sum_{i , j = 1}^N \left[ ( \Delta_1 \ft_2 + \Delta_2 \ft_1 ) a_{i j} + \frac14 (4 \Delta_1 \Delta_2 + \epsilon^2 \ft_1 \ft_2) \fn_{i j} \right] \sign (\im a_{i j}) \, .
\eea
Assuming that the eigenvalues are ordered by increasing imaginary part, using \eqref{sum:sign}, we obtain
\bea
 \label{W:RTTI:SYM:offshell}
  \cW_{( S^2_\epsilon \times S^1) \times \bR^2} (a_i, \fn_i ; \Delta, \ft , \epsilon ) & =
 - \frac{8 \pi^2 \ii}{g_{\text{YM}}^2} \sum_{k = 1}^N a_k \fn_k \\
 & - \frac{1}{2} \sum_{k = 1}^N (2 k - 1 - N)
 \left[ ( \Delta_1 \ft_2 + \Delta_2 \ft_1 ) a_k + \frac14 (4 \Delta_1 \Delta_2 + \epsilon^2 \ft_1 \ft_2) \fn_k \right] ,
\eea
that, using the $A$-gluing parameterization \eqref{A:gluing:W:TTI}, can be more elegantly put in the form
\be
  \cW_{( S^2_\epsilon \times S^1) \times \bR^2} (a_i, \fn_i ; \Delta, \ft , \epsilon ) =
 - 2 \pi \sum_{\sigma = 1}^2 \frac{\cF_{\text{SW}} \big( a_k^{(\sigma)} ; \Delta_i^{(\sigma)} \big)}{\epsilon^{(\sigma)}} \, ,
\ee
where the effective Seiberg-Witten prepotential evaluated in the large $N$ limit \cite[(3.67)]{Hosseini:2018uzp} reads%
\footnote{One needs to rescale $(a_k , \Delta_i ) \to \pi (a_k , \Delta_i )$ to go back to the conventions of \eqref{SW:sum2:sym}.}
\be
 \label{SW:sum2pi:sym}
 \cF_{\text{SW}} ( a_k ; \Delta_i ) = \frac{1}{4 \pi} \sum_{k = 1}^N \left( \frac{8 \pi^2 \ii}{g_{\text{YM}}^2} a_k^2 + ( 2 k - 1 - N ) \Delta_1 \Delta_2 a_k \right) .
\ee
Extremizing \eqref{W:RTTI:SYM:offshell} over the gauge variables $(a_k , \fn_k)$, we find the solution to the generalized BAEs
\bea
 \label{a:n:SYM:RTTI}
 \mathring{a}_{k} & = \ii \frac{g_{\text{YM}}^2}{( 8 \pi)^2} (2 k - 1 - N) \left(4 \Delta_1 \Delta_2 + \epsilon^2 \ft_1 \ft_2 \right) , \\
 \mathring{\fn}_k & = \ii \frac{g_{\text{YM}}^2}{(4 \pi)^2} (2 k - 1 - N ) ( \Delta_1 \ft_2 + \Delta_2 \ft_1 ) \, .
\eea
Notice that \eqref{a:n:SYM:RTTI} is equivalent to
\be
 \frac{\pd \cF_{\text{SW}} ( a_k^{(\sigma)} ; \Delta_i^{(\sigma)} )}{\pd a_k^{(\sigma)}} = 0 \qquad \Rightarrow \qquad
 \mathring{a}_k^{(\sigma)} = \ii \frac{g_\text{YM}^2}{16 \pi^2} ( 2 k - 1 - N ) \Delta_1^{(\sigma)} \Delta_2^{(\sigma)} \, ,
\ee
for $ \sigma = 1, 2$.
Substituting $( \mathring{a}_k , \mathring{\fn}_k )$ into the twisted superpotential \eqref{W:RTTI:SYM:offshell} we find
\be
 \label{W:RTTI:SYM:mid}
  \cW_{( S^2_\epsilon \times S^1) \times \bR^2} ( \Delta, \ft , \epsilon )
 = - \frac{\ii g_{\text{YM}}^2}{384 \pi^2} N ( N^2 - 1) ( \Delta_1 \ft_2 + \Delta_2 \ft_1 ) ( 4 \Delta_1 \Delta_2 + \epsilon^2 \ft_1 \ft_2 ) \, ,
\ee
that can be more elegantly recast in the factorized form
\be
 \label{W:RTTI:SYM:factorized}
\boxed{
  \cW_{( S^2_\epsilon \times S^1) \times \bR^2} (\Delta, \ft , \epsilon )
 = - \ii N (N^2 - 1) \frac{g_{\text{YM}}^2}{192 \pi^2 \epsilon} \left[ \big(\Delta_1^{(1)} \Delta_2^{(1)} \big)^2 - \big( \Delta_1^{(2)} \Delta_2^{(2)} \big)^2 \right] .
}
\ee
In the $\epsilon \to 0$ limit, \eqref{W:RTTI:SYM:factorized} matches \cite[(3.30)]{Hosseini:2018uzp}.
\paragraph*{$ \cW_{( S^2_\epsilon \times S^1) \times \bR^2} (\Delta, \ft, \epsilon)$ and the 4d central charge.}
Comparing \eqref{a:SYM} with \eqref{W:RTTI:SYM:mid}, we note the following large $N$ relation
\be
  \cW_{( S^2_\epsilon \times S^1) \times \bR^2} (\pi \Delta, \ft, \pi \epsilon) = \frac{2 \pi \ii}{27} g_{\text{YM}}^2 \? a (\Delta, \ft, \epsilon) \, ,
\ee
with $a (\Delta , \ft, \epsilon)$ given in \eqref{a:SYM}.

\subsection{Factorization of the index}\label{indexTTI}

In this section we discuss the factorization properties of the refined twisted index. As we already discussed,  we will make the assumption that  the partition function localizes at the solutions to the \emph{generalized} BAEs given in \eqref{gen:BAEs:RTTI}\cite{Hosseini:2018uzp,Jain:2021sdp}. We will see that this assumption leads to the factorization of the index and the correct entropy for a class of dual black holes and black strings.%
\footnote{Our results differ from those in \cite{Jain:2021sdp}, which do not factorize, due to a different twisted superpotential used in the conjectured BAEs.}

We want to factorize $\log Z_{(S^2_\epsilon \times S^1)\times \Sigma_\fg}$ into contributions coming from the North pole and the South pole of the two-sphere $S^2_\epsilon$. 
 We will see that to each fixed point we can associate a block $\cB_3 ( \Delta , \fs, \epsilon)$. As before, we consider two theories,
\begin{enumerate}[label=\roman*)]
 \item $\cN = 1$ $\USp(2N)$ gauge theory for which we find
 \be
  \label{B3:USp(2N)}
  \cB_3 ( \Delta , \fs, \epsilon) \equiv - \frac{\pi}{2} \frac{F_{S^3 \times \Sigma_\fg} (\Delta , \fs)}{\epsilon} \, .
 \ee
 Here, $F_{S^3 \times \Sigma_\fg} (\Delta , \fs)$, see \eqref{FS3xSigma:USp(2N)}, is the free energy of the theory on $S^3 \times \Sigma_\fg$
 that depends on a set of twisted masses $\Delta$ and background magnetic fluxes $\fs$ for the flavor symmetry.
 \item $\cN = 2$ SYM for which we find
 \be
  \label{B3:6d}
  \cB_3 ( \Delta , \fs, \epsilon) \equiv - \frac{2 \pi g_\text{YM}^2}{27} \frac{a (\Delta , \fs)}{\epsilon} \, ,
 \ee
 where $a (\Delta, \fs)$, see \eqref{a4d:twisted:unrefined}, is the trial central charge of the four-dimensional theory
 obtained via compactifying the 6d $(2,0)$ theory on $\Sigma_\fg$ with the mixing parameter $\Delta$ and the flavor flux $\fs$.
\end{enumerate}

The relation between the blocks for the index, \eqref{B3:USp(2N)} and \eqref{B3:6d}, and the blocks for the twisted superpotential, \eqref{B5:USp(2N)} and \eqref{B5:6d}, is simply
\be\label{blocksrelation} 
 \cB_3 ( \Delta , \fs, \epsilon) = \ii \sum_{i=1}^2 \fs_i \frac{\partial  \cB_5 ( \Delta_i , \epsilon)}{\partial \Delta_i} \, ,\ee
 and consequently
 \be
 \label{indexTh00}\log Z_{(S^2_\epsilon \times S^1)\times \Sigma_g}  =
 \ii \sum_{i=1}^2 \fs_i  \frac{ \partial  \cW_{(S^2_\epsilon \times S^1) \times \bR^2}}{\partial \Delta_i}  \, .
\ee

\subsubsection{$\USp(2N)$ gauge theory with matter}

The refined twisted index in the large $N$ limit does \emph{not} depend explicitly on the refinement parameter $\epsilon$
since
\bea
 \label{no:epsilon:logZ}
 \log Z^{\Delta_K , \? \fs_K, \? \ft_K}_{(S^2_\epsilon \times S^1 ) \times \Sigma_\fg} ( a ) & =
 B_1^K \sign(B_2^K) \sum_{\ell = -\frac{|B_2^{K}| - 1}{2}}^{\frac{|B_2^{K}| - 1}{2}}
 \left( \Li_1 \big( e^{\ii (a + \Delta_K + \ell \epsilon )} \big) + \frac{\ii}2 g_1 (  a + \Delta_K + \ell \epsilon ) \right) \\
 & \overset{\eqref{asymp:Li(s)}}{=}
 \frac{\ii}2 B_1^K \sign(B_2^{K}) \sum_{\ell = - \frac{|B_2^{K}| - 1}{2}}^{\frac{|B_2^{K}| - 1}{2}}
 g_1 ( a +  \Delta_K +  \ell \epsilon ) \\
 & = \frac{\ii}2 B_1^K  B_2^{K} \? g_1 (a + \Delta_K ) \, ,
\eea
where recall that
\be
 B_1^K = \fm + \fs_K + \fg - 1 \, , \qquad B_2^K = \fn + \ft_K - 1 \, ,
\ee
and it simply reads \cite[(3.104)]{Hosseini:2018uzp}
\be
 \label{logZ:USp(2N):beg}
 \log Z_{(S^2_\epsilon \times S^1 ) \times \Sigma_\fg} =
 - \frac{\ii}{2} \sum_{k = 1}^N (2 k-1) \left[ (\fs_2 \ft_1 + \fs_1 \ft_2 ) \mathring{a}_k + ( \Delta_1 \fs_2 + \Delta_2 \fs_1 ) \mathring{\fn}_k \right]  .
\ee
Substituting the saddle points $(\mathring{a}_k , \mathring{\fn}_k)$, see \eqref{TTI:a:Bethe} and \eqref{TTI:n:Bethe}, into \eqref{logZ:USp(2N):beg} we find
\be
 \label{RTTI:final:USp(2N)}
 \boxed{
 \log Z_{(S^2_\epsilon \times S^1 ) \times \Sigma_\fg} (\Delta, \fs, \ft, \epsilon)
 = - \frac{\pi}{2 \epsilon} \left[ F_{S^3 \times \Sigma_\fg} \left( \Delta_i + \frac{\epsilon}{2} \ft_i , \fs_i \right) - F_{S^3 \times \Sigma_\fg} \left( \Delta_i - \frac{\epsilon}{2} \ft_i , \fs_i \right) \right] ,
 }
\ee
with $F_{S^3 \times \Sigma_\fg} (\Delta_i , \fs_i)$ being the free energy of the theory on $S^3 \times \Sigma_\fg$ \cite{Crichigno:2018adf},%
\footnote{We can compare with \eqref{FS3xS2:unrefined}, valid for genus zero.
To have a round $S^3$, we should set $Q = 1$.
Recall that the flavor flux through the $\Omega$-deformed $S^2$ was called $\ft_i$ in \eqref{FS3xS2:unrefined} and now, for the Riemann surface $\Sigma_\fg$,
should be renamed to $\fs_i$ with $\sum_{i = 2}^2\fs_i = 2 - 2 \fg$.
Also, $\Delta_i$ should be rescaled by a factor of $\pi$.}
\be
 \label{FS3xSigma:USp(2N)}
 F_{S^3 \times \Sigma_\fg} (\Delta_i , \fs_i ) = - \frac{8 \sqrt{2}}{15 \pi} \frac{N^{5/2}}{\sqrt{8 - N_g}} \sum_{i = 1}^2 \fs_i \frac{\pd (\Delta_1 \Delta_2)^{3/2}}{\pd \Delta_i} \, .
\ee
Recall that
\be
 \sum_{i = 1}^2 \Delta_i = 2 \pi \, , \qquad \sum_{i = 1}^2 \fs_i = 2 - 2 \fg \, ,
 \qquad \sum_{i = 1}^2 \ft_i = 2 \, .
\ee

\paragraph*{Black holes microstates in AdS$_2 \times S^2_\epsilon \times \Sigma_\fg$.}
The refined topologically twisted index \eqref{RTTI:final:USp(2N)} is expected to reproduce the Bekenstein-Hawking entropy
of a class of rotating dyonic black holes in AdS$_6$ in massive type IIA supergravity whose near horizon geometry is a fibration of AdS$_2$ over the twisted space $S^2_\epsilon \times \Sigma_\fg$.
Unfortunately, the most general black holes are still to be constructed and the only known example \cite[Sect.\,6.3.1]{Hosseini:2020wag} was found,
using gauged supergravity of class $\cF$ in four dimensions, when the fluxes through the Riemann surface, using the notations of \cite{Hosseini:2020wag}, are constrained as follows
\be
 s^1 = \frac{2}{3} \, , \qquad s^2 = 0 \, .
\ee
The above choice leaves us with $\fg > 1$.
In \cite{Hosseini:2020wag} the magnetic fluxes along the $S^2_\epsilon$ were denoted by $p^i$, $i=1,2$, satisfying the twisting condition
\be
 p^1 + p^2 = - \frac{2}{3} \, ,
\ee
and the angular momentum by $\cJ$.
Then, the Bekenstein-Hawking entropy reads
\be
 \label{twisted:BH}
 S_{\text{BH}} = \frac{\pi}{9 \sqrt{2} G_{\text{N}}^{(4)}}
 \sqrt{ 1 - 6 p^1 (3 p^1+1) - \sign(6 p^1+1) \sqrt{(2 p^1+1) (6 p^1+1)^3 - 4 \times 3^5 \cJ^2} } \, .
\ee
The above entropy can be obtained by extremizing the Legendre transform of the refined index, \ie\;
\be
 \label{ISCFT:USp(2N)}
 \cI_{(S^2_\epsilon \times S^1 ) \times \Sigma_\fg}  (\Delta, \epsilon) =
 \log Z_{(S^2_\epsilon \times S^1 ) \times \Sigma_\fg} (\Delta, \fs, \ft, \epsilon) - \ii \epsilon J - \Lambda ( \Delta_1 + \Delta_2 - 2 \pi ) \, ,
\ee
with respect to the chemical potentials $(\Delta_1, \Delta_2, \epsilon)$ and the Lagrange multiplier $\Lambda$, that enforces the constraint $\sum_{i = 1}^2 \Delta_i = 2 \pi$.
Define
\be
 \Pi \equiv \sqrt{ \frac{(2 \ft_1 - 3) - \left(\frac{9 \pi }{( \fg - 1) F_{S^5}} J \right)^2 (2\ft_1 - 1)^{-3}}{3 (2 \ft_1 - 1)}} \, ,
\ee
with $F_{S^5}$ being the exact free energy of the $\cN = 1$ $\USp(2N)$ gauge theory on $S^5$ \cite{Jafferis:2012iv}
\be
 \label{F:S5:USp(2N)}
 F_{S^5} = - \frac{9 \sqrt{2} \? \pi}{5} \frac{N^{5/2}}{\sqrt{8 - N_\text{f}}} \, .
\ee
Then, the extrema of \eqref{ISCFT:USp(2N)} are given by
\bea
 \label{ISCFT:USp(2N):crit}
 \mathring{\Delta}_1 & = \frac{\pi}{2} ( 1 + \Pi^{-1} ) \, , \qquad
 \mathring{\Delta}_2 = \frac{\pi}{2} ( 3 - \Pi^{-1} ) \, , \\
 \mathring{\epsilon} & = \ii \frac{9 \sqrt{2} \? \pi^2}{( \fg - 1 ) F_{S^5}} \frac{J}{\Pi ( 2 \ft_1 - 1)^2 \sqrt{\Pi ( 2 \ft_1 - 1 )^2 - 2 \ft_1 ( \ft_1 - 1) + 1}} \, , \\
 \mathring{\Lambda} & = - \frac{( \fg - 1 )}{9 \sqrt{2} \? \pi} F_{S^5} \\
 & \times \sqrt{ 36 \left( 1 - 2 \ft_1 ( \ft_1 - 1 ) \right) + 6 \sign( 2 \ft_1-1) \sqrt{12 (2 \ft_1-3) (2 \ft_1-1)^3 - 3^5 \left( \frac{2 \pi J }{(\fg - 1) F_{S^5}} \right)^2} } \, .
\eea
Plugging \eqref{ISCFT:USp(2N):crit} back into the $\cI$-functional \eqref{ISCFT:USp(2N)} we find
\be
 \cI_{(S^2_\epsilon \times S^1 ) \times \Sigma_\fg} \Big|_{\eqref{ISCFT:USp(2N):crit}} ( \ft_1, J ) = 2 \pi \Lambda \Big|_{\eqref{ISCFT:USp(2N):crit}} ( \ft_1, J ) = S_{\text{BH}} ( p^1, \cJ ) \, ,
\ee
where we used the identification \cite[(7.15)]{Hosseini:2020wag}
\bea
 \fs_i & = - 3 | 1 - \fg | s^i \, , \qquad \ft_i = - 3 p^i \, , \quad i = 1, 2 \, ,\\
 J & = \frac{1}{2 G_\text{N}^{(4)}} \cJ \, ,
\eea
along with the standard AdS$_6$/CFT$_5$ dictionary
\be
 \label{AdS6:CFT5:dict}
 \frac{1}{G_\text{N}^{(6)}} = - \frac{3}{\pi^2} F_{S^5} \qquad \Rightarrow \qquad
 \frac{1}{G_{\text{N}}^{(4)}} = \frac{\vol (\Sigma_\fg)}{G_\text{N}^{(6)}} = - \frac{12 | 1 - \fg |}{\pi} \? F_{S^5} \, .
\ee
This is in complete agreement with \cite[Sect.\,7.1]{Hosseini:2020wag} upon identifying
\be
 \omega_{\text{there}} \equiv - \frac{\ii}{\pi} \epsilon \, , \qquad \chi_{\text{there}}^i \equiv \frac{2}{3 \pi} \Delta_i \, , \quad i = 1, 2 \, .
\ee

\subsubsection{$\cN = 2$ super Yang-Mills}
\label{sssect:RTTI:logZ:sym}

The refined twisted index in the strong 't Hooft coupling limit $\lambda \gg 1$ does not depend explicitly on the refinement parameter $\epsilon$, see \eqref{no:epsilon:logZ},
and it is simply given by \cite[(3.37)]{Hosseini:2018uzp}
\be
 \log Z_{(S^2_\epsilon \times S^1 ) \times \Sigma_\fg} =
 - \frac{\ii}{2} \sum_{k = 1}^N (2 k - 1 - N) \left[ (\fs_2 \ft_1 + \fs_1 \ft_2 ) \mathring{a}_k + ( \Delta_1 \fs_2 + \Delta_2 \fs_1 ) \mathring{\fn}_k \right]  .
\ee
Substituting the saddle points $(\mathring{a}_k , \mathring{\fn}_k)$, see \eqref{a:n:SYM:RTTI}, in the above expression we obtain
\be
 \label{RTTI:sym}
 \boxed{
 \log Z_{(S^2_\epsilon \times S^1 ) \times \Sigma_\fg} ( \Delta, \fs , \ft, \epsilon)
 = - \frac{2 \pi g_{\text{YM}}^2}{27 \epsilon}
 \left[ a \left( \Delta_i + \frac{\epsilon}{2} \ft_i , \fs_i \right) - a \left( \Delta_i - \frac{\epsilon}{2} \ft_i , \fs_i \right) \right] ,
 }
\ee
with $a (\Delta_i , \fs_i)$ being the trial central charge of the four-dimensional theory obtained by
compactifying the 6d $(2,0)$ theory of type $A_{N-1}$ on $\Sigma_\fg$ with a flavor flux $\fs$ \cite{Bah:2012dg} (see also \cite[(C.7)]{Hosseini:2018uzp})%
\footnote{We can also compare with \eqref{a:SYM}, valid for genus zero. We should set $\epsilon = 0$ in \eqref{a:SYM}, rename $\ft_i$ as $\fs_i$ and enforce $\sum_{i = 1}^2\fs_i = 2 - 2 \fg$ for a generic Riemann surface.
Also, $\Delta_i$ should be rescaled by a factor of $\pi$.}
\be
 \label{a4d:twisted:unrefined}
 a (\Delta_i , \fs_i ) = - \frac{9 N (N^2 - 1)}{128 \pi^3} \sum_{i = 1}^2 \fs_i \frac{\pd (\Delta_1 \Delta_2)^{2}}{\pd \Delta_i} \, .
\ee
Recall that
\be
 \sum_{i = 1}^2 \Delta_i = 2 \pi \, , \qquad \sum_{i = 1}^2 \fs_i = 2 - 2 \fg \, ,
 \qquad \sum_{i = 1}^2 \ft_i = 2 \, .
\ee

\paragraph*{Charged Cardy formula.} The refined topologically twisted index \eqref{RTTI:sym} is expected to reproduce the density of states
of a class of rotating dyonic black strings in AdS$_7\times S^4$ in M-theory whose near horizon geometry is a fibration of AdS$_3$ over the twisted space $S^2_\epsilon \times \Sigma_\fg$.
A class of such strings have been constructed in \cite{Hosseini:2020vgl} wherein it was also shown that the gravitational density of states%
\footnote{The actual computation is done by compactifying the black string on a circle with non-zero momentum and reading the entropy of the corresponding black hole,
in the original spirit of \cite{Strominger:1996sh}.} matches the charged Cardy formula for the dual CFT$_2$. We now show that the same result can be derived from \eqref{RTTI:sym}. 

We interpret our index as the partition function of the 6d $\cN=(2,0)$ $A_{N-1}$ theory on $S^2_\epsilon \times \Sigma_\fg\times S^1\times S^1_{(6)}$,
where $S^1_{(6)}$ is the extra circle opening up at strong coupling.
The modulus $\tau$ of the torus $T^2 = S^1 \times S^1_{(6)}$ 
\be
  \tau = \frac{4 \pi \ii}{g_\text{YM}^2} \, ,
 \ee
is identified with the gauge coupling constant of the five-dimensional theory.
The refined topologically twisted index itself can then be identified with the elliptic genus of the two-dimensional CFT
obtained by compactifying the 6d $(2,0)$ theory on $S^2_\epsilon \times \Sigma_\fg$.

The large $N$ index \eqref{RTTI:sym} can be rewritten as
\be
 \label{RTTI:sym:tau}
 \log Z_{(S^2_\epsilon \times S^1 ) \times \Sigma_\fg} ( \Delta, \fs , \ft, \epsilon)
 = - \frac{8 \ii \pi^2}{27 \tau \epsilon}
 \left[ a \left( \Delta_i + \frac{\epsilon}{2} \ft_i , \fs_i \right) - a \left( \Delta_i - \frac{\epsilon}{2} \ft_i , \fs_i \right) \right] .
\ee
The number of supersymmetric ground states $d_\text{micro}$ is thus given by the Fourier transform of \eqref{RTTI:sym:tau} with respect to $(\tau, \Delta, \epsilon)$,
\be
 \label{density}
 d_{\text{micro}} (\fs , \ft ,e_0, q, J) =
 - \frac{\ii}{(2 \pi)^2} \int_{\mathbbm{i} \mathbb{R}} \rd \beta
 \int_{0}^{2\pi} \rd \Delta \? Z (\fs, \ft, \Delta) \? e^{ \beta e_0 - \ii \Delta Q - \ii \epsilon J} \, ,
\ee
with $\beta \equiv -2 \pi \ii \tau$ and the corresponding integration is over the imaginary axis.
In a saddle point approximation, the number of supersymmetric ground states can obtained by extremizing
\bea
 \label{I:SCFT}
 \cI_{(S^2_\epsilon \times S^1 ) \times \Sigma_\fg}  (\beta , \Delta , \epsilon) & \equiv
 - \frac{16 \pi^3}{27 \beta \epsilon}
 \left[ a \left( \Delta_i + \frac{\epsilon}{2} \ft_i , \fs_i \right) - a \left( \Delta_i - \frac{\epsilon}{2} \ft_i , \fs_i \right) \right] \\
 & + \beta e_0 - \ii \Delta_1 Q_1  - \ii \Delta_2 Q_2 - \ii \epsilon J
 - \Lambda ( \Delta_1 + \Delta_2 - 2 \pi ) \, ,
\eea
with respect to $(\beta, \Delta_1, \Delta_2, \epsilon, \Lambda)$ and evaluating it at its extremum
\bea
 \label{dmicro:I:scft}
 \log d_{\text{micro}} (\fs , \ft ,e_0, Q_1, Q_2, J) & = \cI_{(S^2_\epsilon \times S^1 ) \times \Sigma_\fg} \big|_{\text{crit.}} (\fs , \ft ,e_0, Q_1 , Q_2, J) \\
 & = 2 \pi \Lambda \big|_{\text{crit.}} (\fs , \ft ,e_0, Q_1, Q_2, J) \, .
\eea
Here, we introduced the complex Lagrange multiplier $\Lambda$ that imposes the constraint \eqref{democ:SYM:RTTI} among the chemical potentials and two independent electric charges. As mentioned in section \ref{sec:2}, BPS black objects in AdS have constraints among the charges. For all the known entropy functionals, the constraint arises by requiring that the entropy is real.  We will then fix the relation among the charges by requiring that \eqref{dmicro:I:scft} is a real positive quantity, and we will see later that this is consistent with the gravity dual. The extrema of the $\cI$-functional \eqref{I:SCFT} read
\bea
 \label{I:SCFT:crit}
 \mathring{\Lambda} & = - \ii \frac{\frac{\Xi \mathring{\epsilon}}{8 \pi}
 - J \left( \ft_1 \left( ( Q_1 + Q_2 ) \fs_2 - Q_1 \fs_1 \right) + \ft_2 \left( ( Q_1 + Q_2 ) \fs_1 - Q_2 \fs_2 \right)\right)}
 {J \left( \fs_2 ( \ft_2 - 2 \ft_1 ) + \fs_1 ( \ft_1 - 2 \ft_2 ) \right)} \, , \\
 \mathring{\beta} & = - \ii \frac{N (N^2 - 1)}{24} \frac{\mathring{\epsilon}}{J} \? \ft_1 \ft_2 ( \fs_2 \ft_1 + \fs_1 \ft_2 ) \, , \\
 \mathring{\Delta}_1 & = - \frac{8 \pi J \left( \fs_2 \ft_1 + \fs_1 ( \ft_2 - \ft_1 ) \right) - \mathring{\epsilon} ( Q_1 - Q_2 ) \ft_1 \ft_2 ( \fs_2 \ft_1 + \fs_1 \ft_2 )}
 {4 J \left( \fs_2 ( \ft_2 - 2 \ft_1 ) + \fs_1 ( \ft_1 - 2 \ft_2 )\right)} \, , \\
 \mathring{\Delta}_2 & = - \frac{8 \pi  J \left( \fs_2  ( \ft_1 - \ft_2 ) + \fs_1 \ft_2 \right) + \mathring{\epsilon} ( Q_1 - Q_2 ) \ft_1 \ft_2 ( \fs_2 \ft_1 + \fs_1 \ft_2 )}
 {4 J \left( \fs_2 ( \ft_2 - 2 \ft_1 ) + \fs_1 ( \ft_1 - 2 \ft_2 ) \right)} \, , \\
 \mathring{\epsilon} & = 8 \pi | J | \sqrt{\frac{\fs_2^2 \ft_1^2 + \fs_1 \fs_2 \ft_2 \ft_1 + \fs_1^2 \ft_2^2}{\ft_1 \ft_2 ( \fs_2 \ft_1 + \fs_1 \ft_2 ) \Xi}} \, ,
\eea
where we defined, for the ease of notation,
\bea
 \Xi & \equiv
 4 J^2 \left( \fs_1 ( \ft_1 - 2 \ft_2 ) + \fs_2 ( \ft_2 - 2 \ft_1 ) \right) \\
 & + \ft_1 \ft_2 ( \fs_2 \ft_1 + \fs_1 \ft_2 ) \left[ (Q_1 - Q_2)^2 + e_0 \frac{N(N^2 - 1)}{3} \left( \fs_1 ( \ft_1 - 2 \ft_2 ) + \fs_2 ( \ft_2 - 2 \ft_1 ) \right) \right] .
\eea
We will take $\mathring{\epsilon}$ to be purely imaginary.%
\footnote{This will lead to a unitary CFT in two dimensions.}
Thus, for $ d_{\text{micro}}$ to be real we need
\be
 \label{real:I:SCFT}
 \im \mathring{\Lambda} = 0 \, \quad \Rightarrow
 \quad \frac{Q_2}{Q_1} = - \frac{- \fs_1 \ft_1 + ( \fs_2 \ft_1 + \fs_1 \ft_2 )}{- \fs_2 \ft_2 + ( \fs_2 \ft_1 + \fs_1 \ft_2 )} \, .
\ee
Luckily, this is precisely the constraint among charges for  the dual black strings  in \cite{Hosseini:2020vgl}. The physical interpretation of the constraint is that the black string has zero R-charge, as discussed in  \cite{Hosseini:2020vgl}. Finally, the microscopic degeneracy of states can be put in the following form
\be
 \log d_{\text{micro}} (\fs, \ft, e_0 , Q_1, Q_2 , J ) = 2 \pi \sqrt{\frac{c_{\text{CFT}}}{6} \left( e_0 - \frac{(Q_1 - Q_2)^2}{2 k_{FF}} - \frac{J^2}{2 k} \right)} \, ,
\ee
in agreement with the charged Cardy formula \cite[(5.31)]{Hosseini:2020vgl}, where
\bea
 c_{\text{CFT}} & = 2 N (N^2 - 1) \frac{ \fs_1^2 \ft_2^2 + \fs_1 \fs_2 \ft_1 \ft_2 +\fs_2^2 \ft_1^2}{ \fs_1 (2 \ft_2 -\ft_1) + \fs_2 (2 \ft_1 -\ft_2)} \, , \\
 k & = - \frac{N (N^2 - 1)}{24} \ft_1 \ft_2 (\fs_1\ft_2+\fs_2 \ft_1) \, , \\
 k_{FF} & = - \frac{N (N^2 - 1)}{6} ( \fs_1 (\ft_1-2 \ft_2) + \fs_2 (\ft_2- 2 \ft_1)) \, ,
\eea
are, respectively, the exact central charge, the level of the rotational symmetry, and the flavor symmetry level of the two-dimensional $(0,2)$ CFT.

\section{A mixed index on $( S^2_\epsilon \times S^1 ) \times \Sigma_\fg$}\label{sec:5}

The mixed index on $( S^2_\epsilon \times S^1 ) \times \Sigma_\fg$ was first written  down in  \cite{Jain:2021sdp} by gluing Nekrasov's partition functions.
In this section, we describe the supersymmetric background for the CFT partition function,
and we recover the result using supersymmetric localization. 
We begin by describing
the rigid 5d $\mathcal{N}=1$ supergravity background, with topology
$S^{2}\times S^1\times \Sigma_{\fg}$, on which the
CFT lives. The metric on $S^2 \times S^1$ will
be such that the $S^2$ is metrically fibered over the
$S^1$, corresponding to angular momentum. The background
is the same as the 3d background which can be used to compute the
three-dimensional superconformal index. We will use a topological twist on $\Sigma_{\mathfrak{g}}$,
so that the metric on $\Sigma_{\mathfrak{g}}$ is irrelevant, except
for $\mathfrak{g}=0$ where another angular momentum can be introduced
when the metric admits a continuous isometry. After reduction on the
time circle, this type of angular momentum is equivalent to an $\Omega$-background.
Our metric, spinor, and supergravity conventions are mostly as in
\cite{Hosseini:2018uzp}. Spinor conventions are collected in appendix
\ref{sec:Spinor-conventions}, and supersymmetry conventions in appendix
\ref{sec:Supersymmetry-conventions}. Note that all of the spacetimes
we consider are spin, and we therefore do not need to introduce $\text{Spin}_{\mathbb{C}}$
bundles. 

The gravity side of the holography, in the case under consideration,
consists of black hole solutions of 6d matter coupled $F (4)$
gauged supergravity. The ordinary $F(4)$ gauged supergravity,
introduced by Romans in \cite{Romans1986}, was considered in the
context of holography in \cite{Nishimura2000}.
Therein, the boundary supergravity, which is coupled to a 5d
$\mathcal{N}=1$ superconformal theory, can be described using the
ordinary Weyl multiplet of 5d superconformal tensor calculus \cite{Fujita:2001kv,Kugo:2000hn,Bergshoeff2001,deWit:2009de},
described for instance in \cite{deWit:2017cle}. There are hardly
any remnants, on the boundary, of the complexities of the transformations
in Romans supergravity. According to \cite{Nishimura2000}, we can
identify the boundary metric, $\SU(2)_{R}$ gauge field,
and antisymmetric tensor directly with asymptotic values of the bulk
supergravity. These are the fields which are turned on for the simplest
black hole solutions. More complicated solutions come from the matter
coupled supergravity. The boundary theory in that case is coupled
to the Weyl multiplet plus additional background vector multiplets.
The conditions for preserving supersymmetry in the boundary CFT are
best derived by looking at the Weyl multiplet first, and adding the
matter multiplets on top. 

\subsection{Rigid supergravity background}

A rigid bosonic supersymmetric background is a fixed point of the supersymmetry
transformation of the Weyl multiplet \cite{Festuccia2011,Dumitrescu:2012at,Dumitrescu:2012ha}.
We will follow the description of the Weyl multiplet given in \cite{deWit:2017cle},
translated into the notation of \cite{Hosseini:2018uzp} using appendix
\ref{sec:Supersymmetry-conventions}. The relevant supergravity fields
are the vielbein ${e_{\mu}}^{a}$, the antisymmetric tensor $T_{ab}$,
and an $\SU(2)$ R-symmetry gauge field $A_{\mu}^{(\text{R})}$.%
\footnote{The backgrounds we consider are very similar to those discussed in
\cite{Festuccia2020} in the context of four-dimensional $\mathcal{N}=2$ supergravity.
It seems likely that the background in \cite{Festuccia2020} can be
uplifted, and that the analysis can be carried over to the five-dimensional case,
although we were not able to verify this.}

The supersymmetry transformation of the gravitino in the Weyl multiplet reads
\be
 \label{dpsi}
 \delta\psi_{\mu}=\mathcal{D}_{\mu}\xi+\frac{\ii}{4}T_{ab}\left(3\gamma^{ab}\gamma_{\mu}-\gamma_{\mu}\gamma^{ab}\right)\xi-\gamma_{\mu}\tilde{\xi} \, ,
\ee
where 
\be
\mathcal{D}_{\mu}\xi\equiv\partial_{\mu}\xi+\frac{1}{4}{\omega_{\mu}}^{ab}\gamma_{ab}\xi+\xi\left(A_{\mu}^{\left(\text{R}\right)}\right)^{\text{T}}\,.
\ee
Note that 
\be
\frac{\ii}{4}T_{\nu\rho}\left(3\gamma^{\nu\rho}\gamma_{\mu}-\gamma_{\mu}\gamma^{\nu\rho}\right)=\frac{\ii}{2}T_{\nu\rho}\left({\gamma_{\mu}}^{\nu\rho}-4{\delta_{\mu}}^{\nu}\gamma_{\rho}\right)\,,
\ee
so that the antisymmetric tensor field of \cite{Nishimura2000}
can be identified with the one used in our Weyl multiplet as 
\be
B_{\mu\nu}=-\frac{1}{2}T_{\mu\nu} \, .
\ee
The identification of the other fields is equally straightforward.

We will describe in detail a solution to the equation $\delta\psi_{\mu}=0$
corresponding to a manifold with topology $( S^2 \times S^1 ) \times S^2$,
where a topological twist will be applied to the second $S^2$
factor. The solution is simple to derive by considering an ansatz
whereby the Killing spinor is the product of the ones for the 3d superconformal
index, and a constant spinor on the twisted $S^2$. More
general manifolds, with topology $( S^2\times S^1) \times\Sigma_{\mathfrak{g}}$
for $\mathfrak{g}\ge1$, can be included by simply changing the factor
$\sin\eta$ in the metric, see \eqref{vielbein}, to $1$, in the case of $\mathfrak{g}=1$,
or to $\sinh\eta$ in the case of $\mathfrak{g}\ge2$. When $\mathfrak{g}=1$,
$\Sigma_{\fg}$ is a torus, \ie\;a quotient of $\mathbb{R}^{2}$, and the
supergravity fields and equations are independent of its coordinates.
When $\mathfrak{g}\ge2$, the Riemann surface $\Sigma_{\mathfrak{g}}$
is a quotient of the hyperbolic plane, and we can solve the gravitino
equations by considering those on $( S^2\times S^1) \times\mathbb{H}_{2}$
and noting that the Killing spinor is independent of the coordinates
of $\mathbb{H}_{2}$. The definition of the twisted fields, and other
aspects of localization considered below, are mostly independent of
$\mathfrak{g}$. However, when considering a compact $\Sigma_{\mathfrak{g}}$
with $\mathfrak{g}\ge1$, one should turn off the $\Omega$-deformation
parameter $\tilde{\epsilon}_{2}$, since such a manifold does not admit a
continuous isometry with fixed points.

\paragraph*{The background for $\mathfrak{g}=0$.}

We choose the following vielbein on $( S_{\theta,\phi}^{2}\times S_{\tau}^{1} ) \times S_{\eta,\varphi}^{2}$
\bea
 \label{vielbein}
 e^{1} & = r \rd\theta \, ,\qquad && e^{2}=r\sin\theta \? (\rd\phi-\beta\tilde{\epsilon}_{1}\rd\tau ) \, ,\qquad e^{5}=\beta \rd\tau \, ,\\
 e^{3} & = r_{2}\rd\eta \, ,&& e^{4} =r_{2}\sin\eta \? (\rd\varphi-\beta\tilde{\epsilon}_{2}d\tau ) \, .
\eea
The real geometric parameters $\tilde{\epsilon}_{1,2}$ correspond to the real part of the angular momentum fugacities when the partition function on this background is viewed as an index. We also take an R-symmetry connection 
\be
A^{\left(\text{R}\right)}=\left(-\beta\tilde{\epsilon}_{1} \rd \tau + \cos \eta \?  (\beta\tilde{\epsilon}_{2} \rd \tau - \rd \varphi ) \right) \? \tau_{3}\,,
\ee
corresponding to a partial twist on both the sphere and the part of
the spin connection related to the angular momentum on the twisted
sphere, proportional to $\tilde{\epsilon}_{2}$. Lastly, we turn on the antisymmetric
tensor field $T$ 
\be
 T_{12}=\frac{1}{6r} \, .
\ee
Observe that $T$ is covariantly constant. 

One may check that the Killing spinor equation $\delta \psi_\mu = 0$, see \eqref{dpsi}, is satisfied for $\xi$
of the form described below, and that the vanishing of the variation
of the dilatino is likewise guaranteed as long as one sets the supergravity
field $D$ to the value\footnote{For $\mathfrak{g}>1$ one should change the sign of the second term,
while for $\mathfrak{g}=1$ the second term is absent.} 
\be
D = \frac{1}{6 r^{2}}+\frac{1}{16 r_{2}^{2}} \,.
\ee

This background actually preserves $2$ supercharges of the type covered
by our ansatz. In the limit $\tilde{\epsilon}_{1} \to 0$, the number
of supercharges is further enhanced to $4$. The same is true for
$\mathfrak{g}\ge 2$ with the appropriate change of vielbein and with
$\tilde{\epsilon}_{2} \to 0$. For $\mathfrak{g}=1$, the amount of
supersymmetry is double that of $\mathfrak{g}\ne 1$ in every scenario.

\paragraph*{The superalgebra.}

We choose a specific Killing spinor, which in the notation of appendix
\ref{sec:Spinor-conventions} takes the form\footnote{Note that $\xi$ does not satisfy the symplectic Majorana reality
condition \eqref{eq:symplectic_majorana_condition}. This is reminiscent of the situation for the 3d superconformal index.}
\bea
\xi=\frac{1}{2}\begin{pmatrix}0 & \ii e^{\frac{\ii\phi}{2}}\sin\frac{\theta}{2}\\
-\ii e^{-\frac{\ii\phi}{2}}\sin\frac{\theta}{2} & 0\\
e^{-\frac{\ii\phi}{2}}\cos\frac{\theta}{2} & 0\\
0 & e^{\frac{\ii\phi}{2}}\cos\frac{\theta}{2}
\end{pmatrix}\,.
\eea
 We define $\delta$ to be the supersymmetry transformation associated
with $\xi$. The superalgebra generated by $\delta$ contains a number
of bosonic transformations:
\begin{enumerate}
\item An infinitesimal diffeomorphism with parameter $\ii v$, where
\be
v\equiv\frac{1}{\beta}\partial_{\tau} + \Big(\tilde{\epsilon}_{1}+\frac{\ii}{r}\Big) \partial_{\phi}+\tilde{\epsilon}_{2}\partial_{\varphi}\,.
\ee
\item An infinitesimal R-symmetry transformation given by the matrix
\begin{equation}
\Lambda=-\frac{1}{2} \Big( \tilde{\epsilon}_{1}+\frac{\ii}{r} \Big) \sigma_{3}\,,\label{eq:R-symmetry}
\end{equation}
such that $\xi$ transforms as 
\be
\xi\rightarrow\xi\Lambda^{\text{T}}\,.
\ee
\item A gauge transformation which depends on the field realization.
\end{enumerate}
One can explicitly check that there is no Weyl transformation, nor
conformal isometries, present in the square of the supersymmetry preserved
on this background. This is a necessary, but not sufficient, condition
for using localization. 

Note that the condition \eqref{eq:Gauge_fixed_sugra_t} cannot be solved
for the pair $\xi$, $\tilde{\xi}$, hence the background considered here
is not reachable from the gauge fixed supergravity used in \cite{Hosomichi:2012ek,Kallen:2012cs,Kallen:2012va,Qiu:2016dyj}.
For continuity with \cite{Hosseini:2018uzp}, we will nevertheless
continue to use the notation from \cite{Hosomichi:2012ek} for the
matter fields.

\subsection{Localization }

We would like to perform localization for an arbitrary $\mathcal{N}=1$
theory on the background described above. We will describe first the
moduli space arising from such localization. This space is by definition
the vanishing locus for the supersymmetry transformations of all the
fermions in the theory. However, the fields living in hypermultiplets,
assuming that they have generic masses coming from background flavor
multiplets, will not contribute any moduli. We therefore examine here
the moduli coming from the vector multiplets, which are the configurations
for which the transformation of the gaugino vanishes. 

The supersymmetry transformation of the gaugino reads
\be
\delta\lambda_{I}=-\frac{1}{2}\Gamma^{mn}\left(F_{mn}-4\sigma T_{mn}\right)\xi_{I}-\ii D_{m}\sigma\Gamma^{m}\xi_{I}-\ii D_{IJ}\xi^{J}-2\ii\sigma\tilde{\xi}_{I} \, .
\ee
One can show that the following configuration solves the BPS equation
$\delta\lambda=0$
\bea
F^{ (0 )} & =\frac{\mathfrak{n}}{2}\left(\frac{1}{r^{2}}\text{Vol}(S^2 )+\beta\tilde{\epsilon}_{1}\sin\theta\? \rd\theta\wedge \rd\tau\right)
+ \frac{\mathfrak{m}}{2}\left(\frac{1}{ r_{2}^{2}}\text{Vol} (\Sigma_{\fg} ) + \beta\tilde{\epsilon}_{2}\sin\eta\? \rd\eta \wedge \rd\tau \right ) ,\label{eq:Coulomb_branch_moduli_space} \\
\sigma^{ (0 )} & =\frac{\mathfrak{n}}{2r}\,,\qquad D_{12}^{ (0 )}=\frac{\mathfrak{m}}{2 r_{2}^{2}}\,,
\eea
where $\mathfrak{m}$ and $\mathfrak{n}$ are constants taking values in the magnetic weight lattice of $G$.
They correspond to fluxes for dynamical or background gauge fields. The BPS configuration also implicitly includes
a flat $G$ connection commuting with $\mathfrak{m}$ and $\mathfrak{n}$
which, given the topology of the space, can be taken to be a spacetime
independent profile for $A_{\tau}$.%
\footnote{When $\fg\ge1$, there exist additional factors of the moduli space
of flat connections coming from the holonomy on the non-contractible
cycles of $\Sigma_{\fg}$. These do not appear as deformations of the
superalgebra and play no role in localization.}
For generic $\mathfrak{m,n}$, this flat connection is in the
same Cartan subalgebra as $\mathfrak{m}$ and $\mathfrak{n}$ . 

We henceforth work with the Cartan subalgebra defined by the above
BPS configurations. These define the Coulomb branch of BPS pseudo-vacua.
Due to large gauge transformations wrapping the time circle, the Cartan
elements of the flat connection, denoted by $a_{i}^{\text{flat}}$, are compact
\be
 a_{i}^{\text{flat}}\sim a_{i}^{\text{flat}}+\frac{2\pi}{\beta}n\,,\qquad n\in\mathbb{Z} \, .\label{eq:a_periodicity}
\ee
Note also, that unlike the situation in \cite{Hosseini:2018uzp} the
Cartan elements of $\sigma^{\left(0\right)}$ are fixed by the value
of the fluxes, and do not define a separate non-compact direction
of the Coulomb branch moduli space. 

The BPS configurations include $D_{12}\propto\mathfrak{m}$, and
therefore do not satisfy the original reality conditions for the field
$D_{IJ}$. Instead, convergence of the path integral with a measure
defined by the classical action of the gauge theory sets $D_{12}$
to be purely imaginary, or at least to have a bounded real part, if one makes
the standard rotation from Lorentz signature \cite{Hosseini:2018uzp}.
Configurations with auxiliary fields not satisfying the original reality conditions are common in solutions to the BPS equations of twisted gauge theories \cite{Benini:2015noa,Benini:2016hjo,Hosseini:2018uzp}.
One can try to explain their appearance by considering a $\delta$-exact
mixing term between an auxiliary field and a dynamical scalar. Adding
this term, and integrating over the original contour, effectively
sets the auxiliary field to a complex value in the BPS equation. The
need to add such a mixing term, in the context of localization, is
usually attributed to the existence of fermionic zero modes \cite{Bershtein:2015xfa},
or more generally to the need to lift some of the moduli \cite{Witten:1992xu}.
In some cases, they can also be understood from the point of view
of the effective low energy theory \cite{Dabholkar2020}, where they represent a breakdown of localization associated with contributions
from the boundary of field space. We do not have a good understanding
of the arguments for the necessity of including these configurations
as they apply to the situation at hand. However, the consistency of
the calculation implies that they should be included.

\subsubsection{Twisted fields}

To examine the moduli space more closely, and to deduce the
one loop exact effective action, it is convenient to define twisted
fields. Our definition of these fields is very similar to the one
given in \cite{Qiu:2016dyj}, although the non-pseudo-Majorana nature
of our Killing spinor prevents us from using the expressions in that
work directly.

\paragraph*{Vector multiplets.}

We first define the spinor bilinears 
\be
\kappa_{a}\equiv\bar{\xi}\gamma_{a}\xi \, ,\qquad s\equiv\bar{\xi}\xi \, .
\ee
Note that the scalar $s$ vanishes on a great circle in the untwisted
$S^2$, the equator with respect to the fixed points of
$v$, that $\kappa^{a}\kappa_{a}=s^{2}$, and that at the poles $s=\pm1$
and $\kappa_{a}={\delta_{a}}^{5}$. The vanishing of $s$ means that
the localization here is not of the type examined in \cite{Qiu:2016dyj}
or in \cite{Hosseini:2018uzp}. 

We define the following operators acting on a 2-form $\Xi$
\bea
 \Xi_{V} & \equiv\kappa\wedge i_{\kappa}\Xi \, ,\qquad \Xi_{H}\equiv\Xi-\Xi_{V} \, , \label{eq:2-form-operations}\\
 \Xi_{\pm} & \equiv\frac{1}{2}\left(\Xi_{H}\pm\frac{1}{2s}i_{\kappa}\star\Xi_{H}\right) . 
\eea
Note that 
\be
\Xi=\Xi_{V}+\Xi_{+}+\Xi_{-} \, ,
\ee
but that the decomposition is a complex one.

We now define the following twisted fields
\bea
 \Psi_{a}\equiv\bar{\xi}_{I}\gamma_{a}\lambda^{I} \, ,\qquad
 \chi_{ab}\equiv s\bar{\xi}_{I}\gamma_{ab}\lambda^{I}-\left(\kappa_{a}\Psi_{b}-\kappa_{b}\Psi_{a}\right) , \qquad
 \Phi\equiv s\sigma-\ii i_{\kappa}A\,,
\eea
and the auxiliary field 
\bea
 \label{eq:twisted_auxilliary_field}
 H_{ab} \equiv2s^{2}F_{-ab}+\left(s^{2}-1\right)F_{Vab}+2s\bar{\xi}_{I}\gamma_{ab}\xi_{J}D^{IJ}
 +6s\ii\bar{\tilde{\xi}}_{I}\gamma_{ab}\xi^{I}\sigma+\ii\sigma\left(\kappa\wedge ds\right)_{ab} \, . 
\eea
One can show that the change of variables from $\lambda$ to $\Psi_{a}$ and $\chi_{ab}$
is nonsingular, even at the equator. One can also verify that the following
projections hold
\be
 \chi_{V}=\chi_{+}=0 \,, \qquad H_{V}=H_{+}=0 \,.
\ee
There is a perhaps more natural change of variables, and definition
of (anti)self-duality, for the fermions which makes $\chi$ horizontal
only at the poles and also either self-dual or anti-self dual depending
on which pole. In the end, it makes no difference for the localization
computation. 

After the change of variables, we get the following supersymmetry
transformations for the twisted fields
\bea
\delta A & =\ii\Psi \, ,\qquad\delta\Psi=i_{\kappa}F-\ii D\left(s\sigma\right)=\mathcal{L}_{\kappa}A-\ii D\Phi\,,\\
\delta\chi& = H \, , \qquad\delta H=\mathcal{L}_{\kappa}\chi+\ii\left[\Phi,\chi\right]\,,\qquad\delta\Phi=0\,.
\eea
We can compare these expressions to the ones for either the twisted
backgrounds of \cite{Hosseini:2018uzp}, or the contact type backgrounds
of \cite{Qiu:2016dyj}. The only difference is the appearance of powers
of $s$. Had $s$ been not just nowhere vanishing but actually constant,
as was the case for both \cite{Hosseini:2018uzp} and \cite{Qiu:2016dyj},
we would recover the full set of twisted fields defined in those papers.%
\footnote{Note that $\tilde{\xi}$, which is related to the field $t_{IJ}$
of \cite{Qiu:2016dyj}, vanishes for the twisted backgrounds of \cite{Hosseini:2018uzp}.}
As in \cite{Hosseini:2018uzp}, we define 
\be
a\equiv-\ii\Phi^{\left(0\right)}=a^{\text{flat}}\,.
\ee

\paragraph*{Hypermultiplets.}

The off-shell closed hypermultiplet introduced in appendix \ref{sec:Supersymmetry-conventions}
can be twisted in the same manner as the vector multiplet. We define
\bea
q \equiv\xi_{I}q^{I}\,,\qquad\psi_{q}\equiv-\ii\delta q \, , \qquad
\psi_{F} \equiv\mathcal{P}_{F}\psi\,,\qquad F\equiv\delta\psi_{F}\,.
\eea
The operator $\mathcal{P}_{F}$ is a Hermitian spin projector which
selects half of the components of the spinor $\psi$. It is chosen
so as to have rank $2$, to annihilate $\xi_{I}$
\be
\mathcal{P}_{F}\xi_{I}=0\,, \qquad \text{ for }  \quad I = 1, 2\,,
\ee
and to be invariant under $\delta^{2}$, \ie\;
\[
\mathcal{L}_{\kappa}\mathcal{P}_{F}=0\,.
\]
The precise form of $\mathcal{P}_{F}$ then determines the spinor
$F$ in terms of $F^{I}$ and some expression linear in $q^{I}$.
In our chosen coordinates and vielbein, we can take
\bea
{\left(\mathcal{P}_{F}\right)_{\beta}}^{\alpha}=\frac{1}{2}\begin{pmatrix}1-s & 0 & 0 & \ii\sqrt{1-s^{2}}\\
0 & 1-s & -\ii\sqrt{1-s^{2}} & 0\\
0 & \ii\sqrt{1-s^{2}} & 1+s & 0\\
-\ii\sqrt{1-s^{2}} & 0 & 0 & 1+s
\end{pmatrix}\,.
\eea
The properties of $\mathcal{P}_{F}$ guarantee that the action of
$\delta^{2}$ on the twisted fields is uniform, and of course coincides
with the one on the vector multiplet. $\mathcal{P}_{F}$ can also
be written in the convenient form 
\be
\mathcal{P}_{F}=\frac{1}{2}\left(\kappa^{a}\gamma_{a}+s\mathbbm{1}_{4}\right)\gamma_5\,,
\ee
generalizing the expression for the hypermultiplet projections in
\cite{Qiu:2016dyj}.

\subsubsection{Localizing terms}

We can form a localizing term for the vector multiplet using the twisted
fields as in \cite{Hosseini:2018uzp}
\be
\delta\mathcal{V}_{\text{gauge}}\equiv\delta\int \sqrt{g} \? \text{Tr} \Big(2\ii\chi\wedge\star F+\frac{1}{2}\Psi\wedge\star\left(\delta\Psi\right)^{*} \Big) \, .
\label{eq:cohomological_localizing_term}
\ee
The second term in \eqref{eq:cohomological_localizing_term} is positive
semi-definite and vanishes only at 
\begin{gather}
i_{\kappa}F-\ii D\left(s\sigma\right)=0\,.\label{eq:moment_map}
\end{gather}
This equation is indeed compatible with the Coulomb branch moduli
space \eqref{eq:Coulomb_branch_moduli_space}. The imaginary part implies
that $s\sigma$ is the moment map for the action of the imaginary
part of $\kappa$ with respect to the symplectic form $F$, which
in turn implies that $\sigma$ is constant. The real part implies
that $A_{\tau}$ is a moment map for the real part of $\kappa$ with
respect to the same symplectic form.\footnote{This is true in a gauge where $A$ is $\tau$ independent.} 

Had $\xi$ satisfied the symplectic Majorana condition \eqref{eq:symplectic_majorana_condition},
the expression $\bar{\xi}_{I}\gamma_{ab}\xi_{I}D^{IJ}$ would be real
and hence $H_{ab}$ would be integrated over a shifted real contour.
The first term in \eqref{eq:cohomological_localizing_term} would then
yield a Lagrange multiplier imposing the real constraint $F_{+}=0 \,$.
As it stands, however, $\xi$ does \emph{not} satisfy the symplectic Majorana
condition, $\bar{\xi}_{I}\gamma_{ab}\xi_{J}D^{IJ}$ is not real, and
it is not clear whether the first term in \eqref{eq:cohomological_localizing_term}
can be used as a localizing term. Specifically, the contour over which
$H$ should be integrated needs to be somehow invented. Instead of
finding the correct contour, we can consider the standard localizing
term 
\be
 \delta\mathcal{V}_{\text{gauge alt.}}\equiv\int\sqrt{g} \? \text{Tr}\left(\lambda^{I}\left(\delta\lambda\right)_{I}^{*}\right) .
 \label{eq:gauge_localizing_term}
\ee
This term vanishes on the moduli space by construction. The localizing
term for the hypermultiplets is simply the analogue of the one used
in \cite{Hosseini:2018uzp}
\be
\delta\mathcal{V}_{\text{matter}}\equiv\delta\int\sqrt{g}\left(\psi_{q}^{A}\left(\delta\psi_{q}\right)_{A}^{*}+\psi_{F}^{A}\left(\delta\psi_{F}\right)_{A}^{*}+\psi_{F}^{A}\Gamma^{m}D_{m}q_{A}\right) .
\label{eq:matter_localizing_term}
\ee

\subsubsection{Fluctuations}

We would like to compute the effective action for the supersymmetric moduli
of \eqref{eq:Coulomb_branch_moduli_space}, which is given by a one loop calculation.
We will do this using
a mixture of the equivariant index theorem for transversally elliptic
operators, and the results of Nekrasov \cite{Nekrasov:2002qd} on the local contributions to such an index from a neighborhood of a fixed circle for the equivariant
action modeled on $\mathbb{C}^{2}\times S^{1}$.

\paragraph*{Vector multiplets.}

We will use the standard background gauge 
\be
D_{\mu}^{\left(0\right)}a^{\mu}=0 \, ,
\ee
where $D_{\mu}^{ (0 )}$ is the covariant derivative with
respect to a background connection leading to \eqref{eq:Coulomb_branch_moduli_space}.
Specifically, it is a connection which itself satisfies the gauge
condition $\nabla^{\mu}a_{\mu}^{\left(0\right)}=0\,$. Such a representative
is guaranteed to exist. We fix the gauge by adding a term to the action
given by 
\be
\delta_{\text{BRST}}\int\sqrt{g} \? \tilde{c} \?D_{\mu}^{\left(0\right)}a^{\mu} \, .
\ee

We would like to verify that the relevant operator coming from $\mathcal{V}$
is transversally elliptic. According to \cite{Pestun:2007rz,Gomis:2011pf},
the relevant operator is $D_{\text{oe}}\subset\mathcal{V}$, which
is defined as the order zero term acting between the fields $\varphi_{\text{e}}$
and $\varphi_{\text{o}}$. In our case $\varphi_{\text{e}}=a_{\mu}$, while $\varphi_{\text{o}}=\left\{ \chi,c,\bar{c}\right\} $,
where $c,\bar{c}$ are the ghost and anti-ghost in the gauge fixing
multiplet \cite{Pestun:2007rz}. We find $D_{\text{oe}}$ by expressing
$\lambda$ in $\mathcal{V}$ as the unique linear combination of $\chi$
and $\Psi$ derived above. We then expand $\mathcal{V}$ around the
forms in the non-coordinate basis: $a_{a}$, $\chi_{ab}$, $c$, $\bar{c}$.
We denote the corresponding momenta by $p_{a}$. We find that the (leading) matrix
symbol of $D_{\text{oe}}$ is given by\footnote{We have ignored various numerical constants which multiply entire
rows or columns, since these do not affect the invertibility of the
symbol.} 
\bea
\begin{pmatrix}-\frac{p_{4}}{2s^{2}} & \frac{p_{3}}{2s} & -\frac{p_{2}}{2s} & \frac{p_{1}}{2s^{2}} & 0\\
-\frac{p_{3}}{2s^{2}} & -\frac{p_{4}}{2s} & \frac{p_{1}}{2s^{2}} & \frac{p_{2}}{2s} & 0\\
-\frac{p_{2}}{2s^{2}} & \frac{p_{1}}{2s^{2}} & \frac{p_{4}}{2s^{3}} & -\frac{p_{3}}{2s^{3}} & 0\\
\frac{p_{4}}{2s^{3}} & -\frac{p_{3}}{2s^{2}} & \frac{p_{2}}{2s^{2}} & -\frac{p_{1}}{2s^{3}} & 0\\
-\frac{p_{3}}{2s^{3}} & -\frac{p_{4}}{2s^{2}} & \frac{p_{1}}{2s^{3}} & \frac{p_{2}}{2s^{2}} & 0\\
\frac{p_{2}}{2s} & -\frac{p_{1}}{2s} & -\frac{p_{4}}{2s^{2}} & \frac{p_{3}}{2s^{2}} & 0\\
-p_{1}p_{5} & -p_{2}p_{5} & -p_{3}p_{5} & -p_{4}p_{5} & p_{1}^{2}+p_{2}^{2}+p_{3}^{2}+p_{4}^{2}\\
p_{1} & p_{2} & p_{3} & p_{4} & p_{5}
\end{pmatrix}.
\eea
The top $6$ rows correspond to $\chi_{ab}$, row $7$ to $c$, and
row $8$ to $\bar{c}$. One can clearly see that the rows $\left\{ 4,5,6\right\} $
are multiples of the rows $\left\{ 1,2,3\right\} $. This is the reflection
of the fact that $\chi_{ab}$ is actually constrained, and has only
$3$ degrees of freedom. We choose to remove the rows $\left\{ 4,5,6\right\} $,
and to call the remaining $5\times5$ matrix the matrix symbol of
$D_{\text{oe}}$, denoted $\sigma\left(D_{\text{oe}}\right)$. We
can now evaluate the determinant of $\sigma\left(D_{\text{oe}}\right)$.
It is given by 
\be
\det \sigma \left(D_{\text{oe}}\right)=-\frac{1}{8s^{6}}\vec{p}^{\,2} \left(\vec{p}^{\,2}-p_{5}^{2}\right)\left(p_{1}^{2}+p_{3}^{2}+p_{4}^{2}+s^{2}p_{2}^{2}\right) .
\ee
Note that the singular factor of $s^{-6}$ is an artifact of the coordinate
system and our definition of the twisted fermions and plays no physical
role. 

In order to check transversal ellipticity, we must verify that $\det\sigma \left(D_{\text{oe}} \right)$
is nonzero on the part of the tangent space which is orthogonal to
the equivariant action. This is equivalent to evaluating the simultaneous
vanishing locus of $\det\sigma \left(D_{\text{oe}} \right)$ and of
\be
\left\Vert \kappa^{a}p_{a}\right\Vert ^{2}=p_{5}^{2}+ (1-s^{2} ) \? p_{2}^{2} \, .
\ee
One can easily see that this locus is $\vec{p}=0$, independent of
the value of $-1\le s\le1$, and hence the symbol is transversally
elliptic. Note that $\det\sigma\left(D_{\text{oe}}\right)$ is not
transversally elliptic with respect to the equivariant action on the
time circle alone, since the term $\left(p_{1}^{2}+p_{3}^{2}+p_{4}^{2}+s^{2}p_{2}^{2}\right)$
can vanish when $p_{1}=p_{3}=p_{4}=s=0$ with $p_{2}$ arbitrary.
This is the result of having an untwisted $S^2$. In the
fully twisted case considered in \cite{Hosseini:2018uzp}, the corresponding
operator is elliptic once the time circle is taken into account, \ie\;once $p_{5}$ is removed. 

Comparing with the expressions in \cite{Pestun:2007rz}, we see that
$\sigma \left(D_{\text{oe}} \right)$ coincides with the standard self-dual complex on $S^2\times S^2$ at $s=-1$, $\theta=0$,
and with the anti-self-dual complex at $s=1$, $\theta=\pi$, extended
along the $5^{\text{th}}$ direction. 

\paragraph*{Hypermultiplets.}

The symbol for the hypermultiplet localizing term \eqref{eq:matter_localizing_term}
is the $2\times2$ matrix
\be
D_{\text{oe}}^{\text{hyper}}=\hat{\xi}^{\dagger I}\Gamma^{m}D_{m}\xi_{J}\,,
\ee
tensored with $\Omega_{AB}$, where we have taken care to project
the 5d Dirac operator onto the space of $\varphi_{\text{e}}$, spanned by
$q$ and hence by $\xi_{J}$, and to the conjugate space of $\varphi_{\text{o}}$
spanned by $\psi_{F}$ and hence by $\hat{\xi}_{I}$. We can now evaluate 
\bea
\sigma\left(D_{\text{oe}}^{\text{hyper}}\right) & =-\ii\begin{pmatrix}p_{3}+\ii p_{4} & e^{\ii\phi}\left(p_{1}+\ii sp_{2}\right) \\
-e^{-\ii\phi}\left(p_{1}-\ii sp_{2}\right) & p_{3}-\ii p_{4}
\end{pmatrix} ,\\
\det \sigma\left(D_{\text{oe}}^{\text{hyper}}\right) & = - \left(p_{1}^{2}+p_{3}^{2}+p_{4}^{2}+s^{2}p_{2}^{2}\right) .
\eea
At $s=\pm1$, this is equivalent to the symbol for the 4d Dirac operator,
acting between $S^{+}$ and $S^{-}$, and its complex conjugate. $S^{\pm}$
are the positive and negative chirality spin bundles of a four-manifold.
Clearly, $\sigma (D_{\text{oe}}^{\text{hyper}} )$ is transversally
elliptic with respect to $\kappa$. It is not, however, transversally
elliptic with respect to the equivariant action on the time circle
alone, for the same reasons as the symbol for the vector multiplet
analyzed above. Again, this is the result of having an untwisted $S^2$.

\subsection{Derivation of the partition function}
\label{subsec:mixedindexpart}

Here, we set $\beta=r_{2}=1$. Given the facts in
the previous section, it is reasonable to expect that a theory on
this background localizes onto the instanton/anti-instanton complex
at the fixed points. We therefore expect to get a contribution from
the fixed points which is the same as in the twisted case, from the
north pole of the untwisted $S^2$, and the complex conjugate
contribution at the south pole. We must integrate these contributions
along the continuous modulus coming from the flat connection. Note
that there is no continuous modulus coming from the vev of $\sigma$,
since this is fixed by the moment map equation \eqref{eq:moment_map}.
The modulus $a$ is therefore integrated over a shifted real
contour. We must also sum over the localized instantons/anti-instantons
and over the two sets of fluxes $\mathfrak{m},\mathfrak{n}$. 

When evaluating the partition function on our spaces, using either Nekrasov's instanton partition function or the equivariant index theorem,
the effective values of the real part of the angular momentum parameters at the north and south poles of the untwisted two-sphere have a relative minus sign.\footnote{$\tilde{\epsilon}_1$ is the coefficient of the rotation with respect to $\phi$, which rotates the local frame at the north and south poles in opposite directions.}
Define the following complex angular momentum parameter 
\be
\epsilon_{1}^{\text{N/S}}=\tilde{\epsilon}_{1}^{\text{N/S}}-s^{\text{N/S}}\frac{\ii}{r}\,,
\ee
where $\tilde{\epsilon}_{1}^{\text{N/S}}$ is the plus or minus real geometric
parameter $\tilde{\epsilon}_{1}$, and $s^{\text{N/S}}$ takes the values $\mp1$ when
the point in question is localized at the north and south poles of
the untwisted sphere, respectively. The linear combinations $\epsilon_{1}^{\text{N/S}}$ are all that ever appear in the calculation,
whether as  parameters of the equivariant action, or as coefficients multiplying the flux $\mathfrak{n}$. We henceforth work only with this combination, and denote
\be
 \epsilon_{1} \equiv \tilde{\epsilon}_{1}+\frac{\ii}{r} \, .
\ee
There is no analogous shift for $\tilde{\epsilon}_2$, which we now rename to $\epsilon_2$. 

For the case of $\mathfrak{g}=0$, we can simply make use of the 5d
Nekrasov partition function to compute the partition function. The
conjecture in \cite{Nekrasov:2003vi} implies that the full result
can be written as a sum over equivariant fluxes and an integral over
$a$ of $4$ copies of the 5d Nekrasov partition function associated
to the $4$ fixed points. We have derived the fact that the partition function actually
depends on only two fluxes: $\mathfrak{m,n}$ which correspond to
the homology two cycles of the spacetime manifold. We therefore write the result as
\be
Z_{(S^2_{\epsilon_1}  \times S^1 ) \times S^2_{\epsilon_2}}=\sum_{\left\{ \mathfrak{m},\mathfrak{n}\right\} }\oint \rd a \prod_{l=1}^{4}Z_{\bC^2 \times S^1} \big(a^{ (l )} ; \epsilon_{1}^{ (l )},\epsilon_{2}^{ (l )} \big) \, .
\ee
We choose a parametrization of these fluxes which is suited to the present context. In this parametrization, the parameters $a^{\left(l\right)}$ in the fully twisted case would be given by\footnote{The factors of $1/2$ have been inserted in order to match the
quantization conditions on $\mathfrak{m,n}$ with the expression given
in terms of the parameters $\mathfrak{p}_{l}$ (\cf\;\cite[Sect.\,2.7.2]{Hosseini:2018uzp}).}
\[
a^{\left(l\right)}=a-\epsilon_{1}^{\left(l\right)}\frac{\mathfrak{n}}{2}+\epsilon_{2}^{\left(l\right)}\frac{\mathfrak{m}}{2}\,.
\]
The fixed point data is given in table table \ref{tab:epsilon_values}.\footnote{The fixed point data given in \cite{Hosseini:2018uzp} is adapted
to a general toric geometry. We prefer to use a simpler table for
the specific manifold considered here. } The points $1,2$ lie at the north pole of the untwisted sphere,
and $3,4$ at the south pole. 

\begin{table}\label{tab:S2_S2}
\begin{centering}
\begin{tabular}{|c|c|c|c|c|}
\hline 
$l$ & $1$ & $2$ & $3$ & $4$\tabularnewline
\hline 
\hline 
$\epsilon_{1}^{(l )}$ & $\epsilon_{1}$ & $\epsilon_{1}$ & $-\epsilon_{1}$ & $-\epsilon_{1}$\tabularnewline
\hline 
$\epsilon_{2}^{ (l )}$ & $\epsilon_{2}$ & $-\epsilon_{2}$ & $\epsilon_{2}$ & $-\epsilon_{2}$\tabularnewline
\hline 
\end{tabular}
\par\end{centering}
\caption{\label{tab:epsilon_values}Values of the equivariant parameters for
$S^2_{\epsilon_1} \times S^2_{\epsilon_2}$.}
\end{table}

The perturbative part of the Nekrasov partition function can be calculated
using the equivariant index theorem for transversally elliptic operators.
There is an important subtlety in using the equivariant index theorem
in the situation where the symbol is only transversally elliptic and
not elliptic. One must be careful about how to expand the infinite
sum in the index in order to translate it into a determinant \cite{Pestun:2007rz}.
A derivation for the specific case of the superconformal index was
performed in \cite{Drukker:2012sr}. The prescription used in \cite{Drukker:2012sr}, which recovers the result for the superconformal index computed by counting operators, is that half of the effective $a^{\left(l\right)}$, associated to the south pole of the sphere, should be multiplied by an overall $-1$
(\cf\;\cite[(5.15)-(5.17)]{Drukker:2012sr}). Adopting this approach, we change the sign of $a$ and $\epsilon_{1}^{ (l )}$ for points at the south pole, corresponding
in our case to points $3,4$. From the point of view of the full Nekrasov partition function, this prescription presumably coincides with what one gets by including anti-instantons
instead of instantons at one of the poles. We have already seen that the symbols for the fluctuation operators imply that this should be the case. 

The relevant parameters are therefore
\bea
a^{ (l )} & = a-\epsilon_{1}^{ (l )}\frac{\mathfrak{n}}{2}+\epsilon_{2}^{ (l )}\frac{\mathfrak{m}}{2}\,,\qquad && \text{north pole of }S^2_{\epsilon_1}\,,\\
a^{ (l )} & = - a+\epsilon_{1}^{ (l )}\frac{\mathfrak{n}}{2}+\epsilon_{2}^{ (l )}\frac{\mathfrak{m}}{2}\,, && \text{south pole of }S^2_{\epsilon_1} \,.
\eea
Taking the data from table \ref{tab:epsilon_values}, we get the following
set of parameters
\bea
a^{ (1 )} & = a-\epsilon_{1}\frac{\mathfrak{n}}{2}+\epsilon_{2}\frac{\mathfrak{m}}{2}\,,\\
a^{ (2 )} & = a-\epsilon_{1}\frac{\mathfrak{n}}{2}-\epsilon_{2}\frac{\mathfrak{m}}{2}\,,\\
a^{ (3 )} & = - a-\epsilon_{1}\frac{\mathfrak{n}}{2}+\epsilon_{2}\frac{\mathfrak{m}}{2}\,,\\
a^{ (4 )} & = - a-\epsilon_{1}\frac{\mathfrak{n}}{2}-\epsilon_{2}\frac{\mathfrak{m}}{2}\,.
\eea

For hypermultiplets, we can introduce background holonomies corresponding
to fugacities $\Delta$. In addition to the overall sign change noted
above, there is a shift of the origin of $\Delta$ \cite{Hosseini:2018uzp}.
This shift was first discussed in \cite{Okuda:2010ke}. The relevant
shift in the purely twisted case was a uniform shift, which with our
conventions for $\epsilon_{i}$ is
\be
\Delta\rightarrow\Delta-\frac{1}{2} \big(\epsilon_{1}^{ (l )}-\epsilon_{2}^{ (l )} \big)\,.
\ee
In the case at hand, where the sign of $\Delta$, like that of $a$, changes between the north and south poles, it is better to think of the shift as a shift of the overall parameter. Combining with the result for the vectors, we get the following
\bea
\Delta^{ (l )} & = \Delta-\frac{1}{2}\big(\epsilon_{1}^{ (l )}+\epsilon_{2}^{ (l )} \big)-\epsilon_{1}^{ (l )}\frac{\mathfrak{t}}{2}+\epsilon_{2}^{ (l )}\frac{\mathfrak{s}}{2}\,, \qquad && \text{north pole of }S^2_{\epsilon_1} \,, \\
\Delta^{ (l )} & = - \Delta-\frac{1}{2}\big(\epsilon_{1}^{ (l )}+\epsilon_{2}^{ (l )} \big)+\epsilon_{1}^{ (l )}\frac{\mathfrak{t}}{2}+\epsilon_{2}^{ (l )}\frac{\mathfrak{s}}{2}\,, && \text{south pole of }S^2_{\epsilon_1} \,.
\eea

In standard conventions, the origin of flavor fluxes $\mathfrak{s,t}$
would also be at $0$. However, we use conventions in which the background
fluxes for the universal twist are at $\mathfrak{s}=1-\mathfrak{g}$.
For $\mathfrak{g}=0$ we therefore take $\mathfrak{s}$ below to be
the flavor flux on the twisted sphere plus one.
Taking the data from table \ref{tab:epsilon_values}, we get the following set of parameters
\bea
\Delta^{ (1 )} & = \Delta-\epsilon_{1}\frac{\mathfrak{t}+1}{2}+\epsilon_{2}\frac{\mathfrak{s}-2}{2}\,,\\
\Delta^{ (2 )} & = \Delta-\epsilon_{1}\frac{\mathfrak{t}+1}{2}-\epsilon_{2}\frac{\mathfrak{s}-2}{2}\,,\\
\Delta^{ (3 )} & = - \Delta-\epsilon_{1}\frac{\mathfrak{t}-1}{2}+\epsilon_{2}\frac{\mathfrak{s}-2}{2}\,,\\
\Delta^{ (4 )} & = - \Delta-\epsilon_{1}\frac{\mathfrak{t}-1}{2}-\epsilon_{2}\frac{\mathfrak{s}-2}{2}\,.
\eea

For the cases $\mathfrak{g}\ge1$ one does not have isolated fixed
points. Instead, the fixed loci of the action of $v$ are copies of
$\Sigma_{\mathfrak{g}}$. The partition function on these spaces can
be computed from the two different perspectives, leading to the same
result. We consider the situation with $\mathfrak{g}$ arbitrary and
therefore set $\epsilon_{2}$ to $0$.

\subsubsection{The perturbative partition function}
\label{5d}

We are interested primarily in determining the partition function
in the large $N$ limit, hence we assume that we can safely ignore
non-perturbative contributions. The remaining elements of the localization
calculation are the classical and one loop pieces. 

\paragraph*{The classical piece.}

The classical action on the current background is different from the
one indicated in \cite{Hosseini:2018uzp}. As in \cite{Hosseini:2018uzp},
the part of the action containing hypermultiplets will not contribute
a classical term because there are no moduli in this sector of the
theory. A superconformal action for vector multiplets was constructed
in \cite{deWit:2009de}. The action depends on a totally symmetric
tensor $C_{ABC}$ and vector multiplets $V_{A}$. In order to construct
a Yang-Mills-like action for a dynamical vector multiplet, one of
the $V_{A}$ should be taken to be a background abelian multiplet
fixed to a supersymmetric configuration. Such a configuration necessarily
breaks conformal symmetry.\footnote{One may also construct topological Chern-Simons terms which preserve
conformal symmetry.} For the time being, we reinstate $\beta,r,r_{2}$.

The bosonic action for abelian $V_{A}$, translated into our field
variables and with an overall normalization which matches that of
\cite{Hosseini:2018uzp}, is given by \cite{deWit:2009de}
\bea
S =\int\sqrt{g}\,C_{ABC}\?\text{Tr}\Bigg[ & \sigma^{A}\left(\frac{1}{2}F_{\mu\nu}^{B}F^{\mu\nu C}+D_{\mu}\sigma^{B}D^{\mu}\sigma^{C}+\frac{1}{2}D^{IJ,A}D_{IJ}^{B}-6\sigma^{B}F_{\mu\nu}^{C}T^{\mu\nu}\right)\\
 & - \frac{\ii}{12}\varepsilon^{\mu\nu\rho\sigma\tau}A_{\mu}^{A}F_{\nu\rho}^{B}F_{\nu\rho}^{C}+\sigma^{A}\sigma^{B}\sigma^{C}\left(-\frac{1}{12}R+\frac{8}{3}D+13T_{\mu\nu}T^{\mu\nu}\right)\Bigg]\,.
\eea

We will choose the index $A$ to run from $1$ to $2$, with $V_{2}$
the dynamical vector multiplet and $V_{1}$ a background vector multiplet.
We also choose the following symmetric tensor
\be
C_{ABC}=C{\delta_{A}}^{1}{\delta_{B}}^{2}{\delta_{C}}^{2}+\text{symmetrization} \, .
\ee
In order to get a Yang-Mills like term, we set the following background
configuration for $V_{1}$\footnote{In order to preserve the appropriate supersymmetries, the background
configurations are restricted to the ones that yield the dynamical
moduli.}
\be
F^{1}=\frac{1}{2}\left[\frac{1}{r^{2}}\text{Vol} (S^2 )+\beta \tilde \epsilon_{1}\sin\theta \?\rd\theta\wedge \rd\tau\right] , \qquad\sigma^{1}=\frac{1}{2r}\,,
\ee
with all other fields vanishing. In order to get the normalization
for the Yang-Mills term used in \cite{Hosseini:2018uzp}, we will take $C=2r/g_{\text{YM}}^{2}$.

It is now straightforward to evaluate the classical contribution to
the partition function, by replacing both the supergravity fields
and the dynamical vector multiplet fields with their background values.
The result is 
\be\label{ct}
S_{\text{classical}}=-\ii\frac{16\pi^{2}r}{g_{\text{YM}}^{2}}\text{Tr} ( \mathfrak{m} a )\,.
\ee

There are various puzzling issues about this result.
First, \eqref{ct} is not holomorphic in $\epsilon_1$ as we would expect from gluing the classical pieces of the Nekrasov
partition function. It actually coincides with the result of the gluing when $\epsilon_1$ is purely imaginary (see section \ref{sssect:the partition function:SCI}). It would be interesting to understand
if there exist some extra terms in the bosonic action that could reproduce a manifestly holomorphic result.

Secondly, note that due to the periodicity of $a$, see \eqref{eq:a_periodicity},
the classical term is not gauge invariant under large gauge transformations.
This is already apparent at the level of the 5d Nekrasov partition
function. The issue did not come up in the fully twisted case, described
in \cite{Hosseini:2018uzp}, because the classical contributions had
a different structure. There are at least two possibilities to avoid
this. For theories with 5d fixed points, $g_{\text{YM}}$ must be
taken to $\infty$ before evaluating the partition function in order
to flow to the conformal fixed point. The calculation at intermediate
values of $g_{\text{YM}}$ is presumably ill defined. For theories with a 6d fixed point, on the other hand,
it is not completely clear what six-dimensional observable we are computing, and other details of the 6d physics might come to rescue.
 Another possibility
is that the classical part of the action is always ambiguous because,
as is the case for non-conformal theories in 4d, its coefficient can
be changed by changing the scale which is introduced in computing
the one-loop contribution.

\paragraph*{The one loop piece.}

It should be the case that the one loop determinant can be constructed
from the determinants used in the superconformal index, \ie\;the partition
function on $S^2\times S^1$, raised to an appropriate
power, which stems from the degeneracy of zero energy modes on $\Sigma_{\mathfrak{g}}$.
An explanation for this assertion is as follows. The one loop contribution
can be computed using the equivariant index theorem. In the situation
at hand, where the fixed loci with respect to the action of the symmetry
generator are not points but copies of $\Sigma_{\mathfrak{g}}$, the
relevant index theorem is a hybrid of the equivariant index theorem
for the case of discrete fixed points, and the usual Atiyah-Singer
index theorem applied to the fixed loci. The two contributions should
be multiplied to get the result on the product space.\footnote{The validity of these assertions is most apparent in the delocalized
approach of Berline and Vergne to the equivariant index theorem, where
the index is calculated using an integral over equivariant characteristic
classes \cite{berline1996indice}.} This applies directly to the index for the twisted Dirac operator,
while the index for the AHS (instanton) deformation complex reduces
to the twisted de Rahm complex at the fixed loci. The result of the
calculation with the AS index theorem is simply a number, which we
associate with the degeneracy. Hence, the result of the one loop calculation
is the one stemming from the equivariant index theorem on $S^2\times S^1$,
raised to the power of the degeneracy. 

The relevant fluctuation operators for computing the degeneracy can
be easily extracted by evaluating the symbols $\sigma\left(D_{\text{oe}}\right)$
and $\sigma (D_{\text{oe}}^{\text{hyper}} )$ near the fixed
points of the untwisted sphere and restricting to momenta along the
twisted directions only. The degeneracy can then be computed using
the ordinary index theorem for the relevant complex. For a line bundle
of degree $d$ on a Riemann surface of genus $\mathfrak{g}$, the
Atiyah-Singer index theorem yields $d+1-\mathfrak{g}$ for the twisted
Dolbeault complex, which controls the hypermultiplets. A hypermultiplet
valued in a weight $\rho$ of the dynamical gauge group and a weight
$\nu$ of the flavor symmetry group therefore has degeneracy (\cf\;\cite{Benini:2015noa})
\be
-\rho (\mathfrak{m} )-\nu (\mathfrak{s} )+\mathfrak{g}-1\,.
\ee
The overall minus sign comes from the opposite grading of the Dirac and Dolbeault complexes. The origin for the flavor fluxes is at $2\mathfrak{g}-2$, and the shifted expression is 
\be
-\rho (\mathfrak{m} )-\nu (\mathfrak{s} )-\mathfrak{g}+1\,.
\ee

The vector multiplet is controlled by a twisted version of the de
Rahm complex, which is again related to the Dolbeault complex and yields the same result (c.f. \cite{Benini:2015noa}). The degeneracy for a mode proportional to a root $\alpha$ reads
\be
\alpha\left(\mathfrak{m}\right)+1-\mathfrak{g} \,.
\ee

We must also specify the $R$-charge of the chirals appearing in the
superconformal index. To do this, we compare the square of the supercharge
used to localized the SCI with th eone used here. The ratio of the
$\epsilon_{1}$ dependent term between the derivative with respect
to $\phi$ and the R-symmetry transformation acting on $q^{1}$ is
$-2$. In the context of the SCI, where the eigenvalue with respect
to said derivative is usually denoted $j_{3}$, the ratio is $-2/\Delta$,
where $\Delta$ is both the R-charge and the Weyl weight of the bottom
component of a chiral superfield (\cf\;\cite[(5.14)]{Drukker:2012sr}).
Evidently, we should set $\Delta$ to $1$. Vector multiplets
can also be treated as chiral multiplets for the purposes of constructing
the partition function. This is because a chiral multiplet of $R$-charge
$0$ can get a vev and Higgs a vector multiplet, yielding an empty
theory in the IR. It also holds that the product of the one loop determinants
of chiral multiplets with R-charges $\Delta$ and $2-\Delta$ is $1$
\cite{Kapustin:2011jm}. the vector multiplet may therefore be treated
as a chiral multiplet with R-charge $2$. 

\paragraph*{Fermionic zero modes.}
One must still deal with the fermion zero modes which are the superpartners
of the holonomies on $\Sigma_{\mathfrak{g}}$. This can presumably
be done using the approach described in \eg\;\cite{Benini:2016hjo},
but the details are more complicated in the 5d context. We will rely
on the 2d result. 

\subsubsection{The final formula}
\label{sssect:the partition function:SCI}

We are ready to write the partition function. The result was first derived in  \cite{Jain:2021sdp} by gluing Nekrasov partition functions. We set
\be
 \epsilon_1 = \epsilon \, , \qquad \epsilon_2 = 0 \, ,
\ee
in the following. 
As already mentioned, in our conventions for the background fluxes, the universal twist \cite{Benini:2015bwz} corresponds to $\fs = 1 - \fg$.
One could also set the background fluxes on the \emph{untwisted} two-sphere to zero, $\ft = 0$, however, it is vital to keep $\epsilon$ nonzero.
Finally, we define
\be
 B_{\rho_I } \equiv  \rho_I ( \fm ) + \nu_I ( \fs ) + \fg - 1 \, , \qquad B_\alpha \equiv \alpha ( \fm ) - \fg + 1 \, .
\ee
The mixed index can be written as  \cite{Jain:2021sdp}
\bea
 \label{mixedindex}
 Z (\fm , \fn , a ; \fs , \ft , \Delta , \epsilon )& = \frac{1}{|W|} \sum_{\{ \fm , \fn \} \in \Gamma_{\fh}}
 \oint_{\cC} \prod_{i = 1}^{\text{rk}(G)} \frac{\rd x_i}{2 \pi \ii x_i} \?
 \bigg( \det\limits_{i j} \frac{\partial^2  \cW_{(S^2_\epsilon \times S^1) \times \bR^2} (a , \fn ; \Delta, \ft, \epsilon)}{\partial a_i \partial a_j} \bigg)^{\fg } \\
 & \times \exp \left( \frac{16 \pi^2}{g_{\text{YM}}^2 \epsilon} \Tr_{\text{F}} ( \fm a ) \right)
  \prod_{\alpha \in G}
 \left[ \frac{( x^{-\alpha} \zeta^{- \alpha ( \fn )} ; \zeta^2 )_\infty}{( x^\alpha \zeta^{2 - \alpha ( \fn )} ; \zeta^2 )_\infty}
 \big(\! - \zeta x^{\alpha} \big)^{\frac{1}{2} \alpha ( \fn )} \right]^{B_\alpha} \\
 & \times \prod_{I} \prod_{\rho_I \in \fR_I}
 \left[ \frac{( x^{\rho_I} y^{\nu_I} \zeta^{1 - \rho_I ( \fn ) - \nu_I (\ft )} ; \zeta^2 )_\infty}
 {( x^{-\rho_I} y^{-\nu_I} \zeta^{1- \rho_I ( \fn ) - \nu_I (\ft ) } ; \zeta^2 )_\infty}
 (- x^{\rho_I} y^{\nu_I})^{- \frac{1}{2} \left( \rho_I ( \fn ) + \nu_I (\ft ) \right)} \right]^{B_{\rho_I}} \, ,
\eea
with $x = e^{\ii a}$, $y = e^{\ii \Delta}$, $\zeta = e^{\ii \epsilon / 2}$, and $\cW_{(S^2_\epsilon \times S^1) \times \bR^2} (a , \fn ; \epsilon)$ denoting the effective twisted superpotential of the five-dimensional $\cN = 1$ theory on $(S^2_\epsilon \times S^1) \times \bR^2$, whose expression  is the subject of  the next section. 

For $\fg=0$, the partition function can be obtained by gluing the Nekrasov partition function \eqref{ZN} using the rules discussed in section \ref{subsec:mixedindexpart} and sending $\epsilon_2$ to zero  \cite{Jain:2021sdp}. The appearance of the determinant of the effective twisted superpotential for $\fg\ne 0$  is discussed in section \ref{ssect:W:SCI}.  As noticed in section \ref{5d}, the one-loop contributions are obtained by raising   one-loop determinants of the partition function on $S^2_\epsilon \times S^1$, the generalized superconformal index, to integer powers, $B_\alpha$ and $B_\rho$, counting the number of zero-modes on the twisted $\Sigma_\fg$. The one-loop contribution of a chiral multiplet of R-charge $r_K$ to the three-dimensional superconformal index reads \cite{Imamura:2011su,Kapustin:2011jm}
\be
 \log Z_{\chi}^{\Delta_K , \? \ft_K , \? r_K} ( a ) = \left( q^{\frac12( 1 - r_K )} x^{-1} y_K^{-1} \right)^{-\frac{1}{2} (\fn + \ft_K)}
 \frac{\left( x^{-1} y^{-1}_K q^{\frac12 ( 2 - r_K - \fn - \ft_K )} ; q \right)_{\infty}}
 {\left( x y_K q^{\frac12 \left( r_K - \fn - \ft_K \right)} ; q \right)_{\infty}} \, ,
\ee
where $y_K = e^{\ii \Delta_K}$ and $\ft_K$ denote the flavor fugacity and magnetic flux, respectively. Setting $r_K=1$ and $y_K=y$, $\ft_K= \ft$ we recognize the contributions of the hypermultiplets  in \eqref{mixedindex}, while  setting $r_K=2$ and $\Delta_K=2\pi$, $t_K=0$  we recognize the contributions of the vector multiplets.  Finally, the gluing procedure produces a classical term which is manifestly holomorphic in $\epsilon_1$ and coincides with \eqref{ct}  when $\epsilon_1$ is purely imaginary. As already discussed in section \ref{5d}, the role of the classical term, which is not gauge invariant, is not completely clear to us. In any event, it will only contribute to the partition function of the $\cN=2$ super Yang-Mills theory whose interpretation as a six-dimensional index has yet to be clarified.

\subsection{$\cW_{( S^2_\epsilon \times S^1 ) \times \bR^2}$ and its factorization}
\label{ssect:W:SCI}

In this section we derive the effective twisted superpotential of the two-dimensional theory obtained by compactifying a five-dimensional $\cN = 1$ theory on
$S^2_\epsilon \times S^1$. 
\paragraph*{(i) Gluing $\cW_{( S^2_\epsilon \times S^1 ) \times \bR^2}$.}
The two-sphere $S^2_\epsilon$ is \emph{not} twisted.
We shall use the gluing procedure that was detailed in section \ref{ssect:W:Bethe:Gluing}.
For this purpose, we employ the \emph{identity} gluing that follows from the rules given in section \ref{subsec:mixedindexpart} when $\epsilon_2=0$
\bea
 a^{(1)}_k & = a_k - \frac{\epsilon}{2} \fn_k \, , \qquad \quad \! \Delta^{(1)} = \Delta - \frac{\epsilon}{2} ( \ft + 1 ) \, , \qquad ~~ \epsilon_{1}^{(1)} = \epsilon \, , \\
 a^{(2)}_k & = - a_k - \frac{\epsilon}{2} \fn_k \, , \qquad \Delta^{(2)} = - \Delta - \frac{\epsilon}{2} ( \ft - 1 ) \, , \qquad \epsilon_{1}^{(2)} = - \epsilon \, .
\eea
We can then write%
\footnote{The relative minus follows from the relative minus for $a$ in the identity gluing and the fact that $\cW$ appears with a derivative in the partition function, see \eqref{Wdef:SCI}.}
\be
 \label{W:SCI:glue:NS}
  \cW_{(S^2_\epsilon \times S^1) \times \bR^2} ( a , \fn; \Delta, \ft , \epsilon ) =
 \cW_{(\bR^2_{\epsilon} \times S^1 ) \times \bR^2} (a^{(1)} ; \Delta^{( 1 )}, \epsilon^{( 1 )} )
 - \cW_{(\bR^2_{\epsilon} \times S^1 ) \times \bR^2} (a^{(2)} ; \Delta^{( 2 )}, \epsilon^{( 2 )} ) \, ,
\ee
where $ \cW_{(\bR^2_\epsilon \times S^1) \times \bR^2}$ is given in \eqref{W:flat:cl}-\eqref{W:flat:hyper:P}.

Explicitly, the classical Yang-Mills contribution to the effective twisted superpotential is given by
\be
  \cW^{\text{YM}}_{(S^2_\epsilon \times S^1 ) \times \bR^2} ( g_{\text{YM}} , a , \fn ; \epsilon ) =
 - \frac{8 \ii \pi^2}{g_{\text{YM}}^2 \epsilon} \left( \Tr_{\text{F}} ( a )^2 + \frac{\epsilon^2}{4} \Tr_{\text{F}} ( \fn )^2 \right) .
\ee
Next, the contribution of a hypermultiplet to the twisted superpotential, in an expansion around $\epsilon \to 0$, reads 
\be
 \label{W:SCI:hyper:first}
  \cW^{\cH}_{(S^2_\epsilon \times S^1 ) \times \bR^2} ( a , \fn ; \Delta, \ft, \epsilon ) = -
 \sum_{\rho_I \in \fR} \sum_{s = 0}^\infty \frac{( \ii \epsilon )^{s-1}}{(2 \pi)^s}
 w_s \left( \rho_I ( a ) , \rho_I ( \fn ) ; \nu_I ( \Delta) , \nu_I ( \ft) , 1 \right) \, ,
\ee
where
\bea
 \label{ws:def}
 w_s ( a , \fn ; \Delta , \ft , r ) & \equiv
 g_s \left( \pi ( r -  \fn - \ft ) ] \right)
 \left[ \Li_{3-s} (e^{\ii ( a + \Delta ) })
 + \frac{\ii^{3-s}}{2} g_{3-s} ( a + \Delta ) \right] \\
 & + g_{s} \left( \pi ( 2 - r - \fn - \ft ) \right)
 \left[ \Li_{3-s} (e^{- \ii ( a + \Delta ) })
 + \frac{\ii^{3-s}}{2} g_{3-s} ( 2 \pi - a - \Delta) \right] ,
\eea
with $g_s ( a )$ defined in \eqref{Li:inversion:g}.%
\footnote{Note that  the sign of $\re (a+\Delta)$ is opposite in the two terms in \eqref{W:SCI:glue:NS}  and we need to shift accordingly the functions $g_s(a)$  in \eqref{W:flat:cl}-\eqref{W:flat:hyper:P}.}
Observe that $g_s ( a) = 0$ for $s < 0$.
The contribution of a vector multiplet is simply given by
\be
  \cW^{\cV}_{(S^2_\epsilon \times S^1 ) \times \bR^2} ( a , \fn ; \epsilon ) =
 \sum_{\alpha \in G} \sum_{s = 0}^\infty \frac{( \ii \epsilon )^{s-1}}{(2 \pi)^s} w_s \left( \alpha ( a ) , \alpha ( \fn ) ; 2 \pi , 0, 2 \right) \, .
\ee

\paragraph*{(ii) Bethe approach.}

Alternatively, the twisted superpotential  appears in the mixed index as%
\footnote{We dropped the dependence on the flavor parameters $(\Delta, \fs, \ft)$ to avoid clutter.}
\be
 \label{Wdef:SCI}
 Z_{(S^2_\epsilon \times S^1 ) \times \Sigma_\fg} =\frac{1}{|W|} \sum_{\{ \fm , \fn \} \in \Gamma_{\fh}}
 \oint_{\cC} \prod_{i = 1}^{\text{rk}(G)} \frac{\rd x_i}{2 \pi \ii x_i} \?  \exp \bigg( \ii \sum_{k = 1}^{\text{rk}(G)} \fm_k \frac{\pd  \cW_{(S^2_\epsilon \times S^1) \times \bR^2} ( a , \fn ; \epsilon)}{\pd a_k} \bigg) Z_{\text{int}} \big|_{\fm = 0} (a , \fn ; \epsilon) \, ,
\ee
where $Z_{\text{int}}$ is the integrand in \eqref{mixedindex}, as a direct computation shows.

Performing the sum over the gauge magnetic fluxes $\fm$ through the Riemann surface in \eqref{Wdef:SCI},
we find a set of poles $\mathring{a}$ at the Bethe vacua.
It remains to evaluate the sum over the gauge fluxes $\fn$ on $S^2_\epsilon$.
At this stage, we assume that the mixed index localizes at the solutions to the \emph{generalized} BAEs
\bea
 \label{gen:BAEs:SCI}
 1 & = \exp \bigg( \frac{\pd  \cW_{(S^2_\epsilon \times S^1) \times \bR^2} ( a , \fn ; \epsilon)}{\pd a_k} \bigg) \Big|_{a = \mathring{a} , \, \fn = \mathring{\fn}} \, , \\
 1 & = \exp \bigg( \frac{\pd  \cW_{(S^2_\epsilon \times S^1) \times \bR^2} ( a , \fn ; \epsilon)}{\pd \fn_k} \bigg) \Big|_{a = \mathring{a} , \, \fn = \mathring{\fn}} \, .
\eea

\subsubsection{$\USp(2N)$ gauge theory with matter}
\label{sssect:SCI:W:USp(2N)}

We will now evaluate the effective twisted superpotential
of the $\USp(2N)$ theory with matter in the large $N$ limit.
The result will take a simple form using a democratic basis for the flavor chemical potential $\Delta_m$ and the fluxes $(\fs, \ft)$
\bea
 \label{democ:USp(2N):SCI}
 \fs_1 & \equiv \fs_m \, , \qquad \qquad ~~ \quad \fs_2 \equiv 2 (1 - \fg ) - \fs_m \, ,  \qquad \quad \text{ s.t. } \sum_{i = 1}^2 \fs_i = 2 - 2 \fg \, , \\
 \ft_1 & \equiv \ft_m \, , \qquad \qquad \quad \quad \ft_2 \equiv - \ft_m \, ,  \qquad \qquad \qquad \quad  \text{ s.t. } \sum_{i = 1}^2 \ft_i = 0 \, , \\
 \Delta_1 & \equiv \Delta_m + \frac{\epsilon}{2} \, , \qquad \quad \Delta_2 \equiv 2 \pi - \Delta_m + \frac{\epsilon}{2} \, , \qquad \quad \text{ s.t. } \sum_{i = 1}^2 \Delta_i = 2 \pi + \epsilon \, .
\eea

The effective twisted superpotential has the same structure as \eqref{functional:USp(2N)} and we only need to replace $\cF_{\Delta_k , \? \ft_K} ( a  )$, see \eqref{block:S3S2}, with
\be
 \label{W:block:SCI}
  \cW_{(S^2_\epsilon \times S^1) \times \bR^2}^{\Delta_K , \? \ft_k, \? r}( a ) =
 - \sum_{s = 0}^\infty \frac{( \ii \epsilon )^{s-1}}{(2 \pi)^s}
 w_s ( a , \fn ; \Delta_K , \ft_K , r) \, ,
\ee
with $w_s ( a , \fn ; \Delta_K , \ft_K , r)$ given in \eqref{ws:def}.
Let us start with the $ \cW_{(S^2_\epsilon \times S^1) \times \bR^2}^{\cA + \cV_1}$ contribution. Using \eqref{asymp:Li(s)}, and assuming that $| \im a_i | \to \infty$, we obtain
\bea
 \label{SCI:W:HV1:mid}
  \cW_{(S^2_\epsilon \times S^1) \times \bR^2}^{\cA + \cV_1} & =
 - \frac{1}{4 \epsilon} \sum_{i > j}^N \left[ (4 \Delta_1 \Delta_2 + \epsilon^2 \ft_1 \ft_2 ) a_{i j} + \epsilon^2 ( \Delta_1 \ft_2 + \Delta_2 \ft_1 ) \fn_{i j} \right] \sign \left( \im a_{ i j} \right) \\
 & - \frac{1}{4 \epsilon} \sum_{i > j}^N \left[ (4 \Delta_1 \Delta_2 + \epsilon^2 \ft_1 \ft_2 ) a^+_{i j} + \epsilon^2 ( \Delta_1 \ft_2 + \Delta_2 \ft_1 ) \fn_{i j}^+ \right] \sign \left( \im a_{ i j}^+ \right) .
\eea
Assuming also that the eigenvalues are ordered by increasing imaginary part, \ie\;$a_i > a_j$ for $i > j$, and using \eqref{sum:sign},
we can simplify \eqref{SCI:W:HV1:mid} further and write
\be
 \label{SCI:W:HV1:final}
  \cW_{(S^2_\epsilon \times S^1) \times \bR^2}^{\cA + \cV_1} =
 - \frac{1}{4\epsilon} \sum_{k = 1}^N (2 k - 1) \left[  (4 \Delta_1 \Delta_2 + \epsilon^2 \ft_1 \ft_2 ) a_{k} + \epsilon^2 ( \Delta_1 \ft_2 + \Delta_2 \ft_1 ) \fn_{k} \right] .
\ee
The contribution $ \cW_{(S^2_\epsilon \times S^1) \times \bR^2}^{\cF + \cV_2}$ to the twisted superpotential at large $N$
can be computed similarly using \eqref{asymp:Li(s)} and it is given by
\be
 \label{SCI:W:FV2:final}
  \cW_{(S^2_\epsilon \times S^1) \times \bR^2}^{\cF + \cV_2} =
 - \frac{(8 - N_f )}{3 \epsilon} \sum_{k = 1}^N \left( a_k^2 + \frac{3 \epsilon^2}{4} \fn_k^2 \right) a_k \, .
\ee
Putting \eqref{SCI:W:HV1:final} and \eqref{SCI:W:FV2:final} together we obtain the final expression for the twisted superpotential $ \cW_{(S^2_\epsilon \times S^1) \times \bR^2}$,
\bea
 \label{SCI:W:USp(2N):mid}
  \cW_{(S^2_\epsilon \times S^1) \times \bR^2} ( a , \fn ; \Delta, \ft , \epsilon ) & =
 - \frac{(8 - N_f )}{3 \epsilon} \sum_{k = 1}^N \left( a_k^2 + \frac{3 \epsilon^2}{4} \fn_k^2 \right) a_k \\
 & - \frac{1}{4\epsilon} \sum_{k = 1}^N (2 k - 1) \left[  (4 \Delta_1 \Delta_2 + \epsilon^2 \ft_1 \ft_2 ) a_{k} + \epsilon^2 ( \Delta_1 \ft_2 + \Delta_2 \ft_1 ) \fn_{k} \right] ,
\eea
that can be recast in the following factorized form
\be
  \cW_{(S^2_\epsilon \times S^1) \times \bR^2} (a , \fn ; \Delta, \ft , \epsilon) = - 2 \pi \sum_{\sigma = 1}^2 \frac{\cF_{\text{SW}} \big( a_k^{(\sigma)} ; \Delta_i^{(\sigma)} \big)}{\epsilon^{(\sigma)}} \, ,
\ee
with $\cF_{\text{SW}} ( a_k ; \Delta_i )$ given in \eqref{SW:sum2pi:USp(2N)}, using the identity gluing parameterization
\bea
 \label{id:gluing:W:SCI}
 a_k^{(1)} & \equiv a_k - \frac{\epsilon}{2} \fn_k \, , \qquad \Delta_i^{(1)} \equiv \Delta_i - \frac{\epsilon}{2} \ft_i \, , \qquad \epsilon^{(1)} = \epsilon \, , \\
 a_k^{(2)} & \equiv a_k + \frac{\epsilon}{2} \fn_k \, , \qquad \Delta_i^{(2)} \equiv \Delta_i + \frac{\epsilon}{2} \ft_i \, , \qquad \epsilon^{(2)} = \epsilon \, .
\eea
The Bethe solution can be obtained by extremizing \eqref{SCI:W:USp(2N):mid} with respect to the gauge variables $(a_k , \fn_k)$.
We find 
\be
 \label{SCI:n:Bethe}
 \frac{\pd \cW_{(S^2_\epsilon \times S^1) \times \bR^2} (a , \fn)}{\pd \fn_k} \Big|_{\fn_k = \mathring{\fn}_k} = 0 \qquad \Rightarrow \qquad a_k \mathring{\fn}_k = - \frac{(2 k-1)}{2 ( 8 - N_f )} ( \Delta_1 \ft_2 + \Delta_2 \ft_1 )  \, ,
\ee
and
\be
 \begin{aligned}
 \label{SCI:a:Bethe}
 & \frac{\pd  \cW_{(S^2_\epsilon \times S^1) \times \bR^2} (a , \fn) }{\pd a_k} = 0 \Big|_{a_k = \mathring{a}_k , \, \fn_k = \mathring{\fn}_k}  \\
 & \Rightarrow \qquad \mathring{a}_k = \frac{\ii}{2} \sqrt{\frac{2 k-1}{8-N_f}}
 \left( \sqrt{\Big( \Delta_1 + \frac{\epsilon}{2} \ft_1 \Big) \Big( \Delta_2 + \frac{\epsilon}{2} \ft_2 \Big)}
 + \sqrt{\Big( \Delta_1 - \frac{\epsilon}{2} \ft_1 \Big) \Big( \Delta_2 - \frac{\epsilon}{2} \ft_2 \Big)} \right) .
 \end{aligned}
\ee
Observe that \eqref{SCI:n:Bethe} and \eqref{SCI:a:Bethe} are equivalent to
\be
 \frac{\pd \cF_{\text{SW}} ( a_k^{(\sigma)} ; \Delta_i^{(\sigma)} )}{\pd a_k^{(\sigma)}} = 0 \qquad \Rightarrow \qquad
 \mathring{a}_k^{(\sigma)} = \frac{\ii}{\sqrt{8 - N_f}} \sqrt{ ( 2 k - 1 ) \Delta_1^{(\sigma)} \Delta_2^{(\sigma)} } \, ,
\ee
for $ \sigma = 1, 2$. Note that we use the determination $0 < \Delta_i \mp \frac{\epsilon}{2} \ft_i < 2 \pi$, $i = 1,2$.
Plugging the saddle points $(\mathring{a}_k , \mathring{\fn}_k)$ back into the twisted superpotential \eqref{SCI:W:USp(2N):mid} we obtain
\be
 \label{W:SCI:USp(2N)}
 \boxed{
  \cW_{(S^2_\epsilon \times S^1) \times \bR^2} ( \Delta_i , \ft_i , \epsilon )
 = \frac{4 \pi^2 \ii}{27 \epsilon} \left[ F_{S^5} \left( \Delta_i + \frac{\epsilon}2 \ft_i \right) + F_{S^5} \left( \Delta_i - \frac{\epsilon}2 \ft_i \right) \right] ,
 }
\ee
with $F_{S^5} (\Delta_i)$ being, in form,  the free energy of the theory on $S^5$,
\be
 \label{FS5::pi:epsilon:USp(2N)}
 F_{S^5} ( \Delta_i ) = - \frac{9 \sqrt{2}}{5 \pi^2} \frac{N^{5/2}}{\sqrt{8 - N_f}} (\Delta_1 \Delta_2)^{\frac{3}{2}} \, , \qquad
 \sum_{i = 1}^2 \Delta_i = 2 \pi + \epsilon \, .
\ee

\subsubsection{$\cN = 2$ super Yang-Mills}

For $\cN = 2$ super Yang-Mills we  find a dependence on the non-gauge invariant classical term \eqref{ct}, whose interpretation is not clear.
It is nevertheless instructive to check that the twisted superpotential factorizes. 
Define the democratic basis for the chemical potential $\Delta_{i}$ and the fluxes $(\fs_{i}, \ft_i)$, associated to the $\U(1)^2 \subset \SO(5)$ R-symmetry
\bea
 \label{democ:SYM:SCI}
 \fs_1 & \equiv \fs \, , \qquad \qquad ~~ \quad \fs_2 \equiv 2 (1 - \fg ) - \fs \, ,  \qquad \quad \text{ s.t. } \sum_{i = 1}^2 \fs_i = 2 - 2 \fg \, , \\
 \ft_1 & \equiv \ft \, , \qquad \qquad \quad \quad \ft_2 \equiv - \ft \, ,  \qquad \qquad \qquad \quad  \text{ s.t. } \sum_{i = 1}^2 \ft_i = 0 \, , \\
 \Delta_1 & \equiv \Delta + \frac{\epsilon}{2} \, , \qquad \quad \Delta_2 \equiv 2 \pi - \Delta + \frac{\epsilon}{2} \, , \qquad \quad \text{ s.t. } \sum_{i = 1}^2 \Delta_i = 2 \pi + \epsilon \, .
\eea
The twisted superpotential of the theory reads
\be
 \label{W:SCI:SYM}
  \cW_{(S^2_\epsilon \times S^1) \times \bR^2} =
  \cW_{\text{YM}} ( a_k , \fn_k ) +  \cW_{\cH} (a_i, \fn_i ; \Delta , \ft , \epsilon ) +  \cW_{\cV} (a_i, \fn_i , \epsilon ) \, ,
\ee
with
\be
 \begin{aligned}
  \cW_{\text{YM}} & = - \frac{8 \pi^2 \ii}{g_{\text{YM}}^2 \epsilon} \sum_{k = 1}^N \left( a_k^2 + \frac{\epsilon^2}{4} \fn_k^2 \right) \, , \\
  \cW_{\cH} & = \sum_{i , j = 1}^{N}  \cW_{\Delta, \? \ft, \? r = 1} ( a_{i j} ) \, , \qquad
  \cW_{\cV} = - \sum_{i \neq j}^{N}  \cW_{\Delta = 2 \pi , \? \ft = 0, \? r = 2} ( a_{i j} ) \, ,
 \end{aligned}
\ee
and $ \cW_{\Delta, \? \ft, \? r}( a )$ given in \eqref{W:block:SCI}.
In the strong 't Hooft coupling $\lambda \gg 1$, assuming that $|\im (a_i - a_j )| \to \infty$, \eqref{W:SCI:SYM} can be approximated as, using \eqref{asymp:Li(s)},
\bea
 \label{W:SCI:SYM:mid}
  \cW_{( S^2_\epsilon \times S^1) \times \bR^2} (a_i, \fn_i ; \Delta, \ft , \epsilon ) & =
 - \frac{8 \pi^2 \ii}{g_{\text{YM}}^2 \epsilon} \sum_{k = 1}^N \left( a_k^2 + \frac{\epsilon^2}{4} \fn_k^2 \right) \\
 & - \frac{1}{8 \epsilon} \sum_{i , j = 1}^N \left[ ( 4 \Delta_1 \Delta_2 + \epsilon^2 \ft_1 \ft_2 ) a_{i j} + \epsilon^2 ( \Delta_1 \ft_2 + \Delta_2 \ft_1 ) \fn_{i j} \right] \sign ( \im a_{i j} ) \, .
\eea
Assuming that the eigenvalues are ordered by increasing imaginary part,
we can trade the $\sum_{i , j =1}^N$ with $\sum_{k = 1}^N$ using \eqref{sum:sign} to write
\bea
 \label{W:SCI:SYM:final}
  \cW_{( S^2_\epsilon \times S^1) \times \bR^2} (a_i, \fn_i ; \Delta, \ft , \epsilon ) & =
 - \frac{8 \pi^2 \ii}{g_{\text{YM}}^2 \epsilon} \sum_{k = 1}^N \left( a_k^2 + \frac{\epsilon^2}{4} \fn_k^2 \right) \\
 & - \frac{1}{4 \epsilon} \sum_{k = 1}^N (2 k - 1 - N)
 \left[ ( 4 \Delta_1 \Delta_2 + \epsilon^2 \ft_1 \ft_2 ) a_{k} + \epsilon^2 ( \Delta_1 \ft_2 + \Delta_2 \ft_1 ) \fn_{k} \right] ,
\eea
that, using the identity gluing parameterization \eqref{id:gluing:W:SCI}, can be recast in the following factorized form
\be
  \cW_{(S^2_\epsilon \times S^1) \times \bR^2} (a , \fn ; \Delta, \ft , \epsilon) = - 2 \pi \sum_{\sigma = 1}^2 \frac{\cF_{\text{SW}} \big( a_k^{(\sigma)} ; \Delta_i^{(\sigma)} \big)}{\epsilon^{(\sigma)}} \, ,
\ee
with $\cF_{\text{SW}} ( a_k ; \Delta_i )$ given in \eqref{SW:sum2pi:sym}.
We extremize \eqref{W:SCI:SYM:final} over the gauge variables $(a_k , \fn_k)$ to obtain the solution
\bea
 \label{SCI:Bethe:sym}
 \mathring{a}_k & = \ii \frac{g_\text{YM}^2}{64 \pi^2} ( 2 k - 1- N ) ( 4 \Delta_1 \Delta_2 + \epsilon^2 \ft_1 \ft_2 ) \, ,\\
 \mathring{\fn}_k &= \ii \frac{g_\text{YM}^2}{16 \pi^2} (2 k - 1 - N) ( \Delta_1 \ft_2 + \Delta_2 \ft_1 ) \, ,
\eea
to the generalized BAEs \eqref{gen:BAEs:SCI}.
Observe that \eqref{SCI:Bethe:sym} is equivalent to
\be
 \frac{\pd \cF_{\text{SW}} ( a_k^{(\sigma)} ; \Delta_i^{(\sigma)} )}{\pd a_k^{(\sigma)}} = 0 \qquad \Rightarrow \qquad
 \mathring{a}_k^{(\sigma)} = \ii \frac{g_\text{YM}^2}{16 \pi^2} ( 2 k - 1 - N ) \Delta_1^{(\sigma)} \Delta_2^{(\sigma)} \, ,
\ee
for $ \sigma = 1, 2$.
Substituting these values into \eqref{W:SCI:SYM:final} we find
\bea
  \cW_{( S^2_\epsilon \times S^1) \times \bR^2} (\Delta, \ft , \epsilon ) =
 - \frac{\ii g_{\text{YM}}^2}{3 \times 2^9 \pi^2 \epsilon} &
 \left[ ( 4 \Delta_1 \Delta_2 + \epsilon^2 \ft_1 \ft_2 ) + 2 \ii \epsilon ( \Delta_1 \ft_2 + \Delta_2 \ft_1 ) \right] \\
 \times &
 \left[ ( 4 \Delta_1 \Delta_2 + \epsilon^2 \ft_1 \ft_2 ) - 2 \ii \epsilon ( \Delta_1 \ft_2 + \Delta_2 \ft_1 ) \right] .
\eea
Remarkably, this can be recast in the following factorized form
\be
 \label{W:SCI:SYM:factorized}
\boxed{
  \cW_{( S^2_\epsilon \times S^1) \times \bR^2} ( \Delta, \ft , \epsilon )
 = - \ii N (N^2 - 1) \frac{g_{\text{YM}}^2}{192 \pi^2 \epsilon} \left[ \big(\Delta_1^{(1)} \Delta_2^{(1)} \big)^2 + \big( \Delta_1^{(2)} \Delta_2^{(2)} \big)^2 \right] .
}
\ee

\subsection{Factorization of the index}\label{indexSCI}

In this section we will make the assumption   that the mixed index localizes at the solutions to the  generalized BAEs
 \eqref{gen:BAEs:SCI}. Given \eqref{Wdef:SCI}, up to subleading constant factors, the saddle point contribution to the index  is given by
\be   \label{SCI:BA}
 Z_{(S^2_\epsilon \times S^1 ) \times \Sigma_\fg} = 
 Z_{\text{int}} \big|_{\fm = 0} (a , \fn ; \epsilon) \bigg |_ {\fn = \mathring{\fn} ,\? a = \mathring{a}} \, .
\ee
We will  show that this expression factorizes,%
\footnote{Our results differ from those in \cite{Jain:2021sdp}, which do not factorize, due to a different twisted superpotential used in the conjectured BAEs.} and leads to the correct entropy for a class of dual black holes in massive type IIA.
 Moreover, as for the twisted index, we find
 \be
 \label{indexTh000}\log Z_{(S^2_\epsilon \times S^1)\times \Sigma_g}  =
 \ii \sum_{i=1}^2 \fs_i  \frac{ \partial  \cW_{(S^2_\epsilon \times S^1) \times \bR^2}}{\partial \Delta_i}  \, .
\ee

While the mixed index for the Seiberg theory correctly reproduces the entropy of  black holes in massive type IIA, it is not clear what is the interpretation of the mixed index of $\cN=2$ super Yang-Mills in terms of the
six-dimensional UV fixed point and we are not aware  of dual backgrounds where to test the large $N$ result, which, as already mentioned, explicitly depends on the existence of a non-gauge invariant classical term \eqref{ct}.
It is however interesting to observe that the result factorizes, as expected. 

\subsubsection{$\USp(2N)$ gauge theory with matter}
\label{sssect:SCI:USp(2N)}

We are interested in evaluating the mixed index \eqref{SCI:BA} of the $\USp(2N)$ gauge theory in the large $N$ limit.
To this aim, let us first consider the building block
\be
 \label{main:SCI}
 \cZ (a , \fm , \fn ; \Delta, \fs, \ft, \epsilon) =
 \left[ \frac{( x y \zeta^{r - \fn - \ft} ; \zeta^2 )_\infty}{( x^{-1} y^{-1} \zeta^{2 - r - \fn -\ft} ; \zeta^2 )_\infty} (- x y)^{-\frac12(\fn+\ft)}
 \right]^{\fm + \fs + \fg - 1} 
\, ,
\ee
where $x = e^{\ii a}$, $y = e^{\ii \Delta}$, $\zeta = e^{\ii \epsilon/2}$.
From \eqref{Wdef:SCI} and the form of $\cW$ in terms of \eqref{ws:def}, we can immediately write down the following asymptotic expansion
\be
 \label{main:SCI}
 \log \cZ (a , \fm , \fn ; \Delta, \fs, \ft, \epsilon) =
 ( \fm + \fs + \fg - 1 ) \sum_{s = 0}^\infty \frac{(\ii \epsilon )^{s - 1}}{( 2 \pi )^s} v_s ( a , \fn ; \Delta , \ft , r ) \, ,
 \quad \text{ as } \epsilon \to 0 \, ,
\ee
where
\bea
 \label{vs:def}
 v_s ( a , \fn ; \Delta , \ft , r) & \equiv
 g_s \left( \pi ( r -  \fn - \ft ) ] \right)
 \left[ \Li_{2-s} (e^{\ii ( a + \Delta ) })
 + \frac{\ii^{2-s}}{2} g_{2-s} ( a + \Delta ) \right] \\
 & - g_{s} \left( \pi ( 2 - r - \fn - \ft ) \right)
 \left[ \Li_{2-s} (e^{- \ii ( a + \Delta ) })
 + \frac{\ii^{2-s}}{2} g_{2-s} ( 2 \pi - a - \Delta) \right] ,
\eea
with $g_s ( a )$ defined in \eqref{Li:inversion:g}. Note that $g_s ( a) = 0$ for $s < 0$.

Then, $Z_{\text{int}} \big|_{\fm = 0} (a_i , \fn_i ; \Delta_K , \fs_K, \ft_K ,\epsilon)$ in \eqref{SCI:BA} can be written as
\bea
 \label{SCI:functional:USp(2N)}
 \log Z_{\text{int}} \big|_{\fm = 0} (a_i , \fn_i ; \Delta_K , \fs_K, \ft_K ,\epsilon) & =
 \log \cZ_{\cA + \cV_1} (a_i , \fn_i ; \Delta_m, \fs_m, \ft_m, \epsilon) \\
 & + \log \cZ_{\cF + \cV_2} (a_i , \fn_i ; \Delta_f, \fs_f, \ft_f, \epsilon) \, ,
\eea
with
\be
 \begin{aligned}
 \log \cZ_{\cA + \cV_1} & =
 \sum_{i > j}^{N} \big[
 \log \cZ_{\Delta_m ,\? \fs_m , \? \ft_m , \? r = 1} (\pm a_i \pm a_j)
 - \log \cZ_{\Delta_K = 2 \pi , \? \fs_K = 2 - 2 \fg , \? \ft_K = 0 , \? r = 2} (\pm a_i \pm a_j) \big] \\
 & + (N - 1) \log \cZ_{\Delta_m ,\? \fs_m , \? \ft_m , \? r = 1 } ( 0 ) \, , \\
 \log \cZ_{\cF + \cV_2} & =
 \sum_{i = 1}^{N} \bigg[
 \sum_{f = 1}^{N_f} \log \cZ_{\Delta_f ,\? \fs_f , \? \ft_f , \?  r = 1} ( \pm a_i ) 
 - \log \cZ_{\Delta_K = 2 \pi ,\? \fs_K = 2 - 2 \fg , \? \ft_K = 0 , \? r = 2} (\pm 2 a_i)
 \bigg] \, ,
 \end{aligned}
\ee
Here, we introduced the function
\be
 \label{main:SCI:logZ:block}
 \log \cZ_{\Delta_K, \? \fs_K, \? \ft_K,  \? r} (a ) =
 ( \fs_K + \fg - 1 ) \sum_{s = 0}^\infty \frac{(\ii \epsilon )^{s - 1}}{( 2 \pi )^s} v_s ( a , \fn ; \Delta_K , \? \ft_K, \? r) \, ,
 \quad \text{ as } \epsilon \to 0\, .
\ee
Again, the index $K$ labels all the matter fields in the theory and we used the notation
\be
 \log \cZ_{\Delta_K , \? \fs_K , \? \ft_K , \? r} ( \pm a_i ) \equiv
 \log \cZ_{\Delta_K, \? \fs_K, \? \ft_K , \? r} (a_i, \fn_i)
 + \log \cZ_{\Delta_K, \? \fs_K, \? \ft_K , \? r} ( - a_i , - \fn_i ) \, .
\ee
The only piece that survives the large $N$ limit is given by
\bea
 \label{SCI:logZ:HV1:mid}
 \log \cZ_{\cA + \cV_1} & = - \frac{\ii}{\epsilon} \sum_{i > j}^N \left[ ( \Delta_1 \fs_2 + \Delta_2 \fs_1 ) a_{ i j } + \frac{\epsilon^2}{4} ( \fs_2 \ft_1 + \fs_1 \ft_2 ) \fn_{i j} \right] \sign ( \im a_{i j} ) \\
 & - \frac{\ii}{\epsilon} \sum_{i > j}^N \left[ ( \Delta_1 \fs_2 + \Delta_2 \fs_1 ) a_{ i j }^+ + \frac{\epsilon^2}{4} ( \fs_2 \ft_1 + \fs_1 \ft_2 ) \fn_{i j}^+ \right] \sign ( \im a_{i j}^+ ) \, .
\eea
Assuming that the eigenvalues are ordered by increasing imaginary part and using \eqref{sum:sign},
we can simplify \eqref{SCI:logZ:HV1:mid} further and write
\be
 \label{SCI:logZ:HV1:final}
 \log \cZ_{\cA + \cV_1} =
 - \frac{\ii}{\epsilon} \sum_{k = 1}^N (2 k - 1) \left[ ( \Delta_1 \fs_2 + \Delta_2 \fs_1 ) a_{k} + \frac{\epsilon^2}{4} ( \fs_2 \ft_1 + \fs_1 \ft_2 ) \fn_{k} \right] .
\ee
Finally, plugging the Bethe solutions \eqref{SCI:n:Bethe} and \eqref{SCI:a:Bethe} back into \eqref{SCI:BA},
\be
 \log Z_{(S^2_\epsilon \times S^1 ) \times \Sigma_\fg} \overset{N \gg 1}{=} \log \cZ_{\cA + \cV_1} \Big|_{\fn = \mathring{\fn}, \? a = \mathring{a}} \, ,
\ee
we can recast the final result in the following factorized form
\be
 \label{SCI:final:USp(2N)}
 \boxed{
 \log Z_{(S^2_\epsilon \times S^1 ) \times \Sigma_\fg} (\Delta, \fs, \ft, \epsilon)
 = - \frac{\pi}{2 \epsilon} \left[ F_{S^3 \times \Sigma_\fg} \left( \Delta_i + \frac{\epsilon}{2} \ft_i , \fs_i \right) + F_{S^3 \times \Sigma_\fg} \left( \Delta_i - \frac{\epsilon}{2} \ft_i , \fs_i \right) \right] ,
 }
\ee
where $F_{S^3 \times \Sigma_\fg} (\Delta_i , \fs_i)$ is the free energy of the theory on $S^3 \times \Sigma_\fg$, see \eqref{FS3xSigma:USp(2N)}.
Recall that
\be
 \sum_{i = 1}^2 \Delta_i = 2 \pi + \epsilon \, , \qquad \sum_{i = 1}^2 \fs_i = 2 - 2 \fg \, ,
 \qquad \sum_{i = 1}^2 \ft_i = 0 \, .
\ee

\paragraph*{Black holes microstates in AdS$_2 \times S^2_\epsilon \times \Sigma_\fg$.}
In the following we show that the mixed index \eqref{SCI:final:USp(2N)} gives a statistical derivation of the entropy of Kerr-Newman black holes found in \cite[Sect.\,6.3.2 and 6.3.4]{Hosseini:2020wag}. The near horizon geometry is AdS$_2 \times S^2_\epsilon \times \Sigma_\fg$.

The class of black holes we consider is a two-parameter family of solutions
and was found by specializing the magnetic fluxes along the Riemann surface to (in the notations of \cite{Hosseini:2020wag})
\be
 s^1 = \frac{2}{3} \, , \qquad s^2 = 0 \, .
\ee
This choice leads to a compact Riemann surface $\Sigma_{\fg > 1}$ and vanishing magnetic charges along the $S^2_\epsilon$.
The Bekenstein-Hawking entropy reads
\be
 \label{SBH:sugra:KN}
 S_{\text{BH}} ( q_1, q_2, \cJ ) = \frac{\pi}{3 G_{\text{N}}^{(4)}} \sqrt{\frac{108 q_2^2 ( q_1 + q_2 ) + \cJ}{q_1 + 9 q_2}} \, ,
\ee
with the following constraint among the conserved charges
\be
 \label{constraint:Q12:J}
 \cJ = \frac{( \sqrt{72 q_2 \left(q_1+3 q_2\right)+1} - 1 )^2
 \left( 2 ( q_1 + 3 q_2 ) \sqrt{72 q_2 \left(q_1+3 q_2\right)+1} + q_1 - 3 q_2 \right)}
 {6 \left( \sqrt{72 q_2 \left(q_1+3 q_2\right)+1} - 1 - 6 ( q_1 + 3 q_2 ) ( q_1 + 9 q_2 ) \right)} \, .
\ee
Note that the limit $\cJ \to 0$ is singular and these black holes are always rotating.
We shall extremize the $\cI$-functional for $\ft=0$
\be
 \label{SCI:I:func}
 \cI_{(S^2_\epsilon \times S^1 ) \times \Sigma_\fg} (\Delta, \epsilon) = \log Z_{(S^2_\epsilon \times S^1 ) \times \Sigma_\fg} (\Delta, \fs, \ft=0, \epsilon)
 - \ii \epsilon J - \ii \sum_{i =1}^2 \Delta_i Q_i - \Lambda ( \Delta_1 + \Delta_2 - \epsilon - 2 \pi ) \, ,
\ee
with respect to the chemical potentials $(\Delta_1, \Delta_2, \epsilon)$ and the
Lagrange multiplier $\Lambda$, enforcing the constraint $\Delta_1 + \Delta_2 = 2 \pi + \epsilon$, to obtain
\be
 \label{SCI:Euler's theorem}
 \cI_{(S^2_\epsilon \times S^1 ) \times \Sigma_\fg} \Big|_{\text{crit.}} (Q_1, Q_2, J) = 2 \pi \mathring{\Lambda} (Q_1, Q_2, J) \, ,
\ee
by Euler's theorem, and
\be
 \label{zero:SCI:USp(2N)}
 (\mathring{\Lambda} + \ii Q_1 ) ( \mathring{\Lambda} + \ii Q_2 )^3 - \left( \frac{2 ( \fg - 1 ) }{\sqrt{3} \pi} F_{S^5} \right)^2 ( \mathring{\Lambda} - \ii J)^2 = 0 \, .
\ee
Assuming that the charges and the entropy  $2 \pi \mathring{\Lambda}$ are real we can break \eqref{zero:SCI:USp(2N)} into two equations
\be
 \begin{aligned}
 & \im \eqref{zero:SCI:USp(2N)} = \mathring{\Lambda}^2 ( Q_1 + 3 Q_2 ) - Q_2^2 ( 3 Q_1 + Q_2 ) + \bigg(\frac{2 \sqrt{2} ( \fg - 1)}{\sqrt{3} \pi} F_{S^5} \bigg)^2 J = 0 \, , \\
 & \re \eqref{zero:SCI:USp(2N)} = \mathring{\Lambda}^4 - 3 \mathring{\Lambda}^2 Q_2 ( Q_1 + Q_2 ) + \bigg( \frac{2 ( \fg - 1 )}{\sqrt{3} \pi} F_{S^5} \bigg)^2 ( J + \mathring{\Lambda} ) ( J - \mathring{\Lambda}) + Q_1 Q_2^3  = 0 \, .
 \end{aligned}
\ee
Solving the first equation gives us, using \eqref{SCI:Euler's theorem},
\be
 \cI_{(S^2_\epsilon \times S^1 ) \times \Sigma_\fg} \Big|_{\text{crit}} (Q_1, Q_2, J)
 = 2 \pi  \sqrt{\frac{- \left(\frac{2 \sqrt{2} ( \fg - 1 )}{\sqrt{3} \pi} F_{S^5} \right)^2 J + Q_2^2 ( 3 Q_1 + Q_2 )}{Q_1 + 3 Q_2}} \, ,
\ee
while the second equation leads to the constraint \eqref{constraint:Q12:J} among the conserved charges.
We see again, as in section \ref{sssect:RTTI:logZ:sym}, that  the constraint among the conserved charges of BPS black holes arises as a reality
requirement for the entropy functional at the critical point.
To compare with \eqref{constraint:Q12:J}, we used the identification \cite[(7.15)]{Hosseini:2020wag}
\be
 Q_1 = \frac{1}{3 G_{\text{N}}} q_1 \, , \qquad Q_2 = \frac{1}{G_{\text{N}}} q_2 \, , \qquad J = - \frac{1}{2 G_\text{N}^{(4)}} \cJ \, .
\ee
Using also the standard AdS$_6$/CFT$_5$ dictionary \eqref{AdS6:CFT5:dict} we find that
\be
 \cI_{(S^2_\epsilon \times S^1 ) \times \Sigma_\fg} \Big|_{\text{crit.}} (Q_1, Q_2, J) = S_{\text{BH}} ( q_1, q_2, \cJ ) \, .
\ee
Our result is in complete agreement with \cite[Sect.\,7.2]{Hosseini:2020wag} upon identifying
\be
 \omega_{\text{there}} \equiv - \frac{\ii}{\pi} \epsilon \, , \qquad \chi_{\text{there}}^i \equiv \frac{2}{3 \pi} \Delta_i \, , \quad i = 1, 2 \, .
\ee

For completeness, we note that, in the special case $\ft=0$, both our results and those
 in \cite{Jain:2021sdp}, which uses a different twisted superpotential for the conjectured BAEs,  coincide. 
The differences  show up for $\ft\ne0$, where our expression for the index factorizes while the one in \cite{Jain:2021sdp} does not.
To fully test holographically the two sets of different results we would need
a Kerr-Newman black hole solution in massive type IIA
with non-vanishing magnetic charges. We are not aware of any such solution in the literature and it would be interesting to find one.

\subsubsection{$\cN = 2$ super Yang-Mills}

In the strong 't Hooft coupling limit $\lambda \gg 1$, we need to evaluate 
\be
 \label{functional:SYM}
 \log Z_{\text{int}} \big|_{\fm = 0} (a_i , \fn_i ; \Delta , \fs, \ft ,\epsilon) = \sum_{i , j = 1}^N \log \cZ_{\Delta, \? \fs, \? \ft ,\? r = 1}  ( a_{i j} )
 - \sum_{i \neq j}^N \log \cZ_{ \Delta = 2\pi , \? \fs = 2 - 2 \fg , \? \ft = 0 ,\? r = 2} ( a_{ i j} ) \, ,
\ee
with $ \log \cZ_{\Delta, \? \fs , \? \ft ,\? r} ( a )$ given in \eqref{main:SCI:logZ:block}.
Employing the asymptotic formula \eqref{asymp:Li(s)} and assuming that the eigenvalues are ordered by increasing imaginary part, using \eqref{sum:sign},
we find
\be
 \log Z_{\text{int}} \big|_{\fm = 0} (a_i , \fn_i ; \Delta , \fs, \ft ,\epsilon) =
 - \frac{\ii}{\epsilon} \sum_{k = 1}^N (2 k - 1 - N) \left[ ( \Delta_1 \fs_2 + \Delta_2 \fs_1 ) + \frac{\epsilon^2}{4} ( \fs_2 \ft_1 + \fs_1 \ft_2 ) \fn_k \right] .
\ee
Plugging the Bethe solutions \eqref{SCI:Bethe:sym} back into \eqref{SCI:BA},
\be
 \log Z_{(S^2_\epsilon \times S^1 ) \times \Sigma_\fg} \overset{\lambda \gg 1}{=}
 \log Z_{\text{int}} \big|_{\fm = 0} (\mathring{a}_i , \mathring{\fn}_i ; \Delta , \fs, \ft ,\epsilon) \, ,
\ee
we can recast the final result in the following factorized form
\be
 \label{SCI:fina:sym}
 \boxed{
 \log Z_{(S^2_\epsilon \times S^1 ) \times \Sigma_\fg} (\Delta, \fs, \ft, \epsilon)
 = - \frac{2 \pi g_{\text{YM}}^2}{27 \epsilon}
 \left[ a \left( \Delta_i + \frac{\epsilon}{2} \ft_i , \fs_i \right) + a \left( \Delta_i - \frac{\epsilon}{2} \ft_i , \fs_i \right) \right] ,
 }
 \ee
where $a(\Delta, \fs)$ is the trial central charge of the four-dimensional theory obtained by compactifying the 6d $(2,0)$ theory of type $A_{N-1}$ on $\Sigma_\fg$, and is given in \eqref{a4d:twisted:unrefined}.
Recall that
\be
 \sum_{i = 1}^2 \Delta_i = 2 \pi + \epsilon \, , \qquad \sum_{i = 1}^2 \fs_i = 2 - 2 \fg \, ,
 \qquad \sum_{i = 1}^2 \ft_i = 0 \, .
\ee

\section{An index theorem for the twisted and mixed matrix models}
\label{sect:index theorem}

As we have seen in sections \ref{indexTTI} and \ref{indexSCI}, the large $N$ partition functions on $(S^2_\epsilon \times S^1)\times \Sigma_\fg$ satisfy the relation
\be
 \label{indexTh0}\log Z_{(S^2_\epsilon \times S^1)\times \Sigma_g}  =
 \ii \sum_{i=1}^2 \fs_i  \frac{ \partial  \cW_{(S^2_\epsilon \times S^1) \times \bR^2}}{\partial \Delta_i}  \, ,
\ee
where $\fs_i$,  with $\sum_{i=1}^2 \fs_i=2-2 \fg$, are the magnetic fluxes on $\Sigma_\fg$ and $\Delta_i$ the chemical potentials for global symmetries.
This is a five-dimensional generalization of the index theorem for the topologically twisted index discussed and proved for a large class of three- and four-dimensional gauge theories in \cite{Hosseini:2016tor,Hosseini:2016cyf}.%
\footnote{And verified in many other examples \cite{Hosseini:2016ume,Jain:2019euv,Jain:2019lqb}.} 

We have used constrained variables throughout this paper. This is convenient because  the entropy functional is  then a homogeneous function of degree one in $\Delta_i$ and $\epsilon$. The on-shell values of $\log Z_{(S^2_\epsilon \times S^1) \times \Sigma_\fg}$ and $ \cW$ are also homogeneous functions.  When written in terms of general chemical potentials and fluxes, \eqref{indexTh0} takes the form
\be
 \label{indexTh}
 \log Z  =  \ii (1- \fg)
 \left [\frac{2}{\pi}  \cW  +\sum_{i} \left (\frac{\fs_i}{1-  \fg} -\frac{\Delta_i}{\pi}\right ) \frac{ \partial  \cW}{\partial \Delta_i}
 -\frac{\epsilon}{\pi} \frac{\partial  \cW}{\partial \epsilon} \right ] ,
\ee
where, to avoid clutter in the notations, we have suppressed the reference to the manifold,
and we have generalized the formula to the case of an arbitrary number of global symmetries.
This clearly reduces to \eqref{indexTh0} when $ \cW$ is a homogeneous function of $\Delta_i$ and $\epsilon$ of degree two, as it is the case in this paper. 

It is worth  noticing that, due to the particular form of the differential operator appearing on the right hand side, \eqref{indexTh} is parameterization invariant and it can be used both for constrained and unconstrained variables. Indeed, it is a very simple exercise to check that, for the twisted index,  we obtain the same result for   \eqref{indexTh} computed using $ \cW(\Delta_1,\Delta_2,\epsilon)$ and constrained variables $\Delta_1+\Delta_2=2 \pi$ and $\fs_1+\fs_2=2 -2 \fg$ or, solving the constraint first, and using $ \cW_{{\rm unc}} (\Delta, \epsilon) \equiv   \cW(\Delta,2\pi -\Delta,\epsilon)$ and independent variables $\Delta\equiv \Delta_1$ and $\fs\equiv\fs_1$. Similarly, for the mixed index, we could use $ \cW(\Delta_1,\Delta_2,\epsilon)$ and constrained variables $\Delta_1+\Delta_2 =2 \pi+\epsilon$ and $\fs_1+\fs_2=2 -2 \fg$ or, by solving the constraint first, $ \cW_{{\rm unc}} (\Delta, \epsilon) \equiv   \cW(\Delta +\epsilon/2,2\pi -\Delta+\epsilon/2,\epsilon)$ and independent variables $\Delta\equiv \Delta_1-\epsilon/2$ and $\fs\equiv\fs_1$, with the same final result.

The relation \eqref{indexTh} was found to hold also in  \cite{Jain:2021sdp}, that uses a different $ \cW$ compared to this paper and we expect it to hold in general. We now provide an alternative derivation  that can be  generalized to other five-dimensional theories. 
We will prove \eqref{indexTh} using only a few facts about the general structure of the theories and of the large $N$ saddle point.  
We will assume that, in the large $N$ limit, the gauge variables $a$ and fluxes $\fn$ scale with the same positive power of $N$, and the eigenvalues are ordered by increasing imaginary part, {\it i.e.} $\im a_i > \im a_j$ for $i>j$.
We also assume that
\begin{enumerate}[label=(\roman*)]
 \item \label{cond:i}The matter content of the theory is vector-like: if $\rho(a)$ is a weight also $-\rho(a)$ is.
 \item \label{cond:ii} There is {\it long range forces cancellation} between vectors and hypers, which, in practice, translates into the requirement
 that $\sum_{\alpha \in G} \alpha( a ) - \sum_I \sum_{\rho_I \in \fR} \rho_I ( a )$, with $\alpha$ running over the roots and $\rho$ over the weights of the hypermultiplets,
 either cancels or it is subleading in the large $N$ limit.
 This condition is necessary to have a well-defined saddle point.
\end{enumerate}
These conditions are satisfied by  all the theories discussed in this paper and extend to a larger class of quivers that are relevant to the study five- or six-dimensional fixed points.

For the twisted index, the contribution of a hypermultiplet to the twisted superpotential is given in \eqref{W:TTI:general}, which,
using \eqref{asymp:Li(s)} and \eqref{prod:g2:g3}, in the large $N$ limit reads
\be\label{asymTTI}
  \cW (a, \fn, \Delta ; \ft )= -\sum_{s=0}^{3} \frac{1-(-1)^s}{2}\frac{\epsilon^{s-1}}{(2\pi)^s}  g_s(\pi(2-\fn-\ft)) g_{3-s}( a+\Delta )  \sign \left( \im a \right) ,
\ee
while for a vector we should take the opposite sign and set $\Delta=2 \pi$ and $\ft=2$.  To avoid clutter,  we omitted the gauge and flavor representation indices and we used $a$ as a shorthand for $\rho(a)$ or $\alpha(a)$ (and similarly for $\fn$, $\Delta$, and $\ft$). For the mixed index, the contribution of a hypermultiplet to the twisted superpotential is given by \eqref{W:SCI:hyper:first} and \eqref{ws:def}, which, using \eqref{asymp:Li(s)} and \eqref{invf}, can be written  in the large $N$ limit as 
\bea\label{asymSCI}
  \cW (a, \fn, \Delta ; \ft )=  \sum_{s=0}^{3} \frac{\epsilon^{s-1}}{2 (2\pi)^s} &\bigg [ g_s(\pi(r-\fn-\ft))   \\
  &+(-1)^s g_s(\pi(2-r-\fn-\ft)) \bigg ] g_{3-s}( a+\Delta )  \sign \left( \im a\right) ,
\eea
 with $r=1$, while for a vector we should take the opposite sign and set $\Delta=2 \pi$, $\ft=0$ and $r=2$.

\noindent \paragraph*{Proof of \eqref{indexTh}:} At large $N$ there is no contribution to the index from classical pieces and fundamental hypermultiplets.
Using the notations of sections \ref{sec:4} and \ref{sec:5}, the contribution of a hypermultiplet to the logarithm of the partition function integrand is
\be
 \ii \fm \frac{\partial  \cW^\cH}{\partial a} + \ii (\fs -1 +\fg)  \frac{\partial  \cW^\cH}{\partial \Delta} \, ,
\ee
where, in our notations $\fs_1 =\fs$,  we have suppressed the gauge indices in $\fm$ and $a$ and $\cW^\cH$ is one of the functions \eqref{asymTTI} or \eqref{asymSCI}. 
The contribution of the vector multiplet is similarly given by
\be
 \ii \fm \frac{\partial  \cW^\cV}{\partial a} + \ii (1 -\fg)  \frac{\partial  \cW^\cV}{\partial \Delta_g} \Big  |_{\Delta_g=2\pi} \, ,
\ee
where, for convenience, we introduced a fake chemical potential $\Delta_g$ for the vectors, which has to be set to $\Delta=2 \pi$ at the end of the computation. 
By setting $\fm=0$, we obtain the large $N$ contribution  to the index 
\be
 \label{pr1} \ii (\fs -1 +\fg)
 \frac{\partial  \cW^\cH}{\partial \Delta} + \ii (1 -\fg)  \frac{\partial  \cW^\cV}{\partial \Delta_g} \Big  |_{\Delta_g=2\pi}
 \equiv  \ii \fs  \frac{\partial  \cW}{\partial \Delta} + \ii (1-\fg) \left( -  \frac{\partial  \cW^\cH}{\partial \Delta} +\frac{\partial  \cW^\cV}{\partial \Delta_g} \Big  |_{\Delta_g=2\pi} \right) ,
\ee
where $\cW=\cW^\cH+\cW^\cV |_{\Delta_g=2\pi}$. We now promote the number $\pi$ to the independent variable ${\bm \pi}$,
using the same trick as in \cite{Hosseini:2016tor}, and we  write the previous expression as
\be
 \label{pr11} \ii \fs  \frac{\partial  \cW}{\partial \Delta} + \ii (1-\fg) \frac{\partial  \cW}{\partial {\bm \pi}} \, .
\ee
Indeed, observing that
\be
 \frac{\partial  \cW}{\partial {\bm \pi}} =
 \frac{\partial  \cW^\cH}{\partial {\bm \pi}}+\frac{\partial  \cW^\cV}{\partial {\bm \pi}}+ 2 \frac{\partial  \cW^\cV}{\partial \Delta_g}\Big  |_{\Delta_g=2{\bm \pi}} \, ,
\ee 
the difference between \eqref{pr11} and \eqref{pr1}, 
is given by
\be
 \label{pr111} \left ( \frac{\partial}{\partial \Delta} +\frac{\partial}{\partial {\bm \pi}}  \right )  \cW^\cH+ \left ( \frac{\partial}{\partial \Delta_g} +\frac{\partial}{\partial {\bm \pi}}  \right )  \cW^\cV \Big  |_{\Delta_g=2{\bm \pi}} \, .
\ee
All dependences on $\Delta$ and ${\bm \pi}$ are in the functions $g_{3-s}( a+\Delta )$ and the only potential nonvanishing contributions in \eqref{pr111} come from $s=0$ or $s=1$
\be
 \label{pr2} \left ( \frac{\partial}{\partial \Delta} +\frac{\partial}{\partial {\bm \pi}}  \right ) g_{3-s}( a+\Delta ) =  \frac{{\bm \pi}({\bm \pi}-a-\Delta)}{3} \delta_{s,0} -\frac{{\bm \pi}}{3}\delta_{s,1}  \, . \ee
Multiplying by the corresponding polynomials of $\fn$ in \eqref{asymTTI} and \eqref{asymSCI}, we see that \eqref{pr111} is  a linear function of $a$ and $\fn$, multiplied by $\sign (\im a)$.
The constant pieces vanish when summing over all gauge variables by assumption \ref{cond:i}. It is also easy to see, by direct inspection of \eqref{asymTTI} and \eqref{asymSCI}, that the linear terms in $a$ and $\fn$, either directly cancel
or vanish when summing over vectors and hypers by assumption \ref{cond:ii}. 
Now we are ready to complete the proof.
Given  the explicit form of \eqref{asymTTI} and \eqref{asymSCI}, $\cW$ is a homogeneous function of $a,\Delta, \epsilon,{\bm \pi}$ of degree two and therefore
\be
 2 \cW =   a \frac{\partial  \cW}{\partial a} +  \Delta \frac{\partial  \cW}{\partial \Delta} + \epsilon \frac{\partial  \cW}{\partial \epsilon}
 + {\bm \pi} \frac{\partial  \cW}{\partial {\bm \pi}} \, .
\ee
At the saddle point, $\frac{\partial  \cW}{\partial a}=0$ and we can rewrite \eqref{pr11} as
\be 
  \ii (1- \fg)  \left [\frac{2}{\pi}  \cW + \left (\frac{\fs}{1-  \fg} -\frac{\Delta}{\pi}\right ) \frac{ \partial  \cW}{\partial \Delta}  -\frac{\epsilon}{\pi} \frac{\partial  \cW} {\partial \epsilon} \right ] .
\ee
This completes the proof.

\section{Discussion and outlook}
\label{sect:conclusions}

In this paper we have demonstrated the large $N$ factorization properties of a class of five-dimensional supersymmetric partition functions with a view toward holographic applications to the microscopic counting for black holes in AdS. 

There is by now compelling evidence that the large $N$ saddle point of supersymmetric indices that are holographically dual to free energies of AdS black objects can be obtained by gluing elementary objects with the rules discussed in section \ref{sec:2}.  Black hole physics suggests that this should be the case in general. There are many examples in three and four dimensions and we have provided here other examples in five dimensions.
In some cases, this factorization can be proved by using the finite $N$ decomposition in holomorphic blocks, in other cases, as in this paper, a direct computation is needed. 

We mostly focused on the Seiberg theory and the $\cN=2$ super Yang-Mills theory in five dimensions, which can be used to describe the physics of black objects in AdS$_6\times_{w} S^4$ and AdS$_7\times S^4$, but  our results can be easily extended to other more complicated quiver theories. 

In this paper we considered the topologically twisted  index for manifolds of the type $(S^2_\epsilon \times S^1)\times \Sigma_\fg$. In the special case $\fg=0$, we could turn on a second refinement parameter and consider the partition function on $S^2_{\epsilon_1} \times S^2_{\epsilon_2} \times S^1$. It would be interesting to analyse whether  this partition function  factorizes or not. We expect that this is the case, but we have no evidence from holography, since we are not aware of AdS black object solutions with horizon $S^2\times S^2$ and two independent rotational parameters.  

One main open problem remains to justify the assumptions about the saddle point for the twisted and mixed index. Generalizing \cite{Hosseini:2018uzp,Crichigno:2018adf,Jain:2021sdp}, we assumed that the two partition functions localize at the solution of a set of Bethe ansatz equations, in analogy to similar three- and four-dimensional indices. This assumption led us to the correct results for  the entropy of the known black objects in AdS$_6\times_{w} S^4$ and AdS$_7\times S^4$. However, a slightly different assumption  leads to a different result where factorization does not hold  \cite{Jain:2021sdp}. It would be nice to have a first principle derivation of the large $N$ limit of the twisted and mixed index.

\section*{Acknowledgements}

SMH is supported in part by the STFC Consolidated Grant ST/T000791/1, WPI Initiative, MEXT, Japan at IPMU, the University of Tokyo and JSPS KAKENHI Grant-in-Aid (Early-Career Scientists), No.20K14462.
AZ is partially supported by the INFN,  and the MIUR-PRIN contract 2017CC72MK003. The work of IY was financially supported by the European Union's Horizon 2020 research and innovation programme under the Marie Sklodowska-Curie grant agreement No. 754496 - FELLINI.  

\appendix

\section{Refined $S^3_b \times S^2_\epsilon$ partition function}
\label{app:CJW}

In this appendix we derive the one-loop contribution of a hypermultiplet to the refined $S^3_b \times S^2_\epsilon$ partition function.
The contribution of a vector multiplet, as discussed in the main text, can be obtained by simply setting $\Delta_K = 2$ and $\ft_K = 2$.
Following \cite{Crichigno:2018adf}, we uplift on $S^1$ the one-loop contribution of a hypermultiplet to the $S^2_{\epsilon_b} \times S^2_{\epsilon_2}$ partition function
incorporating one unit of flux for the $\U(1)$ principal bundle over $S^2_{\epsilon_b}$, which gives rise to the Hopf fibration $S^1 \hookrightarrow S^3_b \hookrightarrow S^2_{\epsilon_b}$, and write
\bea
 \label{logZ:S3xS2:derivation:1}
  \log Z_{\Delta_K , \ft_K} ( \tilde a ) & =
  \prod_{n \in \bZ}
  \prod_{\ell_1 = - \frac{|\fm_1 + n|-1}{2}}^{\frac{|\fm_1+ n|-1}{2}}
  \prod_{\ell_2 = - \frac{|B^K|-1}{2}}^{\frac{|B^K|-1}{2}}
  \left( \tilde a + 1 - \Delta_K + n + \ell_1 \epsilon_b + \ell_2 \epsilon_2 \right)^{- \sign (\fm_1 + n)\sign (B^K)} \, ,
\eea
where $B^K = \fn + \ft_K - 1$. Using $k \equiv |\fm_1 + n|$, \eqref{logZ:S3xS2:derivation:1} can be put in the form
\be
 \label{logZ:S3xS2:derivation:2}
 \log Z_{\Delta_K , \ft_K} ( \tilde a ) = \prod_{k = 1}^\infty
 \prod_{\ell_1 = - \frac{k-1}{2}}^{\frac{k-1}{2}}
 \prod_{\ell_2 = - \frac{|B^K|-1}{2}}^{\frac{|B^K|-1}{2}}
 \left( \frac{ \tilde a + 1 - \Delta_K - k - \fm_1 + \ell_1 \epsilon_b + \ell_2 \epsilon_2}{\tilde a + 1 - \Delta_K + \ell - \fm_1 + \ell_1 \epsilon_b + \ell_2 \epsilon_2} \right)^{- \sign (B^K)} \, .
\ee
Upon the following change of variables
\bea
 a & \equiv \tilde a -\fm_1 \, , \qquad \qquad \quad \, \epsilon_b \equiv - 2 \? \frac{b-b^{-1}}{b+b^{-1}} \, , \\
 i & \equiv \frac12 ( k - 1 - 2 \ell_1 ) \, , \qquad j \equiv \frac12 ( k - 1 + 2 \ell_1 ) \, ,
\eea
\eqref{logZ:S3xS2:derivation:2} can then be recast as
\bea
 \label{logZ:S3xS2:derivation:2}
 \log Z_{\Delta_K , \ft_K} ( a ) & =
 \prod_{\ell_2 = - \frac{|B^K|-1}{2}}^{\frac{|B^K|-1}{2}}
 \prod_{i , j = 0}^\infty
 \left( - \frac{j b^{-1} + k b + Q + Q ( a + 1 - \Delta_K + \ell_2 \epsilon_2 ) } {j b + k b^{-1} + Q - Q ( a + 1 - \Delta_K + \ell_2 \epsilon_2 )} \right)^{\sign (B^K)} \\
 & = \prod_{\ell_2 = - \frac{|B^K|-1}{2}}^{\frac{|B^K|-1}{2}} S_2 \left( - \ii Q ( a + 1 - \Delta_K + \ell_2 \epsilon_2 ) | b \right)^{\sign (B^K)} \, .
\eea
This is exactly the contribution of a hypermultiplet to the refined $S^3_b \times S^2_\epsilon$ partition function \eqref{FS3bxS2e:main} after we identify $\ell_2 \equiv \ell$ and $\epsilon_2 \equiv \epsilon$.

\section{Asymptotic behavior of $q$-functions}

The double $(p,t)$-factorial is defined as
\be
 ( x ; p , t)_\infty = \prod_{i , j = 0}^{\infty} (1 - x p^i t^j ) \, , \quad \text{ with } \quad  | p | , | t | < 1 \, ,
\ee
where $x = e^{\ii a}$, $p = e^{- \ii \epsilon_1}$ and $t = e^{- \ii \epsilon_2}$.
Then, we can write
\bea
 \label{log(x;p,t):orig}
 \log ( x ; p , t)_{\infty}
 & = \sum_{i,j=0}^{\infty} \log ( 1 - x p^i t^j )
% = - \sum_{i,j=0}^{\infty} \sum_{k=1}^{\infty} \frac{( x p^i t^j )^k}{k} \\
% & = - \sum_{k=1}^{\infty } \frac{x^k}{k} \sum_{i=0}^{\infty } p^{i k} \sum_{j=0}^{\infty} t^{j k}
 = - \sum_{k=1}^{\infty} \frac{1}{( 1-p^k )( 1 - t^k )} \frac{x^k}{k} \, .
\eea
As $\epsilon_{1,2} \to 0$, the double $(p,t)$-factorial has the following asymptotic expansion
\be
 \label{log(x;p,t):GW}
 \log ( x ; p , t)_{\infty} = \sum_{s , l = 0}^{\infty} ( - \ii )^{s +2 l}  \? H_{s, l} \? ( \epsilon_1  \epsilon_2 )^{l -1} ( \epsilon_1 + \epsilon_2 )^s \Li_{3 - s -2 l} ( x ) \, ,
\ee
where the coefficients $H_{s,l}$ can be determined by comparing \eqref{log(x;p,t):orig} and \eqref{log(x;p,t):GW}, as $p,t \to 1$ $(\epsilon_{1,2} \to 0)$.

The $q$-Pochhammer is defined as
\be
 ( x ; q)_\infty = \prod_{i = 0}^{\infty} (1 - x q^i) \, , \quad \text{ with } \quad | q | < 1 \, ,
\ee
where $x = e^{\ii a}$ and $q = e^{\ii \epsilon}$.
Then, we can write
\bea
 \label{log(x;q):orig}
 \log ( x ; q)_{\infty}
 & = \sum_{i=0}^{\infty} \log ( 1 - x q^i )
% = - \sum_{i = 0}^{\infty} \sum_{k=1}^{\infty} \frac{( x q^i )^k}{k} \\
% & = - \sum_{k=1}^{\infty} \frac{x^k}{k} \sum_{i=0}^{\infty } q^{i k}
 = - \sum_{k=1}^{\infty} \frac{1}{( 1- q^k )} \frac{x^k}{k} \, .
\eea
As $\epsilon \to 0$, the $q$-Pochhammer symbol has the following asymptotic expansion
\be
 \label{log(x;q):expand}
 \log ( x ; q)_{\infty} = \sum_{s = 0}^\infty \frac{( \ii \epsilon )^{s-1} B_s}{s!} \? \Li_{2-s}(x) \, ,
\ee
where $B_s$ is the $s$th Bernoulli number.

\section{Supersymmetry conventions}
In this appendix, we describe our conventions for spinors, supersymmetry transformations, and rigid supergravity.

\subsection{\label{sec:Spinor-conventions}Spinor conventions}

We describe $\mathcal{N}=1$ (minimal) Euclidean supersymmetry in five dimensions.%
\footnote{This is usually referred to as $\mathcal{N}=2$ supersymmetry in the
supergravity literature.} The Euclidean spin group is the compact symplectic group $\text{Spin}\left(5\right)\simeq\text{USp}\left(4\right)$.
The minimal spinor has $4$ complex, or $8$ real, components and
sits in the pseudo-real representation $4$ of $\text{USp}\left(4\right)$.
A complex spinor $\xi$ may be usefully represented by a $4\times2$
matrix $\xi_{\alpha I}$, where $\alpha = 1,\ldots,4 $
and $I = 1 ,2$. This presentation is useful because
the action of the $\SU (2 )$ R-symmetry is manifest -- it
acts on the $I$ index in the fundamental representation. Such a spinor
has twice the minimal number of components. It is a complexification
of the minimal spinor, which it is necessary to introduce in order
to write the most general transformations of the Euclidean superalgebra. 

Define the usual Pauli matrices by
\bea
\sigma_{1}=\begin{pmatrix}0 & 1\\
1 & 0
\end{pmatrix},\qquad\sigma_{2}=\begin{pmatrix}0 & -\ii\\
\ii & 0
\end{pmatrix},\qquad\sigma_{3}=\begin{pmatrix}1 & 0\\
0 & -1
\end{pmatrix}\,.
\eea
The 5d Euclidean Clifford algebra is generated by matrices ${\left(\Gamma_{a}\right)_{\beta}}^{\alpha}$
\bea
\Gamma_{1} & = \sigma_{2}\otimes\sigma_{1} \, ,\qquad
\Gamma_{2} = \sigma_{2}\otimes\sigma_{2} \, , \qquad
\Gamma_{3} = \sigma_{2}\otimes\sigma_{3} \, , \qquad
\Gamma_{4} = \sigma_{1}\otimes\mathbbm{1}_{2}\,,\\
\Gamma_{5} & =\Gamma_{1}\Gamma_{2}\Gamma_{3}\Gamma_{4}=\text{diag} (1,1,-1,-1 )\,.
\eea
We also define a charge conjugation matrix
\be
\mathcal{C}=-\ii\mathbbm{1}_{2}\otimes\sigma_{2} \, .
\ee
These satisfy
\bea
 \{ \Gamma_{a},\Gamma_{b} \} & = 2 \delta_{ab} \, , \qquad \qquad \quad
 \Gamma_{a}^{\dagger}=\Gamma_{a} \, , \\
 \mathcal{C} \Gamma_{a}\mathcal{C}^{-1} & = \Gamma_{a}^{\text{T}} \, , \qquad \qquad \qquad \!
 \mathcal{C} = \mathcal{C}^{*} = - \mathcal{C}^{\text{T}} \, , \qquad
 \mathcal{C}^{*}\mathcal{C} = - 1\, , \\
 ( \mathcal{C}\Gamma_{a} )^{\text{T}} & = - \mathcal{C}\Gamma_{a} \, , \qquad
 ( \mathcal{C}\Gamma_{ab} )^{\text{T}} = \mathcal{C}\Gamma_{ab} \, , \\
 \Gamma_{abc} & = \frac{1}{2} (\Gamma_{a}\Gamma_{bc}+\Gamma_{bc}\Gamma_{a} ) \, .
\eea

Define the index raising operator $\varepsilon^{IJ}=-\varepsilon^{JI}$,
such that $\varepsilon^{IJ}=1$ and $\varepsilon^{IJ}\varepsilon_{JK}={\delta^{I}}_{K}$
defines the inverse. We define the Majorana conjugate spinor as
\be
\label{eq:majorana_conjugate}
\bar{\xi}^{\alpha I}\equiv\mathcal{C}^{\alpha\beta}\xi_{\beta J}\varepsilon^{IJ} \, .
\ee
In the action of the Euclidean supersymmetry algebra, one uses only
the Majorana type conjugate, in both the action and the supersymmetry
transformations. We will also use the following convention for spinor bilinears
\be
\bar{\xi_{1}}\xi_{2}\equiv\left(\mathcal{C}\xi_{1}^{I}\right)^{\text{T}}\xi_{2I} =
 \left(\xi_{1I} \right)^{\text{T}}\mathcal{C}\xi_{2}^{I} \, , \qquad
 \bar{\xi_{1}}\gamma^{a}\xi_{2}\equiv\left(\mathcal{C}\xi_{1}^{I}\right)^{\text{T}}\gamma^{a}\xi_{2I} =
 \left(\xi_{1I}\right)^{\text{T}}\mathcal{C}\gamma^{a}\xi_{2}^{I} \,.
\ee
We also have the Fierz identities
\bea
\xi_{1}^{I}\left(\bar{\xi}_{2I}\psi\right) & = - \xi_{2}^{I}\left(\bar{\xi}_{1I}\psi\right)+\frac{1}{2}\psi\left(\bar{\xi}_{2I}\xi_{1}^{I}\right)+\frac{1}{2}\Gamma^{m}\psi\left(\bar{\xi}_{2I}\Gamma_{m}\xi_{1}^{I}\right) , \\
\Gamma^{m}\xi_{1}^{I}\left(\bar{\xi}_{2I}\psi\right) & = - \xi_{1}^{I}\left(\bar{\xi}_{2I}\Gamma^{m}\psi\right)-\xi_{2}^{I}\left(\bar{\xi}_{1I}\Gamma^{m}\psi\right)
- \Gamma^{m}\xi_{2}^{I}\left(\bar{\xi}_{1I}\psi\right)-\Gamma^{m}\psi\left(\bar{\xi}_{1I}\xi_{2}^{I}\right)-\psi\left(\bar{\xi}_{1I}\Gamma^{m}\xi_{2}^{I}\right) .
\eea

A complex spinor $\xi$ may satisfy a symplectic Majorana condition:
a reality condition, compatible with the spin and R-symmetry groups,
which eliminates half of the components. Define the Dirac conjugate
of a spinor to be
\be
\left(\bar{\xi}^{\text{D}}\right)^{\alpha I}\equiv\left(\xi^{*}\right)^{\alpha I}.
\ee
The symplectic Majorana condition is 
\be
\left(\bar{\xi}^{\text{D}}\right)^{\alpha I}=\bar{\xi}^{\alpha I} \, ,
\label{eq:symplectic_majorana_condition}
\ee
where the $\bar{\xi}^{\alpha I}$ is the Majorana conjugate, see \eqref{eq:majorana_conjugate}.

\subsection{\label{sec:Supersymmetry-conventions}Rigid minimal 5d conformal
supergravity}

We will mostly follow the 5d conformal supergravity conventions of
\cite{deWit:2017cle}. From this starting point, one can derive by
partial gauge fixing the 5d supergravity used in \cite{Hosomichi:2012ek},
which was used to define the twisted backgrounds in \cite{Hosseini:2018uzp}.
The conformal supergravity offers a more flexible framework for finding
rigid backgrounds. The supergravity fields of conformal supergravity
are contained in the Weyl multiplet. We refer the reader to \cite{deWit:2017cle}
for details about the relevant Weyl multiplet and its supersymmetry
transformations. We summarize the aspects of the Weyl multiplet needed
in order to construct rigid backgrounds below. In the notation of
\cite{deWit:2017cle}, $\mu,\nu,\ldots$ are spacetime indices, $a,b,\ldots$
are tangent space indices, and $i,j,\ldots$ are $\SU (2 )$ R-symmetry indices. 

We begin by simplifying the transformations of the Weyl multiplet
by taking k-gauge, setting the dilatational gauge field to zero
\be
 b_{\mu}\to 0 \, .
\ee
The remaining bosonic fields in the 5d Weyl multiplet are 
\be
 {e_{\mu}}^{a}\,,\quad{{V_{\mu}}_{i}}^{j}\,,\quad T_{ab}\,,\quad D \, .
 \label{eq:bosonic_supergravity_fields}
\ee
The independent fermionic fields are the gravitino $\psi_{\mu}^{i}$
and the dilatino $\chi^{i}$. In order to bring the notation more
in line with the one used in \cite{Hosseini:2018uzp}, we will rename
$V_{\mu}$ to $2A_{\mu}^{\left(\text{R}\right)}$. We will also replace
the supersymmetry transformation parameters in \cite{deWit:2017cle}
as follows
\be
\epsilon^{\text{there}} = e^{\pi\ii/4} \xi \, , \qquad
\eta^{\text{there}} = - 2 \ii e^{\pi \ii / 4} \tilde{\xi} \, ,
\ee
and rotate the gravitino
\be
\psi^{\text{there}}_{\mu} = 2 e^{\pi \ii / 4} \psi_{\mu} \, .
\ee
We will also use $I,J,\ldots$ as R-symmetry indices. Note that the
gauge fixed supergravity used in \eg\;\cite{Hosomichi:2012ek} includes
an extra symmetric bosonic field $t^{IJ}$ such that
\be
 \tilde{\xi}_{I}={t_{I}}^{J}\xi_{J} \, .
 \label{eq:Gauge_fixed_sugra_t}
\ee

The gravitino variation, in the matrix spinor notation of the previous
section, reads
\be
 \delta\psi_{\mu}=\mathcal{D}_{\mu}\xi+\frac{\ii}{4}T_{ab}\left(3\Gamma^{ab}\Gamma_{\mu}-\Gamma_{\mu}\Gamma^{ab}\right)\xi-\gamma_{\mu}\tilde{\xi} \, ,
\ee
where
\be
\mathcal{D}_{\mu}\xi\equiv\partial_{\mu}\xi+\frac{1}{4}{\omega_{\mu}}^{ab}\Gamma_{ab}\xi+\xi\left(A_{\mu}^{\left(\text{R}\right)}\right)^{\text{T}} \, .
\ee
The variation of the dilatino can be found in \cite{deWit:2017cle}.
A rigid background is a bosonic fixed point of the supersymmetry transformations.
Equivalently, it is a solution to the equations 
\be
\delta_{\xi,\tilde{\xi}}\psi_{\mu}^{i}=\delta_{\xi,\tilde{\xi}}\chi^{i}=0 \, ,
\ee
for some spinors $\xi$, $\tilde{\xi}$ and some configuration of the
fields in \eqref{eq:bosonic_supergravity_fields}. The spinor $\tilde{\xi}$
can always be solved for as
\be
 \tilde{\xi} = \frac{1}{5}\Gamma^{\mu} \left(\mathcal{D}_{\mu}\xi+\frac{\ii}{4}T_{ab}\left(3\Gamma^{ab}\Gamma_{\mu}-\Gamma_{\mu}\Gamma^{ab}\right)\xi \right) .
\ee

Note that other derivations of the 5d superconformal tensor calculus,
such as \cite{Bergshoeff2001} and its followups or \cite{Fujita:2001kv}
and its followups, have a coupling to the antisymmetric tensor field
that is seemingly incompatible with the one used here. de Wit \etal\;explain that this is due to different conventions for the $F (4 )$ algebra.
Moreover, the various versions apparently all produce the
same Poincare supergravity when a vector multiplet is used as a compensator.
There is also an alternative Weyl multiplet called the dilaton Weyl
multiplet which is derived for instance in \cite{Bergshoeff2001}
and contains different auxiliary fields and different transformations.

\subsection{Matter multiplet transformations}

The superalgebra generated by a single Killing spinor pair $\xi$, $\tilde{\xi}$
includes the following bosonic transformations 
\begin{enumerate}
\item An infinitesimal diffeomorphism with parameter $\ii v$, where the
vector $v^{\mu}$ is given by
\be
v^{\mu}\equiv\bar{\xi}\gamma^{\mu}\xi\,.
\ee
\item An infinitesimal R-symmetry transformation acting as the following
matrix on fundamental $\SU (2 )_{R}$ indices
\be
 \Lambda_{IJ}\equiv6\ii\bar{\xi}_{(I}\tilde{\xi}_{J)} \, .
\ee
\item A gauge transformation with parameter $s\sigma-\ii v^{\mu}A_{\mu}$, where $s\equiv\bar{\xi}\xi\, .$
\item A Weyl transformation with parameter $\omega\equiv-2\ii\bar{\xi}\tilde{\xi}$.
\end{enumerate}
We will continue using the notation for matter fields used in \cite{Hosseini:2018uzp}.
The matter fields in \cite{deWit:2017cle} have therefore been renamed
as follows
\bea
 W_{M}^{\text{there}} & = A_{m} \, , \qquad \qquad
 \sigma^{\text{there}} = \sigma \, ,\qquad 
 \Omega_{i}^{\text{there}} = e^{\pi\ii/4} \lambda_{I} \, , \\
 Y_{ij}^{\text{there}} & = - \frac{\ii}{2}D_{IJ} \, , \qquad
 A_{i}^{\text{there}} = q_{I} \, , \qquad
 \zeta^{\text{there}} = e^{3\pi\ii/4}\psi \, ,
\eea
which includes an obvious renaming of the spacetime and R-symmetry indices.

With these definitions, the transformations for vector multiplets
in a rigid bosonic background are 
\bea
 \delta A_{m} & = \ii\bar{\xi}_{I}\Gamma_{m}\lambda^{I} \, , \qquad
 \delta \sigma = - \bar{\xi}_{I}\lambda^{I} \, , \\
 \delta \lambda_{I} & = - \frac{1}{2}\Gamma^{mn}\left(F_{mn}-4\sigma T_{mn}\right)\xi_{I}-\ii D_{m}\sigma\Gamma^{m}\xi_{I}-\ii D_{IJ}\xi^{J}-2\ii\sigma\tilde{\xi}_{I}\,,\\
 \delta D_{IJ} & = - 2 \bar{\xi}_{(I}\gamma^{a}D_{a}\lambda_{J)}-2\left[\sigma,\bar{\xi}_{(I}\lambda_{J)}\right]+2\tilde{\xi}_{(I}\lambda_{J)}+\ii\bar{\xi}_{(I}\Gamma^{ab}T_{ab}\lambda_{J)} \, .
\eea
The vector multiplet is off-shell closed.

The transformations for hypermultiplets in a rigid bosonic background are
\bea
 \delta q_{I}^{A} = - 2 \ii \bar{\xi}_{I}\psi^{A} \, , \qquad
 \delta \psi^{A} = \Gamma^{m}\xi_{I}D_{m}q^{AI}+\ii\xi_{I}\left(\sigma q\right)^{AI}+3\tilde{\xi}_{I}q^{AI} \, .
\eea
Indices $A,B,\ldots$ denote the pseudo-real gauge/flavor symmetry
representation to which the hypermultiplets belong. The invariant
antisymmetric form for this representation is denoted $\Omega_{AB}$,
with $\Omega_{AB}\Omega^{BC}={\delta_{A}}^{C}$. The reality condition on the squark is 
\be
 \left(q^{*}\right)_{A}^{I}=\varepsilon^{IJ}\Omega_{AB}q_{J}^{B} \, .
\ee

The transformations for the hypermultiplet are only closed on-shell.
Off-shell closure can be achieved, for a specific $\xi$, $\tilde{\xi}$
pair by introducing an auxiliary field $F^{AI}$, and a spinor $\hat{\xi}_{I}$ such that 
\be
\bar{\xi}_{I}\hat{\xi}^{J} = 0 \, , \qquad\mathcal{L}_{v}\hat{\xi}+\hat{\xi}\Lambda^{\text{T}} = 0 \, .
\ee
One can check that the solution space for these equations is two-dimensional.
Let 
\be
\hat{\delta}_{\xi,\tilde{\xi}}\psi^{A}=\delta_{\xi,\tilde{\xi}}\psi^{A}+\hat{\xi}_{I}F^{AI}\,,
\ee
and define 
\be
 \delta \big(\hat{\xi}_{I}F^{AI} \big) = \delta_{\xi,\tilde{\xi}}^{2} \? \psi^{A}
 + \ii \big( \mathcal{L}_{v}\psi^{A}+\psi^{A}\Lambda^{\text{T}} \big) \, ,
\ee
then $\hat{\delta}_{\xi,\tilde{\xi}}$ is off-shell closed and satisfies
the correct algebra.

\bibliographystyle{ytphys}

\bibliography{5dPFs}

\end{document}